\documentclass[11pt,a4paper]{article}
\usepackage[dvipdfmx]{graphicx}
\usepackage{bmpsize}
\usepackage{jheppub}

\usepackage{comment}

\newcommand{\al}[1]{\begin{align}#1\end{align}}

\newcommand{\bp}{\begin{pmatrix}}
\newcommand{\ep}{\end{pmatrix}}
\newcommand{\bb}{\begin{bmatrix}}
\newcommand{\eb}{\end{bmatrix}}

\newcommand{\beq}{\begin{equation}}
\newcommand{\eeq}{\end{equation}}
\newcommand{\bea}{\begin{eqnarray}}
\newcommand{\eea}{\end{eqnarray}}

\newcommand{\fulltoday}{\number\day\space \ifcase\month\or
    January\or February\or March\or April\or May\or June\or
    July\or August\or September\or October\or November\or December\fi
    \space\number\year}

 \makeatletter
    
    \@addtoreset{equation}{section}
  \makeatother

\allowdisplaybreaks[3]

\usepackage{calc}
\newcounter{hours}\newcounter{minutes}
\makeatletter                   
\renewcommand*{\thehours}{\two@digits\c@hours}
\renewcommand*{\theminutes}{\two@digits\c@minutes}
\makeatother

\newcommand{\hats}{\widehat{s}}
\newcommand{\hatt}{\widehat{t}}
\newcommand{\hatu}{\widehat{u}}

\definecolor{mygrn}{rgb}{0,0.5,0}

\title{\boldmath Phenomenology of flavorful composite vector bosons in light of $B$ anomalies}

\author[a,b]{Shinya Matsuzaki,}
\author[c,1]{Kenji Nishiwaki,\note{Corresponding author.}}
\author[d]{Ryoutaro Watanabe,}

\affiliation[a]{Institute for Advanced Research, Nagoya University, Nagoya 464-8602, Japan}
\affiliation[b]{Department of Physics, Nagoya University, Nagoya 464-8602, Japan}
\affiliation[c]{School of Physics, Korea Institute for Advanced Study~(KIAS), Seoul 02455, Republic of Korea}
\affiliation[d]{Physique des Particules, Universit\'e de Montr\'eal, \\ C.P. 6128, succ.~centre-ville, Montr\'eal, QC, Canada H3C 3J7}

\emailAdd{synya@hken.phys.nagoya-u.ac.jp}
\emailAdd{nishiken@kias.re.kr}
\emailAdd{watanabe@lps.umontreal.ca}

\abstract{
We analyze the flavor structure of composite vector bosons arising in a model of vectorlike technicolor -- often called hypercolor (HC) --  with eight flavors that form a one-family content of HC fermions. 
Dynamics of the composite vector bosons, referred to as HC $\rho$ in this paper, are formulated together with HC pions by the hidden local symmetry (HLS), in a way analogous to QCD vector mesons.  
Then coupling properties to the standard model (SM) fermions, which respect the HLS gauge symmetry, are described in a way that 
couplings of the HC $\rho$s to the left-handed SM quarks and leptons are given by a well-defined setup as taking the flavor mixing structures into account. 
Under the present scenario, we discuss significant bounds on the model from electroweak precision tests, flavor physics, and collider physics. 
We also try to address $B$ anomalies in processes such as $B \to K^{(*)} \mu^+\mu^-$ and $B \to D^{(*)}\tau\bar\nu$, recently reported by LHCb, Belle, (ATLAS, and CMS in part.)
Then we find that the present model can account for the anomaly in $B \to K^{(*)} \mu^+\mu^-$ consistently with the other constraints while it predicts no significant deviations in $B \to D^{(*)}\tau\bar\nu$ from the SM, 
which can be examined in the future Belle~II experiment. 
The former is archived with the form $C_9 = -C_{10}$ of the Wilson coefficients for effective operators of $b \to s \mu^+\mu^-$, which has been favored by the recent experimental data. 
We also investigate current and future experimental limits at the Large Hadron Collider (LHC) 
and see that possible collider signals come from dijet and ditau, or dimuon resonant searches for the present scenario with TeV mass range. 
To conclude, the present $b \to s \mu^+\mu^-$ anomaly is likely to imply discovery of new vector bosons in the ditau or dimuon channel in the context of the HC $\rho$ model.
Our model can be considered as a UV {completion} of conventional $U(1)'$ models. 
}

\subheader{Report number: KIAS-P17038, UdeM-GPP-TH-17-257}
\arxivnumber{1706.01463}

\begin{document}

\maketitle
\flushbottom

\section{Introduction}
%%%%%%%%%%%%%%%%%%%%%%%%%%%%%%%%%%%%%%%
%%%%%%%%%%%%%%%%%%%%%%%%%%%%%%%%%%%%%%%  
The CERN Large Hadron Collider (LHC) has discovered~\cite{Aad:2012tfa,Chatrchyan:2013lba} a Higgs boson, which is the last element to compensate the particle content predicted in the standard model (SM). 
Indeed, the SM explains particle phenomenologies in good agreement with experiments that include the Higgs coupling measurements. 
Though the structure of the electroweak symmetry breaking (EWSB) in the SM, i.e., the origin of the negative-mass squared for the Higgs field, is still mysterious, 
the present LHC data on the EW interactions and Higgs coupling properties are likely to imply that new physics beyond the SM would have no direct correlation with the dynamical structure of the EWSB. 
This might be why it is hard to see new particles in a TeV scale range. 
{Nevertheless, it would still be worth investigating a bulk of new physics at TeV scale as long as it is reachable at the LHC and insensitive to EW precision variables.}

One interesting key to access such a new physics paradigm involves a vectorlike scenario, in which new particles are charged vector-likely under the SM gauges so that it is fairly insensitive to the EWSB structure. 
Among vectorlike models, one of viable candidates would be embedded in a strongly coupled sector,  
often called hypercolor (HC), or vectorlike-confinement, or chiral-symmetric technicolor~\cite{Kilic:2009mi,Pasechnik:2013bxa,Lebiedowicz:2013fta,Pasechnik:2014ida}.
This class of scenarios predicts variety of new phenomena in TeV scale physics such as presence of new composite particles 
which can be searched at the LHC~\cite{Lebiedowicz:2013fta} and affect the Higgs coupling property~\cite{Pasechnik:2013bxa}. 
Dark matter candidates (HC composite scalars or fermions) are also involved in some models~(see {\it e.g.,} Refs.~\cite{
Pasechnik:2014ida,Appelquist:2014jch,Hochberg:2014kqa,Appelquist:2015yfa,Appelquist:2015zfa,Carmona:2015haa,Hochberg:2015vrg,Harigaya:2016rwr,Kamada:2016ois,Kribs:2016cew,Forestell:2016qhc})~\footnote
{Other HC models have been proposed in a different context, where the origin of the EWSB can be addressed by dynamically induced bosonic seesaw mechanisms~\cite{Calmet:2002rf,Kim:2005qb,Haba:2005jq,Antipin:2014qva,Haba:2015lka,Haba:2015qbz,Ishida:2016ogu,Ishida:2016fbp,Ishida:2017ehu,Haba:2017wwn,Haba:2017quk}. 
See also Refs.~\cite{Hur:2007uz,Hur:2011sv,Heikinheimo:2013fta,Holthausen:2013ota,Abel:2013mya,Kubo:2014ova,Antipin:2014mga,Carone:2015jra,Kubo:2015cna,Hatanaka:2016rek,Molgaard:2016bqf,Abel:2017ujy}
for dynamical EWSB triggered by strongly-coupled hidden sectors. 
}.

In this paper, we begin with extending the vectorlike technicolor (or HC) models in the market~\cite{Kilic:2009mi,Pasechnik:2013bxa,Lebiedowicz:2013fta,Pasechnik:2014ida} 
by introducing eight HC-fermion flavors that form one-family content.  
We concentrate particularly on the flavorful composite vectors embedded in the $SU(8)$-HC flavor 63-plet, which includes color-octet, -triplet (``leptoquark'') and -singlet (similar to $Z', W'$) composite vectors. 
Hereafter, we refer to the composite vectors and pseudoscalars as HC $\rho$ and HC $\pi$, respectively. 
In this scenario, the EWSB is triggered in the same way as in the SM, namely, by the $SU(2)_W$-fundamental Higgs doublet.

Dynamics of composite vectors is formulated {by use of} the hidden local symmetry (HLS)~\cite{Bando:1984ej,Bando:1985rf,Bando:1987ym,Bando:1987br,Harada:2003jx} 
-- the method established in addressing QCD vector mesons 
including off-shell properties and couplings to the external gauge fields --  
based on an extension of nonlinear realization of ``chiral'' $SU(8)_L \times SU(8)_R$ symmetry for the HC fermions whose couplings to the SM gauge fields are vector-like. 
In this case, the SM gauge symmetries $SU(3)_c \times SU(2)_W \times U(1)_Y$ are realized as unbroken symmetries (diagonal subgroup), 
come down from a spontaneous breaking of the extended $SU(3)_{c'} \times SU(2)_{W'} \times U(1)_{Y'} \times [SU(8)]_{\rm HLS}$ gauge symmetries. 
The spontaneously broken gauge-degree of freedom of $SU(8)$ reflects the presence of 63 massive composite gauge bosons (HC $\rho$'s), which in part arise as mixtures of the HLS and the SM gauge bosons.             
Then the couplings of the $SU(8)$-63 plet composite vectors to the SM fermions and SM gauge bosons are unambiguously determined by the HLS-gauge invariance. 
Since some of composite vectors can mix with the SM gauge bosons, the couplings to the SM fermions are limited by electroweak precision tests as fundamental requirements. 
As we will see, the present scenario is safe for flavor universal EW {measurements} such as the oblique parameters, while is severely constrained from flavor-dependent processes, 
in particular, from the forward-backward asymmetry of tau lepton $A_{\rm FB}^{(0,\tau)}$ and $Z$ boson decay to bottom quark pair $R_b$.

We then turn to flavor specific phenomena and investigate constraints from current experiments for flavor and collider physics on {the parameter space of} the HC $\rho$ couplings to the SM fermions. 
The flavorful HC $\rho$ couplings are severely constrained so that the third-generation couplings and their small mixings to the second-generation fermions are only viable. 
In particular, we will see whether the present scenario can accommodate recent anomalies which have been seen in several $B$ meson decays such as $B \to K^{(*)} \mu^+\mu^-$ and $B \to D^{(*)}\tau\bar\nu$. 
A vast amount of works have been made in the last several years on the topics of anomalies in the $b \to s \mu^+ \mu^-$ data 
(see, {\it e.g.}, Refs.~\cite{Descotes-Genon:2013vna,Descotes-Genon:2013wba,Altmannshofer:2013foa,Hiller:2014yaa,Altmannshofer:2014rta} for earlier works)~\footnote
{Explanations by contributions of composite bosonic particles in the context of composite Higgs scenarios were also reported~\cite{Gripaios:2014tna,Niehoff:2015bfa,Niehoff:2015iaa,Carmona:2015ena,Barbieri:2016las,DAmico:2017mtc}. 
}.
Deviations of experimental data in the {ratios} $R_{D^{(*)}} = \mathcal B(B \to D^{(*)}\tau\bar\nu) / \mathcal B(B \to D^{(*)}\ell\bar\nu)$ have also been studied in several specific models 
(see, {\it e.g.}, Refs.~\cite{Sakaki:2013bfa,Dumont:2016xpj,Datta:2012qk,Celis:2012dk,Crivellin:2013wna,Dorsner:2013tla,Freytsis:2015qca,Deshpand:2016cpw,Ivanov:2016qtw,Altmannshofer:2017poe} for earlier works.) 
We emphasize that the present HC scenario naturally includes various and flavorful new vector bosons that can address $B$ physics, as a consequence of the presence of the strong HC sector at high energy scale. 
This is a crucial difference from other low-energy effective models (such as a $U(1)'$ model) in which new vector bosons are introduced without any dynamical reason. 
%added
In particular, we will see that the present HC model can be considered as a UV completion of conventional $U(1)'$ models.

The implications to the collider physics at LHC are also addressed so that the HC $\rho$ mesons with mass of TeV scale can be consistent with the current 13 TeV bounds from {di-jet, di-tau and di-muon searches}. 
Finally, we discuss discovery and/or exclusion potential for the HC $\rho$ mesons at future LHC experiments with higher luminosity 
and then see that the future LHC will be sensitive enough to discover or exclude those TeV mass HC $\rho$ mesons as well as to confirm (in)consistency with the $B$ anomalies.

This paper is structured as follows: 
In Sec.~\ref{model:description} we start with introducing the one-family model of the HC and then derive the low-energy effective Lagrangian including the HC pions and HC rho mesons with respect to the HLS formulation.  
Then we describe the HC rho couplings to the SM fermions and find that there are two types allowed by the HLS gauge invariance --  
one is of direct coupling type (generically dependent on fermion flavors) while another is of flavor-universal type (arising from the mixing between the HC rho mesons and the SM gauge bosons). 
After a brief introduction of the HC pion coupling forms as well, we summarize phenomenological features for the present HC rho model in Sec.~\ref{pheno} -- 
mass splittings of the HC rho mesons, oblique corrections of the EW sector, and flavor-dependent corrections to $A_{\rm FB}^{(0,\tau)}$ and $R_b$. 
These provide fundamental requirements for the present model. 
In Sec.~\ref{sec:flavor_issues}, we discuss the flavor physics issues to give a list of flavor observables relevant for the present scenario of the model. 
The current flavor bounds are then placed on the HC rho mass and the coupling strength to obtain the allowed parameter space. 
Finally, in Sec.~\ref{collider}, we show the HC rho sensitivity at the LHC experiments by taking into account resonance searches from {di-jet, di-tau and di-muon} channels. 
Sec.~\ref{summary} is devoted to summary of this work.
We also provide a couple of Appendices to show details of computations leading to the results described in the text, and some details related to HC pion physics which are subdominant in the present study.

%%%%%%%%%%%%%%%%%%%%%%%%%%%%%%%%%%%%%%%
%%%%%%%%%%%%%%%%%%%%%%%%%%%%%%%%%%%%%%%
%%%%%%%%%%%%%%%%%%%%%%%%%%%%%%%%%%%%%%%
\section{Model Description}
\label{model:description}
%%%%%%%%%%%%%%%%%%%%%%%%%%%%%%%%%%%%%%%
%%%%%%%%%%%%%%%%%%%%%%%%%%%%%%%%%%%%%%%
%%%%%%%%%%%%%%%%%%%%%%%%%%%%%%%%%%%%%%%

\subsection{One-family model of HC} 
%%%%%%%%%%%%%%%%%%%%%
%%%%%%%%%%%%%%%%%%%%%
In this section we introduce the one-family model of the HC {and outline the HC model scenario}. 
The HC gauge group is assumed to be $SU(N_{\rm HC})$ with $N_{\rm HC} \ge 3$, 
where the 8 HC fermions $(F)$ belong to the fundamental representation. 
Among the HC fermions, 
the HC-quarks $Q_{L/R}=(U,D)^T_{L/R}$ carry the QCD charge of $SU(3)_c$ 
and the same EW charges as those of the SM-quark doublets, 
while the HC-leptons $L_{L/R}=(N,E)_{L/R}^T$ are charged only 
under the EW gauges which are the same as those of the SM-lepton doublets. 
The summary of the charge assignment is listed in Table~\ref{tab:1}.  
Thus the HC sector possesses the approximate 
global ``chiral'' $U(8)_{F_L} \times U(8)_{F_R}$ symmetry (explicitly 
broken in part by the SM gauging {and possible vectorlike fermion masses}), among which the $U(1)_{F_A}$ part 
is explicitly broken by the axial anomaly in the same manner as in 
the QCD case.

%%%%%%%%%%%%%% Table [begin] %%%%%%%%%%%%%%
\begin{table}[t] 
\begin{tabular}{|c||c|c|c|c|}
\hline 
\hspace{20pt}&\hspace{10pt} {$SU(N_{\rm HC})$} \hspace{10pt}
&\hspace{10pt} $SU(3)_c$ \hspace{10pt} 
&\hspace{10pt} {$SU(2)_W$}  \hspace{10pt}
&\hspace{10pt} $U(1)_Y$  \hspace{10pt}  \\
\hline \hline 
$Q_{L/R} =  
\left( 
\begin{array}{c}
U \\ 
D
\end{array}
\right)_{L/R} 
$ 
 & $N_{\rm HC}$ & $\mathbf{3}$ & $\mathbf{2}$ & $1/6$ \\ 
\hline 
$L_{L/R} =  
\left( 
\begin{array}{c}
N \\ 
E
\end{array}
\right)_{L/R} 
$ 
 & $N_{\rm HC}$ & $\mathbf{1}$ & $\mathbf{2}$ & $-1/2$ \\ 
\hline 
\end{tabular}
\caption{ 
 The SM charge assignment for eight HC fermions $F_{L/R}=(Q, L)^T_{L/R}$ 
in the one-family model. }
\label{tab:1}
\end{table}
%%%%%%%%%%%%%% Table [end] %%%%%%%%%%%%%%

At the scale $\Lambda_{\rm HC}$, say $\Lambda_{\rm HC}={\cal O}(1-10)$ TeV, 
the HC gauge interaction gets strong 
to develop the nonzero ``chiral'' condensate 
$\langle \bar{F}^{{A}} F^{{B}} \rangle \sim \Lambda_{\rm HC}^3 \cdot \delta^{{AB}}$
({$A$ and $B$} being indices for $SU(8)$ fundamental representations), 
which breaks the ``chiral'' symmetry 
of 8 HC fermions down to the vectorial one: 
$SU(8)_{F_L} \times SU(8)_{F_R} \rightarrow SU(8)_{F_V}$.  
According to the spontaneous breaking, 
the 63 Nambu-Goldstone~{(NG)} bosons emerge, which will be pseudoscalars 
by the explicit breaking terms including the SM gauge interactions 
and possibly present vectorlike fermion mass terms like $m_F^0 \bar{F} F$, 
as discussed in Refs.~\cite{Kilic:2009mi,Pasechnik:2013bxa,Lebiedowicz:2013fta,Pasechnik:2014ida}.

By naively scaling the hadron spectroscopy in QCD, 
we may find 63 composite vectors (HC $\rho$ mesons) as 
the next-to-lightest HC hadrons~\footnote{The lightest $f_0(500)$ scalar meson in QCD may be a mixture of $\bar{q}q$ 
and $qq \overline{qq}$ due to some feature specific 
to the three-flavor $(u,d,s)$ structure, 
according to recent analyses, {\it e.g.}, see Ref.~\cite{Fariborz:2009cq}.}~\footnote{If one happens to take some special value of $N_{\rm HC}$, 
e.g. $N_{\rm HC}=3$, 
then the one-family HC dynamics would turn  
from the ordinary QCD to be quite different, so-called, 
the walking gauge theory having the 
almost nonrunning gauge coupling (i.e. approximate scale invariance). 
Then the lightest hadron spectrum would include a dilaton 
with the mass as small as the HC pion mass, arising 
as the (approximate) spontaneous breaking of the scale invariance~\cite{Aoki:2016wnc}. 
In the present study, 
for simplicity 
we will disregard the possibility of light dilaton.  
}.  
Thus the low-energy effective theory of the HC sector 
would be constructed 
from the 63 HC pions ($\sim \bar{F}^{{A}} i \gamma_5 F^{{B}}$)  
and {also 63 HC} rho mesons ($\sim \bar{F}^{{A}} \gamma_\mu F^{{B}}$). 
Then the HC rho couplings to the SM particles arise 
indirectly from mixing between the SM gauge bosons (``indirect couplings''), 
and directly by an extended HC theory (``direct couplings''),  
which could be generated from extended (vector, or scalar) 
interactions communicating the HC and the SM fermion sectors 
(which would be like a generalized extended technicolor scenario).  
Note that the latter direct coupling can generically be flavor-dependent. 
Both types of couplings are unambiguously formulated 
by the HLS formalism, which is {the main target of the {subsections} below}.

\subsection{HLS formulation} 
%%%%%%%%%%%%%%%%%%%%%
%%%%%%%%%%%%%%%%%%%%%

In this section we formulate the effective Lagrangian 
including the HC vectors along with the HC pions, arising 
from the one-family model of the HC introduced in the previous section.

The effective Lagrangian for those vectors can be formulated 
based on the HLS formalism, which has succeeded in QCD rho meson 
physics~\cite{Bando:1984ej,Bando:1985rf,Bando:1987ym,Bando:1987br,Harada:2003jx}, 
where the $\rho$'s are introduced as the gauge bosons of the HLS. 
Based on the nonlinear realization of 
the HLS and the ``chiral'' $SU(8)_{F_L} \times SU(8)_{F_R}$ symmetry,  
the Lagrangian is written as~\footnote{We have imposed the $C$ and $P$ invariance as well, and 
assumed that the $P$ invariance is violated only 
through the weak interactions when the HC pions and $\rho$ mesons 
couple to the SM fermions through the external weak gauges 
(See also the $\rho$ couplings to the SM fermions in Eq.(\ref{rho-ff}), which 
take the left-handed form).   
}   
\begin{equation} 
 {\cal L} 
= - \frac{1}{2} {\rm tr}[ \rho_{\mu\nu}^2] + f_\pi^2 {\rm tr}[\hat{\alpha}_{\perp \mu}^2] 
+ \frac{m_\rho^2}{g_\rho^2} {\rm tr}[\hat{\alpha}_{||\mu}^2] 
+ \cdots 
\,, \label{Lag:HLS}
\end{equation}
in a manner invariant under the $SU(8)_{F_L} \times SU(8)_{F_R} \times [SU(8)_{F_V}]_{\rm HLS}$ symmetries, where we define{
\begin{eqnarray} 
 \rho_{\mu\nu} 
 &=& \partial_\mu \rho_\nu - \partial_\nu \rho_\mu - i g_\rho [\rho_\mu, \rho_\nu] \,, 
 \notag \\[0.5em] 
 \hat{\alpha}_{\perp \mu} 
 &=& \frac{D_\mu \xi_R \cdot \xi_R^\dag - D_\mu \xi_L \cdot \xi_L^\dag}{2i} 
 \,,\quad\quad  
 \hat{\alpha}_{|| \mu} 
 = 
 \frac{D_\mu \xi_R \cdot \xi_R^\dag + D_\mu \xi_L \cdot \xi_L^\dag}{2i} \,,
 \label{alpha} \\[0.5em] 
 D_\mu \xi_{R(L)} 
 &=& \partial_\mu \xi_{R(L)} - i g_\rho  \rho_\mu \xi_{R(L)} +i \xi_{R(L)} {\cal R}_\mu({\cal L}_\mu) 
 \,, \notag 
\end{eqnarray}
with the HLS gauge coupling $g_\rho$, the HC pion decay constant $f_\pi$, and the external gauge fields ${\cal R}_\mu$ and ${\cal L}_\mu$ that are associated by gauging the ``chiral'' symmetry.}  
Ellipses include terms of higher derivative orders~\cite{Tanabashi:1993np,Harada:2003jx}.  
Under the HLS and the ``chiral'' symmetry, 
the transformation properties for basic variables 
-- $\xi_{L,R}$ (nonlinear bases), $\rho_\mu$ (HLS field), and $\hat{\alpha}_{\perp \mu}$, $\hat{\alpha}_{|| \mu}$ (covariantized Maurer--Cartan one forms) --  are described as
\begin{align}
 \xi_{L} & \to h(x) \cdot \xi_{L} \cdot g_{L}^\dag(x)  \,, &\hspace{-3em}  \xi_{R} & \to h(x) \cdot \xi_{R} \cdot g_{R}^\dag(x)  \,, \notag \\
 \rho_\mu &\to h(x) \cdot \rho_\mu \cdot h^\dag(x) + \frac{i}{g_\rho} h(x) \cdot \partial_\mu h^\dag(x)  \,, &\hspace{-3em} \rho_{\mu\nu} &\to h(x) \cdot \rho_{\mu\nu} \cdot h^\dag(x) \,, \label{HLS:trans:list} \\
 \hat{\alpha}_{\perp {\mu}} &\to h(x) \cdot \hat{\alpha}_{\perp {\mu}} \cdot h^\dag(x) \,, &\hspace{-3em} \hat{\alpha}_{|| {\mu}} &\to h(x) \cdot \hat{\alpha}_{|| {\mu}} \cdot h^\dag(x) \,, \notag
\end{align}
where $h(x) \in [SU(8)_{F_V}]_{\rm HLS}$ and $g_{R,L}(x) \in [SU(8)_{F_{R,L}}
]_{\rm gauged}$. 
The HLS is thus the gauge degree of freedom independent of the SM-external gauges 
and has spontaneously been broken together with the ``chiral'' 
symmetry in terms of the nonlinear realization: 
$\langle \xi_{L,R} \rangle = 1$. The nonlinear bases $\xi_{L}$ and $\xi_{R}$ can be parametrized by 
the {NG} bosons $\pi$ for the ``chiral'' symmetry and ${\cal P}$ for the HLS. 
Hence, they are parametrized as 
${\xi_{\,^R_L}}=e^{i {\cal P}/f_{\cal P}} \cdot  e^{\pm i\pi/f_\pi}$, where the HLS decay constant $f_{\cal P}$ is 
related to the HC rho mass as $m_\rho = g_\rho f_{\cal P}$ and then the ${\cal P}$s are eaten by the HLS 
gauge boson $\rho_\mu$ due to the Higgs mechanism.

Note that, by construction, at the leading order of derivative expansion  
the HLS completely forbids triple vector vertices  
involving the external gauge fields along with HLS, 
such as ${\cal G}-{\cal G}-\rho$ and ${\cal G}-\rho-\rho$ (${\cal G}={\cal L}, {\cal R}$). 
When going beyond the leading order of derivative expansion, {\it e.g.}, at ${\cal O}(p^4)$ one will encounter mixing terms such as ${\rm tr}[\hat{\cal G}_{\mu\nu} \rho^{\mu\nu}]$ 
where $\hat{\cal G}_{\mu\nu}=\xi_{L/R} {\cal G}_{\mu\nu} \xi_{L/R}^\dag$~\cite{Tanabashi:1993np,Harada:2003jx}. 
The size of effects is on the order of one-loop, ${\cal O}(1/(4\pi)^2)$, and thus should potentially be small as long as the derivative expansion works as in the case of the chiral perturbation for pions.

Hereafter, we shall focus on the 63 composite $\rho$ vectors, {namely, the HC rho mesons that couple to the SM fermions.} 
Possible interference effects from the HC pions on the vector meson physics will be addressed later on.

\subsection{Particle assignment for HC $\rho$ and HC $\pi$} 
%%%%%%%%%%%%%%%%%%%%%
%%%%%%%%%%%%%%%%%%%%%

\begin{table}[t] 
\begin{tabular}{c|c|c|c}
\hline 
%\hspace{15pt} technipion \hspace{15pt} & 
\hspace{15pt} composite vector \hspace{15pt} & 
\hspace{15pt} constituent \hspace{15pt} & 
\hspace{15pt} color \hspace{15pt} & 
\hspace{15pt} isospin \hspace{15pt} 
\\  
\hline \hline 
%$\theta_a^{i}$ & 
$\rho_{(8) a}^{{\alpha}}$ &  
$\frac{1}{\sqrt{2}}\bar{Q} \gamma_\mu \lambda^{{a}} \tau^{{\alpha}} Q$ & octet & triplet \\  
%$\theta_a^0$ & 
$\rho_{(8) a}^0$ & 
$ \frac{1}{2\sqrt{2}} \bar{Q} \gamma_\mu \lambda^{{a}} Q$ & octet & singlet \\ 
%$T_c^{i}$ $(\bar{T}_c^{i})$ & 
\hline 
$\rho_{(3) c}^{{\alpha}}$ $\left(\bar{\rho}_{(3) c}^{{\alpha}}\right)$ & 
$\frac{1}{\sqrt{2}} \bar{Q}_c \gamma_\mu \tau^{{\alpha}} L$ (h.c.) & triplet & triplet  \\ 
%$T_c^0$ ($\bar{T}_c^0$) & 
$\rho_{(3) c}^0$ $\left(\bar{\rho}_{(3) c}^0\right)$ & 
$ \frac{1}{2 \sqrt{2}} \bar{Q}_c \gamma_\mu L $ (h.c.)   & triplet & singlet \\ 
\hline 
%$P^i $ & 
$\rho_{(1)'}^{{\alpha}}$ & 
$\frac{1}{2 \sqrt{3}} (\bar{Q} \gamma_\mu \tau^{{\alpha}} Q - 3 \bar{L} \gamma_\mu \tau^{{\alpha}} L)$ & singlet & triplet \\  
%$P^0$ & 
$\rho_{(1)'}^0$ &   
$\frac{1}{4 \sqrt{3}} (\bar{Q} \gamma_\mu Q - 3 \bar{L} \gamma_\mu L)$  & singlet & singlet \\  
%$\Pi^i $ & 
\hline 
$\rho_{(1)}^{{\alpha}}$ & 
$\frac{1}{2} (\bar{Q} \gamma_\mu\tau^{{\alpha}} Q + \bar{L} \gamma_\mu \tau^{{\alpha}} L)$ & singlet & triplet \\   
\hline 
\end{tabular} 
\caption{ The HC rho mesons and 
their associated constituent HC quarks 
$Q_c=(U,D)_c$ and leptons $L=(N,E)$. 
Here $\lambda^{{a}}$ ($a=1,\cdots,8$) are the Gell-Mann matrices, 
$\tau^{{\alpha}}$ $SU(2)$ generators defined as 
$\tau^{{\alpha}}=\sigma^{{\alpha}}/2$ ($\alpha=1,2,3$) with the Pauli matrices $\sigma^{{\alpha}}$, and 
the label $c$ stands for the QCD-three colors, $c=r,g,b$. 
The numbers attached in lower scripts $(1,3,8)$ correspond to 
the representations under the QCD color, i.e., singlet, triplet and octet 
for $(1,3,8)$.
{Since the unbroken color symmetry acts as an internal degree of freedom, four color-octet (real), four color-triplet (complex), and seven color-singlet (real) states are observed as distinguishable particles.}
}
\label{tab:TR} 
\end{table}

The HC $\rho$ fields are constructed from the underlying HC fermions 
as listed in Table~\ref{tab:TR} and expanded 
by the $SU(8)$ group generators (with the Lorentz vector label $\mu$ suppressed)~\footnote{The way of embedding the $\rho$ fields into the one-family $8 \times 8$ matrix form 
has been chosen to be on the basis of allowing to mix with the SM gauge bosons, 
so-called the Drell-Yan basis in light of the LHC production.}:   
\begin{eqnarray} 
\sum_{A=1}^{63} \rho^A \cdot T^A 
&=& 
\sum_{{\alpha}=1}^3 \sum_{a=1}^8 \rho^{{\alpha}}_{(8)a} \cdot T_{(8)a}^{{\alpha}} 
+ 
\sum_{a=1}^8 \rho_{(8)a}^0 \cdot T_{(8) a}  
\nonumber \\ 
&& 
+ 
\sum_{c=r,g,b} \sum_{{\alpha}=1}^3
\left[ {\rho_{(3)c}^{[1]\alpha} \cdot T_{(3) c}^{[1]\alpha} + 
       \rho_{(3)c}^{[2]\alpha} \cdot T_{(3) c}^{[2]\alpha}} \right] 
+
\sum_{c=r,g,b}
\left[ {\rho_{(3)c}^{[1]0} \cdot T_{(3) c}^{[1]} + 
       \rho_{(3)c}^{[2]0} \cdot T_{(3) c}^{[2]}} \right] 
\nonumber \\ 
&&
+ 
\sum_{{\alpha}=1}^3 \rho^{{\alpha}}_{(1)} \cdot T_{(1)}^{{\alpha}}  
+ 
\sum_{{\alpha}=1}^3 \rho^{{\alpha}}_{(1)'} \cdot T^{{\alpha}}_{(1)'} + \rho^0_{(1)'} \cdot T_{(1)'} 
\,,
\end{eqnarray} 
with  
\begin{eqnarray} 
 T_{(8) a}^{{\alpha}}
 &=& 
\frac{1}{\sqrt{2}}  \left( 
\begin{array}{c|c} 
  \tau^{\alpha} \otimes \lambda^{{a}}  &  \\ 
  \hline 
   & {\bf 0_{2 \times 2}} 
\end{array}
\right)  \,, 
\qquad 
T_{(8)a} 
= 
\frac{1}{2 \sqrt{2}}
\left( 
\begin{array}{c|c} 
  {\bf 1}_{2\times 2} \otimes \lambda^{{a}}  &  \\ 
  \hline 
   & {\bf 0_{2 \times 2}}  
\end{array}
\right)  \,, \nonumber \\
%%%%%%%%%%%%%%%%%%%%%
{T^{[1]\alpha}_{(3) c}} 
&=&  
{
\frac{1}{\sqrt{2}}
\left( 
\begin{array}{c|c} 
 & \tau^{\alpha} \otimes {\bf e}_c    \\ 
  \hline 
\tau^{\alpha} \otimes {\bf e}_c^\dag  &  
\end{array}
\right) 
\,, \qquad
T^{[2]\alpha}_{(3) c} 
 =  
\frac{1}{\sqrt{2}}
\left( 
\begin{array}{c|c} 
 & -i  \tau^{\alpha} \otimes {\bf e}_c    \\ 
  \hline 
 i  \tau^{\alpha} \otimes {\bf e}_c^\dag  &  
\end{array}
\right)
\,,} \nonumber \\
%%%%
{T^{[1]}_{(3) c}} 
&=&  
{
\frac{1}{2\sqrt{2}}
\left( 
\begin{array}{c|c} 
 & {\bf 1}_{2 \times 2} \otimes {\bf e}_c    \\ 
  \hline 
{\bf 1}_{2 \times 2} \otimes {\bf e}_c^\dag  &  
\end{array}
\right) 
\,, \qquad
T^{[2]}_{(3) c} 
 =  
\frac{1}{2\sqrt{2}}
\left( 
\begin{array}{c|c} 
 & -i {\bf 1}_{2 \times 2} \otimes {\bf e}_c    \\ 
  \hline 
 i {\bf 1}_{2 \times 2} \otimes {\bf e}_c^\dag  &  
\end{array}
\right)
\,,} \nonumber \\
%%%%%%%%%%%%%%%%%%%%%
 T_{(1)}^{{\alpha}} 
 &=& 
\frac{1}{2} \left( 
\begin{array}{c|c} 
  \tau^{{\alpha}} \otimes {\bf 1}_{3\times 3}  &  \\ 
  \hline 
   & \tau^{{\alpha}} 
\end{array}
\right)  \,, 
\qquad 
T^{{\alpha}}_{(1)'} 
= 
\frac{1}{2 \sqrt{3}}
\left( 
\begin{array}{c|c} 
  \tau^{{\alpha}} \otimes {\bf 1}_{3\times 3}  &  \\ 
  \hline 
   & -3 \cdot \tau^{{\alpha}} 
\end{array}
\right)  
\,, \nonumber \\
T_{(1)'}
&=& 
\frac{1}{4 \sqrt{3}}
\left( 
\begin{array}{c|c} 
  {\bf 1}_{6 \times 6}  &  \\ 
  \hline 
   & -3\cdot {\bf 1}_{2 \times 2} 
\end{array}
\right) 
\,, 
\end{eqnarray}
where $\tau^{{\alpha}} = \sigma^{{\alpha}}/2$ ($\sigma^{{\alpha}}$: Pauli {matrices}), {$\lambda^a$ and ${\bf e}_c$ represent the Gell-Mann matrices and three-dimensional unit vectors in color space, respectively,}
and the generator $T^A$ is normalized as ${\rm tr}[T^A T^B]=\delta^{AB}/2$.
%%%
For color-triplet components (leptoquarks), we define the following eigenforms which discriminate ${\bf 3}$ and ${\bf \bar{3}}$ states of the $SU(3)_c$ gauge group,
\al{
T^{\alpha}_{(3) c} &\equiv \frac{1}{\sqrt{2}} \left( T^{[1]\alpha}_{(3) c} +i T^{[2]\alpha}_{(3) c} \right) 
=
\left( 
\begin{array}{c|c} 
 & \tau^{\alpha} \otimes {\bf e}_c    \\ 
  \hline 
 {\bf 0_{2 \times 6}} &  
\end{array}
\right)\,, & 
T^{\alpha}_{(\bar{3}) c} &\equiv \left( T^{\alpha}_{(3) c} \right)^\dagger, \notag \\
%%%%%
T_{(3) c} &\equiv \frac{1}{\sqrt{2}} \left( T^{[1]}_{(3) c} +i T^{[2]}_{(3) c} \right) 
=
\frac{1}{2}
\left( 
\begin{array}{c|c} 
 & {\bf 1_{2 \times 2}} \otimes {\bf e}_c    \\ 
  \hline 
 {\bf 0_{2 \times 6}} &  
\end{array}
\right)\,, & 
T_{(\bar{3}) c} &\equiv \left( T_{(3) c} \right)^\dagger,  \label{eq:LQ_relation_1}
}
\vspace{-6mm}
\al{
\rho^\alpha_{(3) c} &\equiv \frac{1}{\sqrt{2}} \left( \rho^{[1]\alpha}_{(3) c} -i \rho^{[2]\alpha}_{(3) c} \right), \quad
\rho^0_{(3) c} \equiv \frac{1}{\sqrt{2}} \left( \rho^{[1]0}_{(3) c} -i \rho^{[2]0}_{(3) c} \right), \quad
\bar{\rho}^\alpha_{(3) c} \equiv \left( \rho^\alpha_{(3) c} \right)^\dagger, \quad
\bar{\rho}^0_{(3) c} \equiv \left( \rho^0_{(3) c} \right)^\dagger.
\nonumber
}
Also, the following relations hold:
\al{
\rho_{(3)c}^{[1]\alpha} \cdot T_{(3) c}^{[1]\alpha} + \rho_{(3)c}^{[2]\alpha} \cdot T_{(3) c}^{[2]\alpha}
=
\rho_{(3)c}^{\alpha} \cdot T_{(3) c}^{\alpha} + \bar{\rho}_{(3)c}^{\alpha} \cdot T_{(\bar{3}) c}^{\alpha}, \notag \\
%%%
\rho_{(3)c}^{[1]0} \cdot T_{(3) c}^{[1]} + \rho_{(3)c}^{[2]0} \cdot T_{(3) c}^{[2]\alpha}
=
\rho_{(3)c}^{0} \cdot T_{(3) c} + \bar{\rho}_{(3)c}^{0} \cdot T_{(\bar{3}) c},
	\label{eq:LQ_relation_2}
}
which are useful for rewriting the above decomposition in the eigenforms.
The following normalization conditions of the ${\bf 3}$ and ${\bf \bar{3}}$ states of the $SU(3)_c$ gauge group in the new {bases} are also useful:
${\rm tr}[T_{(3)c}^{i} T_{(\bar{3})c'}^{j}]=\delta^{ij}\delta_{cc'}/2$, 
${\rm tr}[T_{(3) c}^{0} T_{(\bar{3})c'}^{0}]=\delta_{cc'}/2$,
$\text{(others)} = 0$.

We can express the {vector fields $\rho$} {of $8 \times 8$ matrix with block matrices} as 
\begin{equation} 
{\rho = 
\left( 
\begin{array}{cc} 
(\rho_{QQ})_{6\times 6} & {(\rho_{QL})_{6 \times 2}} \\ 
{(\rho_{LQ})_{2 \times 6}} & (\rho_{LL})_{2 \times 2} 
\end{array}
\right)}
\,, \label{rho:para}  
\end{equation}  
where 
\begin{align}
 \rho_{QQ} = 
 &
 \left[ \sqrt 2 \rho_{(8)a}^{{\alpha}} \left( {\tau^\alpha} \otimes {\lambda^{{a}} \over 2} \right) 
 + {1 \over \sqrt 2} \rho_{(8)a}^0 \left( {\bf 1}_{2\times2} \otimes {\lambda^{{a}} \over 2} \right) \right] \notag \\[0.5em]
 & 
 + \left[ {1 \over 2} \rho_{(1)}^{{\alpha}} \left( {\tau^\alpha} \otimes {\bf 1}_{3\times3} \right) 
 + {1 \over 2\sqrt 3} \rho_{(1)'}^{{\alpha}} \left( {\tau^\alpha} \otimes {\bf 1}_{3\times3} \right) 
 + {1\over 4 \sqrt 3} \rho_{(1)'}^0 \Big( {\bf 1}_{2\times2} \otimes {\bf 1}_{3\times3} \Big) \right] \,, \notag \\[1em] 
 \rho_{LL} = 
 &
 {1 \over 2} \rho_{(1)}^{{\alpha}} \left( {\tau^\alpha} \right) - {\sqrt 3 \over 2} \rho_{(1)'}^{{\alpha}} \left( {\tau^\alpha} \right) - {\sqrt 3 \over 4} \rho_{(1)'}^0 \Big( {\bf 1}_{2\times2} \Big) \,, \notag \\[1em]
 \rho_{QL} = 
 &
 \rho_{(3){c}}^{{\alpha}} \left( {\tau^\alpha} \otimes {\bf e}_c \right) + {1 \over 2} \rho_{(3) {c}}^0 \Big( {\bf 1}_{2\times2} \otimes {\bf e}_c \Big) \,, \notag \\[1em]
 {\rho_{LQ}} = & \Big( {\rho_{QL}} \Big)^\dag \,{.}
 \label{rho:assign}  
\end{align}
%and ${\bf e}_c$ is the three-dimensional unit vector on the QCD color space.
{Thus, distinct particles consist of four color-octet (real), four color-triplet (complex), and seven color-singlet (real) states.}
The {$SU(2)_W$} charge, in terms of which the HC rho field is decomposed as in Eq.(\ref{rho:para}), is identified with the one in the SM.  
The SM fermions will carry corresponding $SU(2)_W$ charges so that they couple to the HC rho once a relationship for the external SM gauge fields is given,  
as will be seen later.

In a manner similar to the HC rho mesons in Eq.(\ref{rho:assign}), 
the HC pions are embedded into the $8 \times 8$ matrix of $SU(8)$ as 
\begin{eqnarray} 
\pi &=& 
\left( 
\begin{array}{cc} 
 (\pi_{QQ})_{6 \times 6} & {(\pi_{QL})_{6 \times 2}} \\ 
 {(\pi_{LQ})_{2 \times 6}} & (\pi_{LL})_{2 \times 2}  
\end{array}
\right) 
%\,, \qquad 
%\pi_{LQ} = \pi_{QL}^\dag
\,, \nonumber \\ 
\pi_{QQ} &=& \left[ \sqrt{2} \pi_{(8)} + \frac{1}{\sqrt{2}} \pi_{(8)}^0   \right]  
+ \left(  \frac{1}{2} \pi_{(1)}  + \frac{1}{2\sqrt{3}} \pi_{(1)'}  
+ \frac{1}{4 \sqrt{3}} \pi^0_{(1)'}  \right) \otimes {\bf 1}_{3\times 3} 
\,, \nonumber \\ 
\pi_{QL} &=& \pi_{(3)} + \frac{1}{2} \pi^0_{(3)}
\,, \nonumber \\ 
\pi_{LQ} &=& (\pi_{QL})^\dag = 
\bar{\pi}_{(3)}  + \frac{1}{2} \bar{\pi}^{0}_{(3)}  
\,, \nonumber \\ 
\pi_{LL} &=& 
 \left( \frac{1}{2} \pi_{(1)}  - \frac{\sqrt{3}}{2} \pi_{(1)'} 
- \frac{\sqrt{3}}{4} \pi^{0}_{(1)'} \right) 
 \,,\nonumber \\
%%%%%
\pi_{(8)} &=& \pi_{(8) a}^{{\alpha}} \left( \tau^{{\alpha}} \otimes \frac{\lambda^a}{2} \right)
\,, \qquad 
\pi^{0}_{(8)} = \pi_{(8) a}^{0} \left( {\bf 1}_{2 \times 2} \otimes \frac{\lambda^a}{2} \right)
\,, \nonumber \\ 
\pi_{(3)} &=& \pi_{(3)c}^{{\alpha}} \left( \tau^{{\alpha}} \otimes {\bf e}_c \right)
\,, \qquad 
\pi^{0}_{(3)} = \pi_{(3)c}^{0} \left( {\bf 1}_{2 \times 2} \otimes {\bf e}_c \right)
\,, \nonumber \\ 
\pi_{(1)'} &=&  \pi_{(1)'}^{{\alpha}} \left( \tau^{{\alpha}} \right)
\,, \qquad 
\pi_{(1)'}^{0} = \pi_{(1)'}^{0} \left( {\bf 1}_{2 \times 2} \right) 
\,, \nonumber \\ 
\pi_{(1)} &=& \pi^{{\alpha}}_{(1)} \left( \tau^{{\alpha}} \right)
\,. 
\label{pi-list}
%\nonumber 
\end{eqnarray}
%%%%%%%%%%%%%%%%%%%%%%%%%%%%

\subsection{Short sketch for masses of HC $\rho$ and HC $\pi$ \label{sec:mass_rho_and_pi}}

As evident from Eq.(\ref{Lag:HLS}), the HC $\rho$ meson masses are degenerate due to the global $SU(8)_{F_V}$ symmetry {unless the external gauge fields are present}. 
With the SM gauges tuned on, however, some of the HC $\rho$ mesons charged under the SM gauge fields will get shifts to masses, due to the mixing with the SM gauge bosons, (see {Eq.(\ref{mass:mixing})} in the later discussion.) 
{Note that the massless property of the gluon and the photon is intact under the mass mixings because they are protected by the residual $SU(3)_c \times U(1)_{\text{em}}$ gauge symmetries after the Higgs mechanism works.}
Most sizable corrections will arise from mixing with the QCD gluon, 
which makes the mass of the color-octet isosinglet $\rho_{(8)}^0$ lifted up by amount of ${\cal O}(4\pi \alpha_s/g_\rho) = {\cal O}(0.1)$ for  $g_\rho = {\cal O}(10)$ and $\alpha_s=g_s^2/(4\pi) \simeq 0.1$.

The HC pions listed in Eq.(\ref{pi-list}) get massive through the explicit breaking effect outside of the HC dynamics. 
One may make them massive by introducing some Dirac masses ($m_F^0$) for HC fermions 
{(like $m_F^0 \bar{F}F$)}.  
To realize the HC pions as {NG} bosons for the spontaneous breaking of ``chiral'' symmetry, such an explicit HC fermion mass should be small {so that}
{$M_\pi \simeq {\cal O}(\Lambda_{\rm HC}/(4 \pi)) \simeq {\cal O}(f_\pi)${.}}
This implies $M_\pi \sim {\cal O}(100)$ GeV for $\Lambda_{\rm HC} \sim {\cal O}(1)$ TeV. 
However, the present model can have {another} source which would allow some HC pions to get more massive: 
since the present HC theory consists of the one-family content with the number of HC fermions $N_F=8$,   
the masses of HC pions having the SM charges could be enhanced by the amplification of the explicit breaking effect, as discussed in~\cite{Matsuzaki:2015sya} and references therein.  
The enhancement will then be most eminent for QCD colored pions, $\pi_{(3)}$ and $\pi_{(8)}$ due to the relatively large QCD coupling strength. 
Following \cite{Matsuzaki:2015sya}, we evaluate the size of colored HC pion masses from the QCD gluon exchange contribution as  
$
 M_{\pi_{(3),(8)}}^2 \sim C_2 \alpha_s(M_{\pi}) \Lambda_{\rm HC}^2 {\rm ln}
 \frac{\Lambda_{\rm UV}^2}{\Lambda_{\rm HC}^2}   
$, 
with {$C_2={4 \over 3} \, (3)$} for color-triplet (octet) HC pions, where $\Lambda_{\rm UV}$ denotes some ultraviolet high-energy scale up to which the HC theory is valid.  
Taking $\alpha_s(M_\pi) \sim 0.1$ and $\Lambda_{\rm UV} \sim 10^{16}$ GeV, for example, 
we thus estimate the $\pi_{(3)}$ and $\pi_{(8)}$ masses as $M_{\pi_{(3)}} \sim 3$ TeV and $M_{\pi_{(8)}} \sim 4$ TeV, respectively, for $\Lambda_{\rm HC}\sim 1$ TeV.

In a similar way, the EW gauge interaction makes masses of EW-charged HC pions lifted up. 
This effect becomes operative for the $\pi_{(1)}^{\pm,3}$ and $\pi_{(1)'}^{\pm,3}$ pions to yield $M_{\pi_{(1),(1)'}^{\pm,3}} \sim 2\,\text{TeV}$ for $\Lambda_{\rm HC} \sim 1\,\text{TeV}$ and $\Lambda_{\rm UV} \sim 10^{16}\,\text{GeV}$ as a benchmark.
{Hereafter, The indices `$\pm$' and `$3$' discriminate components of $SU(2)_W$ triplets.
The index `$0$' emphasizes that the designated states are $SU(2)_W$ singlets.}

Thus, the sizes of the HC pion masses are roughly expected as 
\begin{eqnarray}
{M_{\pi_{(1)'}^0}} &\sim&  
{ 
{\cal O}(f_\pi) ={\cal O}(100) \, {\rm GeV} 
}
\,, \notag \\ 
M_{\pi_{(1)'}^{\pm,3}} &\sim& 2 \, {\rm TeV} 
\,, \notag \\ 
M_{\pi_{(1)}^{\pm,3}} &\sim& {2} \, {\rm TeV} 
\,, \notag \\
M_{\pi_{(3)}^{\pm,3,0}} &\sim& 3 \,{\rm TeV} 
\,, \notag \\ 
M_{\pi_{(8)}^{\pm,3,0}} &\sim& 4 \, {\rm TeV} 
\,, \label{pi:masses}
\end{eqnarray}
for $\Lambda_{\rm HC} \sim 1\,\text{TeV}$ and $\Lambda_{\rm UV} \sim 10^{16}\,\text{GeV}$.  
This is {the} significant feature for the HC pion in our model particularly when we discuss collider bound on the HC rho mesons.
Hereafter we shall take the above HC pion spectroscopy as a benchmark in the present study on the HC rho meson physics. 
{Most of HC pions thus become as heavy as HC rhos, which 
implies the loss of pseudo NG nature in the full quantum field theory, 
though at the classical level 
the ``chiral" symmetry is conserved approximately enough 
due to the smallness of the gauge couplings. 
As was discussed in Ref.~\cite{Matsuzaki:2015sya}, this happens due to the amplification 
of explicit breaking effect, induced from non-perturbative 
feature of the QCD with many flavors (nearly conformal/scale-invariant dynamics). 
One might then think of integrating out heavy HC pions 
to get a reduced low-energy description characterized by 
smaller ``chiral" manifold than the original class of $SU(8)$. 
However, one cannot do it because heavy HC pions in the mass eigenstate 
basis involve all HC quark and leptons when they are formed, 
as seen from Eq.(\ref{pi-list}) (or in the same way of embedding as in 
Table~\ref{tab:TR} for HC rhos.). Thus one needs to construct the ``chiral" $SU(8)$ 
manifold to describe HC pions, which are manifestly NG bosons at the classical 
level of the ``chiral" $SU(8)$ dynamics, {even though it becomes so heavy at the quantum level.}}

\subsection{Couplings to SM particles \label{sec:rho_coupling_to_SMfermion}} 
%%%%%%%%%%%%%%%%%%%%%
%%%%%%%%%%%%%%%%%%%%%

\subsubsection{Direct $V$-$f_L$-$f_L$ coupling terms: extended HC-origin}
%%%%
The SM fermion fields are written as an eight-dimensional vector on the base of the fundamental representation of $SU(8)$,   
\begin{equation} 
f_{L}= \left( \begin{array}{c} q \\ l \end{array} \right)_{L} 
\,,\quad\quad
f_{R}= \left( \begin{array}{c} q \\ l \end{array} \right)_{R} 
\,,
\label{eq:fLR_SU8form}
\end{equation} 
where $q$ and $l$ are $SU(2)_{F_L,F_R}$ doublets for the quark and lepton fields.  
The SM-covariant derivatives that act on the $f$-fermion multiplets are then expressed as the $ 8 \times 8$ matrix forms: 
\begin{align} 
 D_\mu f_{L} &= {{\bf 1}_{8 \times 8}} \cdot (\partial_\mu f_{L}) - i \, [{\cal L}_\mu^f]_{8 \times 8} \cdot f_{L} \,, \nonumber \\ 
 D_\mu f_{R} &= {{\bf 1}_{8 \times 8}} \cdot (\partial_\mu f_{R}) - i \, [{\cal R}_\mu^f]_{8 \times 8} \cdot f_{R} \,, \label{cov:SM-f}
\end{align}
with
\begin{eqnarray} 
%%%
\left[{\cal L}_\mu^f\right]_{8 \times 8} &=& 
\left( 
\begin{array}{c|c} 
{{\bf 1}_{2 \times 2} \otimes g_s G_\mu^a \frac{\lambda^a}{2}}
+ {\left( g_W  W_\mu \tau^{\alpha} 
+  \frac{1}{6} g_Y B_\mu \right) \otimes {\bf 1}_{3\times 3}}
& {{\bf 0}_{6 \times 2}} \\ 
\hline 
{{\bf 0}_{2 \times 6}} 
& 
g_W W_\mu^{{\alpha}} \tau^{{\alpha}} 
- \frac{1}{2} g_Y B_\mu \cdot {\bf 1}_{2 \times 2} 
\end{array}
\right)   \nonumber \\
&=&
{\sqrt{2} g_s G^a_\mu T_{(8) a} + \frac{2}{\sqrt{3}} g_Y B_\mu T_{(1)'} + 2 g_W W^\alpha_\mu T^{\alpha}_{(1)}}, \nonumber \\
%%% 
\left[{\cal R}_\mu^f\right]_{8 \times 8} &=& 
\left( 
\begin{array}{c|c} 
{{\bf 1}_{2 \times 2} \otimes g_s G_\mu^a \frac{\lambda^a}{2}}
+  {g_Y\, Q_{\rm em}^q \, B_\mu \otimes {\bf 1}_{3\times 3}}
& {{\bf 0}_{6 \times 2}} \\ 
\hline 
{{\bf 0}_{2 \times 6}}
& 
g_Y \, Q_{\rm em}^l \, B_\mu  
\end{array}
\right) 
\,,
\label{eq:SMcovD_SU8form}
\end{eqnarray} 
where $G_\mu, W_\mu$ and $B_\mu$ are the $SU(3)_c \times SU(2)_W \times U(1)_Y$ gauge fields along with the gauge couplings $g_s$, $g_W$ and $g_Y$, respectively; and 
$Q_{\rm em}^{q,l}$ is the electromagnetic (EM) charge defined as   
\begin{equation}  
 Q_{\rm em}^q 
= \left( 
\begin{array}{cc} 
2/3 & 0 \\ 
 0 & -1/3 
\end{array}
\right)
\,, \qquad 
Q_{\rm em}^l 
= \left(
\begin{array}{cc} 
0 & 0 \\ 
 0 & -1 
\end{array}
\right)
\,. 
\end{equation}
{As clearly seen from Eqs.(\ref{cov:SM-f}) and (\ref{eq:SMcovD_SU8form}), the global $SU(8)_{F_L} 
\times SU(8)_{F_R}$ symmetry is explicitly broken 
in the SM fermion sector by the SM gauges, hence of course 
the $SU(8)$ is not a good symmetry for the SM fermions. 
Just for convenience to address SM fermion couplings to the HC rhos 
which form the $SU(8)$ adjoint representation,   
{we have introduced the $SU(8)_{F_L/F_R}$-multiplet form for 
the SM fermions as in Eq.(\ref{eq:fLR_SU8form}).}}

The covariant derivatives for the HC fermions can also be written 
in terms of the $8 \times 8$ matrix form. 
We may relate the charges of the HC fermions  
with those of the SM quark and lepton charges, 
involving 
the HC-quark and -lepton numbers. 
Then the nonlinear bases $\xi_{L,R}$ in Eq.(\ref{HLS:trans:list})
transform under the HLS and the SM gauge group ${\cal G}= SU(3)_c 
\times SU(2)_W \times U(1)_Y$ as 
\begin{equation} 
 \xi_{L} \to h(x) \cdot \xi_{L} \cdot [g_{L}^\dag(x)]_{\cal G} 
 \,, \quad\quad
 \xi_{R} \to h(x) \cdot \xi_{R} \cdot [g_{R}^\dag(x)]_{\cal G} 
 \,.  
\end{equation}
From Table~\ref{tab:1}, 
one thus finds that 
the external gauge fields ${\cal L}_\mu$ and ${\cal R}_\mu$, 
coupled to the nonlinear bases $\xi_{L,R}$ as 
in Eq.(\ref{alpha}), are identified with those coupled to 
the SM fermions as described in Eq.(\ref{cov:SM-f}): 
\begin{align} 
 {\cal L}_\mu &= {\cal L}_\mu^f 
 \,,  
 &
 {\cal R}_\mu &= {\cal L}_\mu^f 
 \,, \nonumber \\ 
 {\rm i.e.,}~~  {\cal V}_\mu &= \frac{{\cal R}_\mu + {\cal L}_\mu}{2} = {\cal L}_\mu^f 
 \,,  
 &
 {\cal A}_\mu &= \frac{{\cal R}_\mu - {\cal L}_\mu}{2} =0\,. \hspace{5em}
 \label{cov:techni}
\end{align} 
It is useful to expand $\hat{\alpha}_{|| \mu}$ and $\hat{\alpha}_{\perp \mu}$ in Eq.(\ref{alpha}) in powers of the HC pion fields $\pi$ with the unitary gauge for the HLS $({\cal P}\equiv 0)$: 
\begin{equation} 
\hat{\alpha}_{|| \mu}= {\cal V}_\mu - g_\rho \rho_\mu 
- \frac{i}{2 f_\pi^2} \left[ \partial_\mu \pi, \pi \right] - \frac{i}{f_\pi} \left[ {\cal A}_\mu, \pi \right] + \cdots 
\,, \label{alpha:expand}
\end{equation} 
and 
\begin{equation} 
 \hat{\alpha}_{\perp \mu} 
= 
\frac{\partial_\mu \pi}{f_\pi} + {\cal A}_\mu - \frac{i}{f_\pi} \left[ {\cal V}_\mu, \pi \right] -
\frac{1}{6 f_\pi^3} \left[ \pi, \left[ \pi, \partial_\mu \pi \right] \right] + \cdots 
\,, \label{alpha:expand:2}
\end{equation} 
with {${\cal V}_\mu = ({\cal R}_\mu + {\cal L}_\mu)/2$ and ${\cal A}_\mu = ({\cal R}_\mu - {\cal L}_\mu)/2$.}
Then the 1-forms in Eqs.(\ref{alpha:expand}) and (\ref{alpha:expand:2}) are represented as  
\begin{equation} 
 \hat{\alpha}_{||\mu} = {\cal L}_\mu^f - g_\rho \rho_\mu + \cdots 
\,, \qquad 
\hat{\alpha}_{\perp \mu} 
= 0 + \cdots 
\,. \label{alpha:expand:3} 
\end{equation}

We may define the dressed fields for the left-handed SM fermions, 
\begin{equation} 
\Psi_L \equiv \xi_L \cdot f_L 
\,, \qquad 
\psi_L \equiv \xi_R \cdot f_L
\,, 
\end{equation}
which transform as 
\begin{equation} 
\Psi_L \to h(x) \cdot \Psi_L 
\,, \qquad 
\psi_L \to h(x) \cdot \psi_L
\,. 
\end{equation}  
These transformations allow us to  
write down the HC $\rho$ couplings to the left-handed SM fermions 
in the HLS-invariant way as 
\begin{equation} 
{\cal L}_{\rho ff} = 
g^{{ij}}_{1L} 
\left( 
\bar{\Psi}^{{i}}_L \gamma^\mu \hat{\alpha}_{||\mu} \Psi^{{j}}_L
\right) 
+
g^{{ij}}_{2L} 
\left( 
\bar{\Psi}^{{i}}_L \gamma^\mu \hat{\alpha}_{||\mu} \psi^{{j}}_L
+ {\rm h.c.}
\right)  
+ 
g^{{ij}}_{3L} 
\left( 
\bar{\psi}^{{i}}_L \gamma^\mu \hat{\alpha}_{||\mu} \psi^{{j}}_L
\right) 
\,,  \label{rho-ff} 
\end{equation}
where {$i$ and $j$} label the generations of the SM fermions $({i,j} = 1,2,3)$.

Using Eqs.(\ref{cov:SM-f}) and (\ref{alpha:expand:3}), one can thus extract the HC $\rho$ and $V_\text{SM}$ (SM gauge boson) couplings to the left-handed SM fermions. 
As a result, we have
\begin{align} 
 {\cal L}_{V f_Lf_L}^\text{direct} =\,
 & 
 g_L^{{ij}} \cdot \bar{q}_L^{\,{i}} \gamma_\mu 
 \left[ g_s G^{\mu}_a \left( {\bf 1}_{2\times 2} \otimes \frac{\lambda_a}{2} \right) + \left( g_W W^{{\alpha}\,\mu}\, \frac{\sigma^{{\alpha}}}{2}  + \frac{g_Y}{6} B^\mu\, {\bf 1}_{2\times2} \right) \otimes {\bf 1}_{3 \times3} - g_\rho \rho^\mu_{QQ} \right] q_L^{{j}} \notag \\
 &
 + g_L^{{ij}} \cdot \bar{l}_L^{{\,i}} \gamma_\mu \left[ g_W W^{{\alpha}\,\mu}\, \frac{\sigma^{{\alpha}}}{2} - \frac{g_Y}{2} B^\mu\, {\bf 1}_{2 \times 2} - g_\rho \rho^\mu_{LL} \right] l_L^{{j}} \notag \\
 &
 - g_L^{{ij}} g_\rho \cdot \bigg[ \bar{q}_L^{{\,i}} \gamma_\mu\, \rho_{QL}^\mu\, l_L^{{j}} + {\rm h.c.} \bigg] \,, 
 \label{couplings:rhoff:II}
\end{align}
where $\rho^\mu_{QQ}$, $\rho^\mu_{LL}$, and $\rho^\mu_{QL}$ are combinations of the HC $\rho$ mesons as defined in Eq.\eqref{rho:assign} and $g_L^{{ij}} = (g_{1L}+ 2 g_{2L} + g_{3L})^{{ij}}$. 
Note that the $V_\text{SM}$-$f_L$-$f_L$ term in Eq.\eqref{couplings:rhoff:II} is not the normal SM interactions but additional contributions in this model.

The HLS invariance actually allows one to write down vector couplings other than those in Eq.(\ref{couplings:rhoff:II}), 
which would take the form like 
\begin{equation} 
 h_L^{{ij}} \bar{\Psi}_L^{{i}} \gamma_\mu \hat{\alpha}_\perp^\mu \Psi_L^{{j}} 
 \,, \label{h-coupling}
 \end{equation}
with the generation-dependent coupling {$h_L^{ij}$}.  
As seen from Eq.(\ref{alpha:expand:3}), however, the 1-form $\hat{\alpha}_{\mu \perp}$ goes to vanish in the unitary gauge of the HLS; $\xi_{L/R} \to 1$ up to HC pion terms $\ni \partial_\mu \pi/f_\pi + \cdots$. 
%%%This coupling term would thus be relevant 
%%%only when the HC pions can have flavorful couplings to the SM fermions, which is not the case in this article.  
{Since HC rhos and pions are 
generically composed of identical HC fermion bilinears, 
they are expected to develop {the couplings to the SM fermions} in the same form, 
which could be generated from an extended HC theory. 
%As will be argued later, to be phenomenologically viable 
%the flavor index is set to 
%third-generation-philic form for both HC rhos and pions 
%(see Eqs.(\ref{rho-ff}) and (\ref{eq:amplitude_pi0toGG})). 
{Explicit modeling of the extended HC sector is beyond the scope of current analysis.}}
We will briefly address possible effects from those HC pion couplings in the later section.

\subsubsection{$\boldmath{\rho}$\! - \!$V_\text{SM}$ mixing structures and induced-indirect couplings to SM fermions}
%%%%
In addition to the direct interactions of Eq.(\ref{couplings:rhoff:II}), the HC $\rho$ mesons also have interactions induced by the mixing with the SM gauge bosons. 
The mixing term is involved in the mass matrix of the vector boson, which is written by 
\begin{align} 
 \frac{m_\rho^2}{g_\rho^2} {\rm tr}[ \hat{\alpha}_{|| \mu}^2] 
 & \supset \frac{m_\rho^2}{g_\rho^2} {\rm tr} \left[({\cal L}_\mu^f - g_\rho \rho_\mu)^2 \right] \notag \\
 & =  \frac{1}{2} \frac{m_\rho^2}{g_\rho^2} \Bigg[
	g_\rho^2 \left( \rho^{\alpha\,\mu}_{(8) a} \right)^2 +
	g_\rho^2 \left( \rho^{\alpha\,\mu}_{(1)'} \right)^2 +
	\left( g_\rho \rho^{0\,\mu}_{(8) a} -\sqrt{2} g_s G^{a\,\mu} \right)^2 +
	\left( g_\rho \rho^{\alpha\,\mu}_{(1)} - 2 g_W W^{\alpha\,\mu} \right)^2 \notag \\
 &\phantom{ =  \frac{1}{2} \frac{m_\rho^2}{g_\rho^2} \Bigg[} \!\!\! +
	\left( g_\rho \rho^{0\,\mu}_{(1)'} - \frac{2}{\sqrt{3}} g_Y B^{\mu} \right)^2 + 2 g_\rho^2
	\left( \overline{\rho}^{\alpha\,\mu}_{(3) c} \rho^{\alpha}_{(3) c\,\mu} +
	       \overline{\rho}^{0\,\mu}_{(3) c} \rho^{0}_{(3) c\,\mu} \right)
    \Bigg],
 \label{mass:mixing}
\end{align} 
where we used the relations in Eqs.(\ref{eq:LQ_relation_1}), (\ref{eq:LQ_relation_2}), (\ref{cov:SM-f}) and the normalization of the $SU(8)$ generators as $\text{tr}\left[ T^A T^B \right] = \delta^{AB}/2$. 
Note that the mixing form is manifestly custodial-symmetric.

At first, we illustrate the case for charged bosons, $W^\pm$, $\rho_{(1)}^\pm$, and $\rho_{(1)'}^\pm$, where $\rho_X^\pm \equiv (\rho_{X}^1 \mp i\rho_{X}^2)/\sqrt 2$. 
The corresponding matrix element is then written by 
\begin{align} 
 \begin{pmatrix} W^+ & \rho_{(1)}^+ & \rho_{(1)'}^+ \end{pmatrix} 
 \begin{pmatrix}  m_W^2 +4r_g^2m_\rho^2  &  -2r_gm_\rho^2 & 0  \\  -2r_gm_\rho^2  & m_\rho^2 & 0 \\ 0 & 0 & m_\rho^2 \end{pmatrix} 
 \begin{pmatrix} W^- \\ \rho_{(1)}^- \\ \rho_{(1)'}^- \end{pmatrix} \,, 
\label{massmat:CC}
\end{align}
with $r_g = g_W/g_\rho$ and $m_W =g_W\, v_\text{VEV}/2$, where $v_\text{VEV}$ is the Higgs VEV {which is the same one as in the SM.}  
Expanding the above mass matrix with respect to $r_g$, one can obtain the mass eigenvalues as
\begin{align} 
 M_W^2 & = m_W^2 \left(1 - \frac{4r_g^2}{1 - r_m^2} + {\cal O}(r_g^4) \right) \,, \notag \\ 
 M_{\rho_{(1)}^\pm}^2 & = m_\rho^2 \left(1 + \frac{4r_g^2}{1 - r_m^2} + {\cal O}(r_g^4) \right) \,, \quad M_{\rho_{(1)'}^\pm}^2 = m_\rho^2 \,,
 \label{EQ:massbasis}
\end{align}
with $r_m = m_W/m_\rho$ and then the mass eigenstates $\widetilde{W}$, $\widetilde{\rho}_{(1)}$, and $\widetilde{\rho}_{(1)'}$ as 
\begin{align} 
 W^\mu &\simeq \widetilde{W}^\mu -2r_g 
  \widetilde{\rho}_{(1)}^\mu \,, 
  \\
 \rho_{(1)}^\mu &\simeq 
 \widetilde{\rho}_{(1)}^\mu + 2r_g 
 \widetilde{W}^\mu \,, \\
 \rho_{(1)'}^\mu &= \widetilde{\rho}_{(1)'}^\mu \,, 
\end{align}
(`$\pm$' is omitted above), to the nontrivial leading order in expansion 
for $r_g \ll1$ and $r_m \ll1 $.

Next, the neutral vector fields have the following mass matrix: 
\al{
\frac{1}{2}
\begin{pmatrix} W^3 & B & \rho^{3}_{(1)} & \rho^{0}_{(1)'} \end{pmatrix}
\mathcal{M}_{\text{neutral}}
\begin{pmatrix} W^3 \\ B \\ \rho^{3}_{(1)} &\\\rho^{0}_{(1)'} \end{pmatrix} +
\frac{1}{2} m_\rho^2 \left( \rho^{3}_{(1)'} \right)^2,
}
with
\al{
\mathcal{M}_{\text{neutral}}
=
\begin{pmatrix}
m_W^2 + 4 \, r_g^2 m_\rho^2 & -t_W m_W^2 & -2 r_g m_\rho^2 & 0 \\
-t_W m_W^2 & t_W^2 \left( m_W^2 + \frac{4}{3} \, r_g^2 m_\rho^2 \right) & 0 & -\frac{2}{\sqrt{3}} t_W r_g m_\rho^2 \\
-2 r_g m_\rho^2 & 0 & m_\rho^2 & 0 \\
0 & -\frac{2}{\sqrt{3}} t_W r_g m_\rho^2 & 0 & m_\rho^2
\label{massmat:NC}
\end{pmatrix}.
}
Here, our notation on the electroweak sector is as follows: $1/e^2 = 1/g_W^2 + 1/g_Y^2$, $s_W = e/g_W$, $c_W = e/g_Y$, $t_W = s_W/c_W$. 
One immediately finds that the above mass matrix has zero determinant, which ensures the existence of the massless photon. 
The mass eigenvalues are then obtained and expanded as follows:
\al{
M_A^2 &= 0, \notag \\
M_Z^2 &= m_Z^2 \left( 1 - \frac{4 c_W^2 (1 + t_W^4/3) r_g^2}{1 - r_m^2/c_W^2} + {{\cal O}}(r_g^4) \right), \notag \\
M_{\rho^{3}_{(1)}}^2  &= m_\rho^2 \left( 1 + \frac{
2 \left\{ - 4 r_m^2 t_W^2 + (3 + t_W^2) 
+ \sqrt{16 r_m^4 t_W^4 - 4 r_m^2 t_W^2 (3-t_W^2) + (3-t_W^2)^2}  \right\} 
r_g^2}{3(1 - r_m^2/c_W^2)} + {{\cal O}} (r_g^4) \right), \notag \\
M_{\rho^{0}_{(1)'}}^2  &= m_\rho^2 \left( 1
- \frac{
2 \left\{ - 4 r_m^2 t_W^2 + (3 + t_W^2) 
- \sqrt{16 r_m^4 t_W^4 - 4 r_m^2 t_W^2 (3-t_W^2) + (3-t_W^2)^2}  \right\} 
r_g^2}{3(1 - r_m^2/c_W^2)}
 + {{\cal O}}(r_g^4) \right), \notag \\ 
M_{\rho^{3}_{(1)'}}^2  &= m_\rho^2 \,,
}
with $m_Z = m_W/c_W$. 
The $Z$ mass ($M_Z$) multiplied by $\cos\theta_W$ 
in the above appears to differ from 
the $W$ mass ($M_W$) in Eq.(\ref{EQ:massbasis}) at the order of ${\cal O}(r_g^2)$. 
The deviation can actually be absorbed  by redefinition of the Weinberg angle, 
so that no sizable correction to the $\rho$ parameter is present, 
consistently with the custodial symmetric mixing {form} in Eq.(\ref{mass:mixing}). 
(This will also be clearly seen when the $\rho$ parameter is explicitly evaluated below.)

The mass eigenstate $A$ and $Z$ boson fields arise from the gauge eigenstate fields $(W^3,B,\rho_{(1)}^3,\rho_{(1)'}^0)$ as 
\al{
A_\mu &\simeq s_W  W_\mu^3 + c_W  B_\mu 
+ \left( 2 s_W r_g \right) \rho_{(1) \mu}^3 
+ \left( \frac{2}{\sqrt{3}} s_W r_g\right) 
\rho_{(1)'\mu}^0, 
\label{EQ:Amassbasis} \\ 
Z_\mu &\simeq 
c_W  W_\mu^3 
- s_W  B_\mu  
+ \left( 2 c_W r_g\right)  \rho_{(1)\mu}^3 
+ \left( \frac{2}{\sqrt{3}} s_W t_W r_g \right) \rho_{(1)'\mu}^0 \,,    
\label{EQ:Zmassbasis}
}
to the nontrivial leading order of $r_g$ and $r_m$. 
One can easily see that, to the leading order of the {($r_g$, $r_m$)}-perturbation, the weak mixing structure for the neutral $A$ and $Z$ gauge bosons are precisely the same as in the SM. 
As will be shown later, the vector mixing in the present model does not give {significant corrections to SM interactions of the EW sector so that oblique corrections are negligible up to the {order} of $r_g^2$.}

Finally, we examine the colored part written by 
\al{
\frac{1}{2} \frac{m_\rho^2}{g_\rho^2}
\begin{pmatrix} G_a & \rho^{0}_{(8)a} \end{pmatrix}
\begin{pmatrix}
g_s^2 & -\sqrt{2} g_s g_\rho \\ -\sqrt{2} g_s g_\rho & 2 g_\rho^2
\end{pmatrix}
\begin{pmatrix} G_a \\ \rho^{0}_{(8)a} \end{pmatrix},
}
which holds zero determinant {and thus guarantees} the massless property of the gluon.
The mass eigenstates $\tilde{G}_a$ and $\tilde{\rho}_{(8)a}$ are given as
\al{
G_a^\mu &= \frac{g_\rho \widetilde{G}_a^\mu + \sqrt{2} g_s \widetilde{\rho}_{(8)a}^\mu}{\sqrt{g_\rho^2 + 2g_s^2}},&
\rho_{(8)a}^\mu &= \frac{\sqrt{2} g_s \widetilde{G}_a^\mu - g_\rho \widetilde{\rho}_{(8)a}^\mu}{\sqrt{g_\rho^2 + 2g_s^2}},  
}
with
\al{
M_{G}^2 &= 0,&
M_{\rho_{(8)}}^2 &= m_\rho^2 (1 + 2 r_{g_s}^2).
}
Here, the ratio $r_{g_s}$ is defined as $g_s/g_\rho$.

The indirect couplings of the HC $\rho$ mesons to SM fermions thus arise from the above flavor-universal $V_{\rm SM}$-$\rho$ mixings in the mass {eigenstates}.
As seen from the expressions of the mixings, such flavor-universal couplings are suppressed for $r_g \ll 1$, namely, for large $g_\rho$, which is also required for the oblique corrections to be negligible. 
This is, indeed, inferred from the QCD case. (See the later sections.)
On the other hand, flavor-specific couplings of the HC $\rho$ mesons to SM fermions are also given with the form $g_\rho g_L^{ij}$ {as in Eq.(\ref{couplings:rhoff:II})} and then it {can significantly contribute} to variety of flavor processes, as we will see in the next section.

\subsubsection{Couplings including HC $\pi$}
%%%%

From the chiral Lagrangian in Eq.(\ref{Lag:HLS}) with the concrete form of the covariantized Maurer--Cartan one forms in Eqs.(\ref{alpha:expand}) and (\ref{alpha:expand:2}), 
we find that the following types of HC pion coupling terms emerge after the expansion (up to the quartic order in fields): 
$\rho$-$\pi$-$\pi$,
${\cal V}$-$\pi$-$\pi$,
${\cal V}$-${\cal V}$-$\pi$-$\pi$ and
$\pi$-$\pi$-$\pi$-$\pi$. 
Their interaction forms  easily read 
\al{
{\cal L}_{\text{$\rho$-$\pi$-$\pi$}}
	&= a g_\rho i \, {\rm tr}\left[ [\partial_\mu \pi, \pi] \rho^\mu \right],
	\label{eq:rho-pi-pi_form}\\
{\cal L}_{\text{${\cal V}$-$\pi$-$\pi$}}
	&= 2 i \left( 1 - \frac{a}{2} \right) \, {\rm tr}\left[ [\partial_\mu \pi, \pi] {\cal V}^\mu \right], 
	\label{eq:V-pi-pi_form}\\
{\cal L}_{\text{${\cal V}$-${\cal V}$-$\pi$-$\pi$}}
	&= - {\rm tr} \left\{ \left[ {\cal V}_\mu, \pi \right] \left[ {\cal V}^\mu, \pi \right] \right\}, \\
{\cal L}_{\text{$\pi$-$\pi$-$\pi$-$\pi$}}
	&= - \frac{3}{f_\pi} {\rm tr} \left\{ (\partial_\mu \pi) \left[ \pi, \left[ \pi, \partial^\mu \pi \right] \right] \right\},
}
with $a \equiv {m_\rho^2}/({g_\rho^2 f_\pi^2})$. 
The specific choice, $a = 2$, turns out {to make} the ${\cal V}$-$\pi$-$\pi$ term vanishes at the leading order. 
This is referred to as the vector dominance in which the chiral perturbation theory reproduces experimental results regarding QCD. 
For the present study, we may therefore assume the vector dominance scenario and then see that the $\rho$-$\pi$-$\pi$ coupling $g_{\rho\pi\pi}$ is completely set by the $g_\rho$: $g_{\rho \pi\pi} = g_\rho$. 
As a reference number, we take~\footnote
{The dependence of the $\rho$-$\pi$-$\pi$ coupling $g_{\rho\pi\pi}$ 
on the number of constituent fermions ($N_F$) 
would not be significant, although the overall coefficient 
of $\rho$-$\pi$-$\pi$ vertex scales with $1/\sqrt{N_F/2}$ 
as discussed in Refs.~\cite{Fukano:2015zua,Fukano:2015hga}. 
}
\begin{equation} 
 g_\rho|_\textrm{vector dominance}  = 
 g_{\rho \pi\pi}|_{\rm QCD} \simeq  6 
 \,. \label{grho:value}
\end{equation}
Note that this reference size of $g_\rho$ is consistent with the perturbative expansion in the charged and neutral gauge boson sectors as shown above. 
(The full set of concrete forms of the $\rho$-$\pi$-$\pi$ couplings are provided in appendix~\ref{appendix:rho-pi-pi}.)

We also note that taking into account of chiral anomalies results in the presence of (covariantized) Wess--Zumino--Witten~(WZW) terms~\cite{Wess:1971yu,Witten:1983tw}, 
which contain three-point interactions of $\rho$-$\rho$-$\pi$, $\rho$-${\cal V}$-$\pi$ and ${\cal V}$-${\cal V}$-$\pi$. 
The HC pion couplings to the SM fermions would be provided by an extended-HC theory, 
as written down in Eq.(\ref{h-coupling}). 
Phenomenology regarding those couplings 
will briefly be mentioned later when the collider signatures are discussed.

%%%%%%%%%%%%%%%%%%%%%%%%%%%%%%%%%%%%%%%
%%%%%%%%%%%%%%%%%%%%%%%%%%%%%%%%%%%%%%%
%%%%%%%%%%%%%%%%%%%%%%%%%%%%%%%%%%%%%%%
\section{Phenomenological features of HC rho mesons} 
\label{pheno}
%%%%%%%%%%%%%%%%%%%%%%%%%%%%%%%%%%%%%%%
%%%%%%%%%%%%%%%%%%%%%%%%%%%%%%%%%%%%%%%
%%%%%%%%%%%%%%%%%%%%%%%%%%%%%%%%%%%%%%%

As explained in the previous section, the HC composite $\rho$ mesons arise from the {$SU(N_{\rm HC})$} HC gauge theory with 8 new HC fermions at the strong scale of ${\cal O}(1-10)$ TeV.  
The total 63 {(or 15 distinct)} massive composite vectors, $\rho^\mu$, listed in Table~\ref{tab:TR} have {identical masses $m_\rho$ in the gauge basis.}
Those 63 HC $\rho$ meson properties are characterized as follows: 
\begin{itemize}
 \item the HC $\rho$ mesons are categorized into three types for $SU(3)_c$: 
 octet ($\rho_{(8)}^{{\alpha},0}$), triplet ($\rho_{(3)}^{{\alpha},0}$), and singlet ($\rho_{(1)^\prime}^{{\alpha},0}$, $\rho_{(1)}^{{\alpha}}$) states with triplet(${\alpha}$)/singlet($0$) isospins of {$SU(2)_W$}, 
 \item $\rho_{(3)}$ couples to quark-lepton pairs, referred to as leptoquark~\cite{Buchmuller:1986zs},  
 \item no {direct} interaction to gluon even though $\rho_{(8)}$ and $\rho_{(3)}$ are charged under $SU(3)_c$ . 
\end{itemize}
The new interactions of $V^\mu (=\rho^\mu, V_\text{SM}^\mu)$ to the SM fermions are described in \eqref{couplings:rhoff:II}. 
There also exists the $\rho^\mu$ - $V_\text{SM}^\mu$ mixing in the mass matrix. 
In the following part, we will discuss 
phenomenological features of the present HC $\rho$ model, 
derived from the above fundamental interaction properties.

\subsection{Mass splitting of HC $\rho$ mesons} 
%%%%%%%%%%%%%%%%%%%%%
%%%%%%%%%%%%%%%%%%%%%

As seen in the previous section, one of the charged HC $\rho$ meson, $\rho^\pm_{(1)}$, gets the correction on the mass due to the $\rho$ - $V_\text{SM}$ mixing, 
while another one $\rho^\pm_{(1)'}$ does not change its mass. 
Applying the same analysis, the masses of all the HC $\rho$ mesons in the mass eigen-basis are obtained as 
\begin{align} 
 \hspace{-4em}
 M_{\rho_{(1)'}^{\pm,3}}^2 = M_{\rho_{(8)}^{\pm,3}}^2 = M_{\rho_{(3)}^{\pm,3}}^2 = M_{\rho_{(3)}^{0}}^2 
 &= m_\rho^2 \,, \notag \\
 M_{\rho_{(1)'}^{0}}^2
 &\simeq m_\rho^2 \,, \notag \\
 M_{\rho_{(1)}^{\pm}}^2 
 &\simeq (1 + 4r_g^2) \,m_\rho^2 \,, \notag \\
 M_{\rho_{(1)}^{3}}^2 
 &\simeq (1 + 4r_g^2) \,m_\rho^2 \,, \notag \\
 M_{\rho_{(8)}^{0}}^2 
 &= (1 + 2r_{g_s}^2) \,m_\rho^2 \,, 
 	\label{eq:rho_mass_splitting_summary}
\end{align}
where $r_{g} = g_W / g_\rho$ and $r_{g_s} = g_s / g_\rho$. 
Recall that we now take $g_\rho =6$ [in Eq.(\ref{grho:value})] as a reference number by following the discussion in the previous section.
Thus, one finds that the masses of the HC $\rho$ mesons are almost degenerated even in the mass {eigenbasis}.

\subsection{Flavor universal corrections to the EW sector
 \label{sec:Higgs_couplings}} 
%%%%%%%%%%%%%%%%%%%%%
%%%%%%%%%%%%%%%%%%%%%

The mass mixing forms in Eq.(\ref{massmat:CC}) and (\ref{massmat:NC}) imply the existence of non standard EW couplings to SM fermions. 
In the case of the large $g_\rho$, as in the reference point inferred from the QCD case (See Eq.(\ref{grho:value})), the EW gauge fields $(W^{\pm, 3}, B)$ can be treated as external gauge fields. 
In that case, one can directly read off the current correlators ($\Pi$: transversely polarized gauge boson amplitudes) composed of SM fermions from the mass matrices in Eqs.(\ref{massmat:CC}) and (\ref{massmat:NC}). 
{Taking into account new physics contributions from the $\rho_{(1)}^{\pm, 3}$ and $\rho_{(1)'}^0$ exchanges,} one thus finds 
\begin{eqnarray} 
\Pi_{W^+W^-}(Q^2) 
&\simeq& {- \frac{s_W^2}{e^2} \left[ m_W^2 + 4 r_g^2 m_\rho^2 \left( \frac{Q^2}{m_\rho^2 + Q^2} \right) \right]}
\,, \notag \\[0.5em]
\Pi_{W^3W^3}(Q^2) 
&\simeq& {- \frac{s_W^2}{e^2} \left[ m_W^2 + 4 r_g^2 m_\rho^2 \left( \frac{Q^2}{m_\rho^2 + Q^2} \right) \right]}
\,, \notag \\[0.5em] 
\Pi_{W^3B}(Q^2) 
&\simeq& {-\frac{s_W c_W}{e^2} \left[ - t_W m_W^2 \right]}
\,, \notag \\[0.5em] 
\Pi_{BB}(Q^2) 
&\simeq& {-\frac{c_W^2}{e^2} \left[ t_W^2 m_W^2 + \frac{4}{3} t_W^2 r_g^2 m_\rho^2 \left( \frac{Q^2}{m_\rho^2 + Q^2} \right) \right]}
\,, \label{Pis}
\end{eqnarray}   
where {$Q^2 \equiv -q^2$ is the space-like external momentum squared,} and terms of ${\cal O}(r_g^2)$ corrections have only been included. 
With these correlators at hand, one can extract 
{the flavor-universal EW (oblique) parameters}, 
called $\hat{S}, \hat{T}, W$ and $Y$~\cite{Barbieri:2003pr}: 
 \begin{eqnarray} 
\hat{S} &=& 
\frac{e^2}{s_W^2} 
\frac{d \Pi_{W^3B}(Q^2)}{d(-Q^2)}\Bigg|_{Q^2=0} 
\,, \nonumber \\[0.5em]
\hat{T} &=& 
\frac{e^2}{s^2_W m_W^2} 
\left[  \Pi_{W^3W^3}(0) - \Pi_{{W^+ W^-}}(0) \right] 
\,, \nonumber \\[0.5em]
W &=& \frac{e^2 m_W^2}{2 s_W^2} 
\frac{d^2 \Pi_{{W^3W^3}}(Q^2)}{d(-Q^2)^2} \Bigg|_{Q^2=0} 
\,, \nonumber \\[0.5em]
Y &=& \frac{e^2 m_W^2}{2 c_W^2} 
 \frac{d^2 \Pi_{BB}(Q^2)}{d(-Q^2)^2} \Bigg|_{Q^2=0} 
\,,
\end{eqnarray} 
which are converted into the Peskin-Takeuchi $S$ and $T$ parameters~\cite{Peskin:1991sw} as~\cite{Chivukula:2004af}  
\begin{eqnarray} 
\alpha_{\rm em} S 
&=&  
4 s_W^2 (\hat{S} - Y - W) 
\,, \nonumber \\ 
\alpha_{\rm em} T 
&=&  
\hat{T} - \frac{s_W^2}{c_W^2}  Y  
\,, 
\end{eqnarray}
where $\alpha_{\rm em}=e^2/(4 \pi)$. 
From Eq.(\ref{Pis}) we obtain the oblique parameters as 
\begin{eqnarray}
 \hat{S} &\simeq& 0 \,, \notag \\ 
 \hat{T} &\simeq& 0 \,, \notag \\ 
 W &\simeq& {4 r_m^2 \cdot r_g^2} \,, \notag \\ 
 Y &\simeq& {\frac{4 t_W^2}{3}} r_m^2 \cdot r_g^2 
 \,, 
\end{eqnarray}
where $r_m =m_W/m_\rho$ and $r_g =g_W/g_\rho$. 
Therefore, the $S$ and $T$ parameters in the present HC $\rho$ model are suppressed by a factor $r_m^2 \cdot r_g^2$. 
Up to the order of ${\cal O}(r_x^2)$ for $x=m,g$, we have
 \begin{equation} 
  \alpha_{\rm em} S \simeq \alpha_{\rm em} T \simeq 0 
  \,. 
 \end{equation}
The result implies that the HC $\rho$ mesons are insensitive to the flavor-universal EW sector.  

\vspace{5em}

\subsection{Key contributions of $\rho$-$f_L$-$f_L$ couplings} 
%%%%%%%%%%%%%%%%%%%%%
%%%%%%%%%%%%%%%%%%%%%

Besides the (negligible) flavor-universal interactions of the HC $\rho$, there are lots of flavor-specific interactions {in} the $\rho$ - $f_L$ - $f_L$ terms as described in Eq.\eqref{couplings:rhoff:II}.
Here we summarize and extract significant terms relevant for flavor and collider physics.

\subsubsection{$\rho_{(1)}$ - $f$ - $f^{(\prime)}$ : }
Decomposing the Lagrangian of Eq.\eqref{couplings:rhoff:II}, we have 
\begin{align}
 &
 \label{eq:decrhouu1}
 \mathcal L_{\rho_{1}uu} = -g_\rho g_L^{ij} \bar u_L^i \gamma_\mu u_L^j \, \left[ +{1 \over 4}~\rho_{(1)3}^\mu +{1 \over 4\sqrt 3}~\rho_{(1)'3}^\mu +{1 \over 4\sqrt 3}~\rho_{(1)'0}^\mu \right] \,, \\[0.5em]
 &
 \mathcal L_{\rho_{1}dd} = -g_\rho g_L^{ij} \bar d_L^i \gamma_\mu d_L^j \, \left[ -{1 \over 4}~\rho_{(1)3}^\mu -{1 \over 4\sqrt 3}~\rho_{(1)'3}^\mu +{1 \over 4\sqrt 3}~\rho_{(1)'0}^\mu \right] \,, \\[0.5em]
 &
 \mathcal L_{\rho_{1}\ell\ell} = -g_\rho g_L^{ij} \bar e_L^i \gamma_\mu e_L^j \, \left[ -{1 \over 4}~\rho_{(1)3}^\mu +{\sqrt 3 \over 4}~\rho_{(1)'3}^\mu -{\sqrt 3 \over 4}~\rho_{(1)'0}^\mu \right] \,, \\[0.5em]
 &
 \mathcal L_{\rho_{1}\nu\nu} = -g_\rho g_L^{ij} \bar \nu_L^i \gamma_\mu \nu_L^j \, \left[ +{1 \over 4}~\rho_{(1)3}^\mu -{\sqrt 3 \over 4}~\rho_{(1)'3}^\mu -{\sqrt 3 \over 4}~\rho_{(1)'0}^\mu \right] \,,  
\end{align}
for neutral HC $\rho$ and 
\begin{align}
 &
 \mathcal L_{\rho_{1}ud} = -g_\rho g_L^{ij} \bar u_L^i \gamma_\mu d_L^j \, \left[ {1\over 2 \sqrt 2} \rho_{(1)+}^\mu + {1\over 2 \sqrt 6} \rho_{(1)'+}^\mu  \right] + \text{h.c.} \,, \\[0.5em] 
 &
 \mathcal L_{\rho_{1}\nu\ell} = -g_\rho g_L^{ij} \bar \nu_L^i \gamma_\mu e_L^j \, \left[ {1\over 2 \sqrt 2} \rho_{(1)+}^\mu - {\sqrt 3 \over 2 \sqrt 2} \rho_{(1)'+}^\mu  \right] + \text{h.c.} \,, 
\end{align}
for charged HC $\rho$. 
We also have terms from the indirect interaction of Eq.\eqref{mass:mixing}, but contributions are suppressed due to large $g_\rho$ ($= 6$ as in Eq.(\ref{grho:value})). 
Combining the two contributions, it is typically given such as $g_\rho \left(g_L^{ij} + \delta^{ij} {{\cal O}} (r_g^2) \right)$ for neutral HC $\rho$ whereas $g_\rho \left(g_L^{ij} + V_\text{CKM}^{ij} {{\cal O}}(r_g^2) \right)$ for charged HC $\rho$. 
For almost all case, we simply neglect the suppressed terms of $\rho_{(1,8)}$ - $f$ - $f^{(\prime)}$ in the following study and will get back to this point for $\bar B \to D^{(*)} \tau\bar\nu$.

\subsubsection{$\rho_{(3)}$ - $f$ - $f^{(\prime)}$ : }
As for leptoquark-type HC $\rho$, we have 
\begin{align}
 &
 \mathcal L_{\rho_{3}d\ell} = -g_\rho g_L^{ij} \bar d_L^i \gamma_\mu e_L^j \, \left[ -{1 \over 2}~\rho_{(3)3}^\mu +{1 \over 2}~\rho_{(3)0}^\mu \right] + \text{h.c.} \,,  \\[0.5em]
 &
 \mathcal L_{\rho_{3}u\ell} = -g_\rho g_L^{ij} \bar u_L^i \gamma_\mu e_L^j \, \left[ +{1 \over \sqrt 2}~\rho_{(3)+}^\mu \right] + \text{h.c.} \,, \\[0.5em]
 &
 \mathcal L_{\rho_{3}d\nu} = -g_\rho g_L^{ij} \bar d_L^i \gamma_\mu \nu_L^j \, \left[ + {1 \over \sqrt 2}~\rho_{(3)-}^\mu \right] + \text{h.c.} \,, \\[0.5em] 
 &
 \mathcal L_{\rho_{3}u\nu} = -g_\rho g_L^{ij} \bar u_L^i \gamma_\mu \nu_L^j \, \left[ +{1 \over 2}~\rho_{(3)3}^\mu +{1 \over 2}~\rho_{(3)0}^\mu \right] + \text{h.c.} \,,  
\end{align}
where $\rho_{(3)0}$ and $\rho_{(3)3}$ have $+{2 \over 3}$ electric charges whereas $\rho_{(3)\pm} = (\rho_{(3)1} \mp i\rho_{(3)2})/\sqrt 2$ have $+{5 \over 3}$ and $-{1 \over 3}$, respectively. 
{Here, we suppressed the indices of the $SU(3)_c$ fundamental representations.}
The leptoquark-type HC $\rho$ has no indirect term from the beginning.

\subsubsection{$\rho_{(8)}$ - $f$ - $f^{(\prime)}$ : }
Finally, $\rho_{(8)}$ couples to quarks with the following form. 
\begin{align}
 &
 \mathcal L_{\rho_{8}uu} = -g_\rho g_L^{ij} \bar u_L^i \gamma_\mu \left(\lambda^a \over 2\right) u_L^j \left[ + {1\over\sqrt 2} \rho_{(8)3}^{a\mu} + {1\over \sqrt 2} \rho_{(8)0}^{a\mu} \right] \,, \\[0.5em]
 &
 \mathcal L_{\rho_{8}dd} = -g_\rho g_L^{ij} \bar d_L^i \gamma_\mu \left(\lambda^a \over 2\right) d_L^j \left[ - {1\over\sqrt 2} \rho_{(8)3}^{a\mu} + {1\over \sqrt 2} \rho_{(8)0}^{a\mu} \right] \,, \\[0.5em]
 &
 \label{eq:decrhouu8}
 \mathcal L_{\rho_{8}ud} = -g_\rho g_L^{ij} \bar u_L^i \gamma_\mu \left(\lambda^a \over 2\right) d_L^j \left[ + \rho_{(8)+}^{a\mu} \right] \,. 
\end{align}

Note that all the above $\rho-f-f'$ couplings accompany with $g_L^{ij}$, the size of which will actually be constrained by {flavor-dependent} EW precision tests as seen below.

\subsection{Deviations of flavor-dependent $V_\text{SM}$-$f_L$-$f_L$ couplings from the SM} 
%%%%%%%%%%%%%%%%%%%%%
%%%%%%%%%%%%%%%%%%%%%

The flavor-dependent couplings with the strength of $g_L^{ij}$ in Eq.\eqref{couplings:rhoff:II} give rise to corrections to the SM gauge-fermion-fermion coupling properties. 
The effective vertex for such a $V_\text{SM}$-$f$-$f$ interaction consists of two sources: {the direct term of $V_\text{SM}$-$f$-$f$ and the term of $\rho$-$f$-$f$ through the $\rho$-$V_\text{SM}$ mixing.} 
It is typically described by 
\begin{align} 
\Gamma_\text{$V_\text{SM}$-$f^i$-$f^j$}(Q^2) 
 &= 
 g_L^{ij}\, g_{V_\text{SM}}~\text{(`direct')} + g_L^{ij} g_\rho \cdot \frac{1}{q^2 - m_\rho^2} \cdot \frac{m_\rho^2}{g_\rho^2} \cdot g_\rho\, g_{V_\text{SM}}~\text{(`mixing')} \notag \\
 &=
 g_L^{ij}\, g_{V_\text{SM}} \frac{Q^2}{m_\rho^2 + Q^2} \simeq - g_L^{ij}\, g_{V_\text{SM}} \frac{m_{V_\text{SM}}^2}{m_\rho^2} \,, 
 \label{shift-WZ-coupling}
\end{align}
at the low energy scale ($q^2 = -Q^2 \simeq m_{V_\text{SM}}^2$). 
This is the salient feature of the HLS formulation: the composite vector bosons as gauge bosons in HLS are correlated with the SM gauge bosons $V_\text{SM} (=G,W,B)$, 
through the nonlinear realization.

As a consequence, the corrections in the EW sector are obtained as  
\begin{align}
 \Delta {\cal L}_\text{$W/Z$-$f$-$f$}^\text{total} = 
- \sum_{\psi=q,l} \bar{\psi}_L^i \gamma^\mu \Bigg[ 
&
\Delta_{W}^{ij}\, \frac{e}{\sqrt{2}s_W} \left( W_\mu^+ I^+ + W_\mu^- I^- \right) 
\notag \\
&
+ \Delta_{Z}^{ij}\, \frac{e}{s_Wc_W} \left( I^3 - s_W^2 Q_{\rm em}^\psi \right) Z_\mu \Bigg] \psi_L^j \,, 
\label{eff:L:ff}
\end{align} 
with the usual notations of 
$I^\pm = \tau^1 \pm i \tau^2$, and $I^3 = \tau^3$, 
where the effective corrections $\Delta_{W/Z}$ are written by 
\begin{equation} 
 \Delta_{W/Z}^{ij} 
 \simeq g_{L}^{ij} \cdot \frac{m_{W/Z}^2}{m_\rho^2}  
 ~\simeq~ g_{L}^{ij} (1+8r_g^2) \cdot \frac{M_{W/Z}^2}{M_{\rho_{(1)}^\pm}^2}  
 ~\simeq~ g_{L}^{ij} \cdot \frac{M_{W/Z}^2}{M_{\rho_{(1)}^\pm}^2}  \,. 
\label{delta:WZ} 
\end{equation}
Note again that $m_V$ is the mass in the gauge basis whereas $M_V$ in the mass-eigen basis. 
For the final form of the above result, we keep the term up to {${{\cal O}}(r_x^2)$ for $x=m,g$.}

A key point of this model is that the coupling $g_L^{ij}$ in the shift parameter of Eq.(\ref{delta:WZ}) share with the HC $\rho$ couplings to the SM fermions [in Eq.(\ref{couplings:rhoff:II})] {\it as a consequence of the HLS formulation, which allows to introduce the SM gauges as a remnant of the spontaneous breaking of the (gauged)``chiral'' and {hidden local} symmetries}. 
This crucial feature puts severe constraints on both diagonal/off-diagonal components of ${g_L^{ij}}$ in the gauge basis, as we will see later.

\subsection{Flavor-dependent constraints from the EW sector  \label{sec:Fundamental_requirements}} 
%%%%%%%%%%%%%%%%%%%%%
%%%%%%%%%%%%%%%%%%%%%
In this subsection we show constraints on the flavorful coupling strength $\Delta_{W/Z}^{ij}$ 
$(g_L^{ij})$ in Eq.(\ref{eff:L:ff}), required from 
flavor-dependent EW precision tests. 
We also discuss {a} reasonable setup for the parameters, 
followed by flavor and collider limits in the next sections.

As can be seen in Eq.\eqref{couplings:rhoff:II}, our model involves lots of new interactions at the tree level, {most of} which are obviously already disfavored. 
In particular, it is easily expected that couplings {to} the first and second generations are severely constrained. 
To avoid such matters as well as to address the flavor anomalies in $B$ decays, the reasonable setup may be given as\footnote{{The flavor-blind choice such as $g_L^{ij} = \text{diag}[1,1,1]$,
where flavor-changing interactions {emerge} in the leptoquark part
in mass eigenbasis,
is disfavored mainly due to severe constraints
from the LHC direct searches of new resonances.}
{Note again that {the} third-generation-philic form is shared with both HC rhos and pions. Its realization by an extended HC sector could be interesting as a future study.}} 
\begin{equation} 
 g_L^{ij} = {\begin{pmatrix} 0 & 0 & 0 \\ 0 & 0 & 0 \\ 0 & 0 & g_L^{33}  \end{pmatrix}^{ij} \,. }
\label{gL33}
\end{equation}
Even when we put the above assumption, constraints on $g_L^{33}$ from precision {measurements} of the electroweak sector must be concerned, 
{which is rephrased in a way that the $Z$-boson couplings to the SM fermions were measured very precisely and a sizable deviation from the SM is immediately disfavored.} 
{In the present model, the form of the $Z$-$f$-$f$ couplings is given as
\al{
\mathcal{L}_{\text{$Z$-$f$-$f$}}
=
\left( \frac{e}{s_Wc_W} \right)
\sum_{\psi = q, l}
\bar{\psi}^{i} \gamma^\mu
\left[
(g^{\psi}_L)^{ij} P_L +
(g^{\psi}_R)^{ij} P_R
\right] Z_\mu
\psi^j,
}
with
\al{
(g^{\psi}_L)^{ij} =
(\delta^{ij} - \Delta_{Z}^{ij}) \left( I^3 - s_W^2 Q_{\rm em}^\psi \right),\quad
(g^{\psi}_R)^{ij} =
 \delta^{ij} \left( - s_W^2 Q_{\rm em}^\psi \right){,}
	\label{eq:definition_gLR}
}
{{where} $P_{L,R}$ are the chiral projectors defined as $P_{\,^L_R} = (1\mp\gamma^5)/2$.}

The deviation from the SM for the $Z$ couplings to the left-handed tau lepton in Eq.(\ref{eff:L:ff}) is severely constrained by the forward-backward asymmetry, $A_{\rm FB}^{(0,\tau)}$.}
The asymmetry $A_{\rm FB}^{(0,l)}$ (for unpolarized electron-positron beams) is defined as $A_{\rm FB}^{(0,l)} = 3 A_l A_e/4$ with {$A_l = (|g_{lL}|^2 - |g_{lR}|^2)/(|g_{lL}|^2 + {|g_{lR}|^2)}$~\cite{ALEPH:2005ab}}. 
In the present model, left-handed and right-handed couplings of the $Z$ boson are read off from Eq.(\ref{eq:definition_gLR}) as 
\begin{eqnarray} 
 g_{lL} = (1 - \Delta_{Z}^{33} \delta^{l\tau}) (I^3_l - s_W^2 Q_{\rm em}^l) 
\,,~~~
 g_{lR} = - s_W^2  Q_{\rm em}^l 
\,,  \label{gV:gA} 
\end{eqnarray}
where we ignored the mixing effect {of the lepton fields by rotating gauge to mass eigen-bases.}
The present experimental value $A_{\rm FB}^{(0,\tau)}|_{\rm exp} = 0.0188 \pm 0.0017$ and the SM prediction $A_{\rm FB}^{(0,l)}|_{\rm SM} = 0.01622 \pm 0.00009$~\cite{Olive:2016xmw} still allow the small discrepancy,  
\begin{align}
  {\left.A_{\rm FB}^{(0,\tau)}\right|_{\rm exp} \over \left.A_{\rm FB}^{(0,\tau)}\right|_{\rm SM}} = 1.159 \pm 0.105 \,, 
\end{align}
within the uncertainty. 
This is then compared with {the HC-rho model value}
\begin{align}
  {\left.A_{\rm FB}^{(0,\tau)}\right|_{\text{HC} \rho,\Delta^{33}_{Z} \neq 0} \over \left.A_{\rm FB}^{(0,\tau)}\right|_{\text{HC} \rho,\Delta^{33}_{Z} = 0}} 
  \simeq 
  1 -7.524 \Delta^{33}_{Z} -4.745 (\Delta^{33}_{Z})^2 \,, 
\end{align}
where we take $s_W^2 = 0.2336$ in the numerical analysis~\footnote
{The Weinberg angle (so-called the $Z$ mass shell value) is derived from the relation
\begin{equation*}
\frac{G_F}{\sqrt{2}} = \frac{\pi \alpha_{\rm em}(M_Z)|_{\overline{\text{MS}}}}{2 s_W^2 (1-s_W^2) M_Z^2},
\end{equation*}
with the input variables $G_F = 1.1663787 \times 10^{-5}\,\text{GeV}^{-2}$, $M_Z = 91.1876\,\text{GeV}$, $\alpha_{\rm em}(M_Z)|_{\overline{\text{MS}}}^{-1} = 127.950$~\cite{Olive:2016xmw}.
}. 
As a result, the allowed range is obtained as 
\al{
& 1\sigma~:~ -3.6 \times 10^{-2} < (\Delta^{33}_{Z})|^{1\sigma}_{\text{FBA}} < -7.2 \times 10^{-3} \,, \\
& 2\sigma~:~ -5.1 \times 10^{-2} < (\Delta^{33}_{Z})|^{2\sigma}_{\text{FBA}} < +6.7 \times 10^{-3} \,.
}

The shifts for the $Z$-boson onshell couplings also contribute to the $Z \to {\rm hadrons}$ decay amplitudes.
The most precise measurement has been done {by $R_b = {\Gamma(Z \to b\bar{b})}/{\Gamma(Z \to {\rm hadrons})}$.} 
It is described as {$R_b \simeq (|g_{bL}|^2 + |g_{bR}|^2)/{\sum_{q=u,d,s,c,b} \left( |g_{qL}|^2 + |g_{qR}|^2  \right)}$} neglecting the tiny phase-space difference,    
where the $Z$ boson couplings to quarks $(g_{q_L,q_R})$ in the present model are read off from Eqs.(\ref{eq:definition_gLR}): 
\begin{eqnarray}
 g_{qL} = (1 - \Delta_{Z}^{33}\, \delta^{qb}) ({I^3_{q}} - s^2_W  Q_{\rm em}^q) 
 \,,~~~
 g_{qR} = - s^2_W Q_{\rm em}^q 
 \,, \label{gL:gR}
\end{eqnarray}
{where we again ignored the mixing effect {of the quark fields via the flavor rotation in the mass-eigen basis, {(which is indeed the case for our scenario as seen in the next section.)}}}
{The present experimental value of $R_b|_{\rm exp} = 0.21629 \pm 0.00066$ and the SM prediction} $R_b|_{\rm SM}=0.21579 \pm 0.00003$~\cite{Olive:2016xmw} allow the tiny discrepancy, 
\begin{align}
  {\left. R_b \right|_{\rm exp} \over \left. R_b \right|_{\rm SM}} = 1.0023 \pm 0.0031 \,.  
\end{align}
Comparing this with {the HC-rho model value}
\begin{align}
  {\left. R_b \right|_{\text{HC} \rho,\Delta^{33}_{Z} \neq 0} \over \left. R_b \right|_{\text{HC} \rho,\Delta^{33}_{Z} = 0}} 
  \simeq
  1 -1.509 \Delta^{33}_{Z} +0.113 (\Delta^{33}_{Z})^2 \,, 
\end{align}
we have 
\al{
& 1\sigma~:~ -3.6 \times 10^{-3} < (\Delta^{33}_{Z})|^{1\sigma}_{R_b} < +4.9 \times 10^{-4} \,, \\
& 2\sigma~:~ -5.6 \times 10^{-3} < (\Delta^{33}_{Z})|^{2\sigma}_{R_b} < +2.5 \times 10^{-3} \,.
}

Combing the above constraints, the allowed range for $g_L^{33}$ is obtained as 
\begin{equation} 
 {-0.72 \times \left( \frac{m_\rho}{1 \,{\rm TeV}} \right)^2 < g_L^{33} < 0.25 \times \left( \frac{m_\rho}{1 \,{\rm TeV}} \right)^2 \,,} 
 \label{gL33:cons}
\end{equation}
at {$95\%$ {C.L.}.} 
As we will discuss later, the $1\sigma$ range points to a negative value for $g_L^{33}$ which excludes {a low-mass range for the HC rho mass scale to be consistent with flavor limits}.

To address flavor-changing processes, we {may} introduce the mixing structure between the second and third generations of the down-type quarks and leptons {as will be seen} in Eq.(\ref{LDrotation}).
From the flavor issues discussed in the next section, the down-quark mixing angle is required to be as small as $10^{-3} - 10^{-2}$ and thus $R_b$ is the most relevant EW limit.
As for the lepton mixing angle, on the other hand, we will consider the two cases $\theta_L \lesssim \pi/4$ and $\theta_L \sim \pi/2$. 
In the former case, $A_{\rm FB}^{(0,\tau)}$ is sufficiently significant. 
However, in the latter case, a constraint on $\Delta_Z^{33}$ from $A_{\rm FB}^{(0,\mu)}$ becomes important. 
The experimental data and the SM prediction are given as $A_{\rm FB}^{(0,\mu)}|_{\rm exp} = 0.0169 \pm 0.0013$ and $A_{\rm FB}^{(0,\mu)}|_{\rm SM} = 0.01622 \pm 0.00009$~\cite{Olive:2016xmw}, 
which in turn provide the constraint as
\begin{equation}
{-1.6 \, (-2.7) \times 10^{-2} < (\Delta^{33}_{Z})|^{1(2)\sigma}_{\mu\text{FBA}} < 5.1 \, (15.6) \times 10^{-3}.}
\end{equation}
We will get back to this result later.

%%%%%%%%%%%%%%%%%%%%%%%%%%%%%%%%%%%%%%%
%%%%%%%%%%%%%%%%%%%%%%%%%%%%%%%%%%%%%%%
%%%%%%%%%%%%%%%%%%%%%%%%%%%%%%%%%%%%%%%
\section{Contribution to flavor observables \label{sec:flavor_issues}} 
%%%%%%%%%%%%%%%%%%%%%%%%%%%%%%%%%%%%%%%
%%%%%%%%%%%%%%%%%%%%%%%%%%%%%%%%%%%%%%%
%%%%%%%%%%%%%%%%%%%%%%%%%%%%%%%%%%%%%%%

The HC $\rho$ mesons provide rich phenomenologies 
{addressing} flavor physics as will be discussed in this section.
One easily finds that the interaction terms of $\rho$-$f$-$f$ in Eq.\eqref{couplings:rhoff:II} and Eqs.\eqref{eq:decrhouu1}--\eqref{eq:decrhouu8} produce flavor changing neutral currents~(FCNC) that are severely constrained.

With the assumption {$g_L^{33}\neq 0$ (and $\text{others} = 0$) as in} Eq.\eqref{gL33}, there is no FCNC term in the gauge basis. 
This setup, however, still causes FCNCs in transforming from the gauge basis to the mass basis:  
\al{
{(u_L)^i   = U^{iI} (u'_L)^I,\quad
(d_L)^i   = D^{iI} (d'_L)^I,\quad
(e_L)^i   = L^{iI} (e'_L)^I,\quad
(\nu_L)^i = L^{iI} (\nu'_L)^I,}
	\label{eq:definition_masseigenstates}
}
where $U$, $D$, and $L$ are three-by-three unitary matrices and the spinors with the prime symbol denote the fermions in the mass basis\footnote
{Ambiguities can remain in the transformation of the right-handed neutrinos when (active) neutrinos are massive. 
Here, we consider massless neutrinos, where no ambiguity remains. 
}. 
{The capital latin indices $I,J$ identify the mass eigenstates.}
The CKM matrix element is then given by $V_{\text{CKM}} \equiv U^\dag ({\bf 1} + \Delta_W) D \simeq U^\dag D$
{with $\Delta_W^{33} \le {\cal O}(10^{-3})$ taken into account}. 
According to the literature~\cite{Bhattacharya:2016mcc}, in order to address several {\it flavor anomalies} recently reported in measurements of $\bar B \to K \mu^+\mu^-$ (and $\bar D^{(*)}\tau\bar\nu$) 
as well as to avoid severe constraints of FCNCs in the first and second generations, the mixing structures of $D$ and $L$ are reasonably parametrized by 
\al{
D =
\begin{pmatrix}
{1} & 0 & 0 \\
0 &  \cos{\theta_D} & \sin{\theta_D} \\
0 & -\sin{\theta_D} & \cos{\theta_D} 
\end{pmatrix}, 
\quad\quad
L =
\begin{pmatrix}
{1} & 0 & 0 \\
0 &  \cos{\theta_L} & \sin{\theta_L} \\
0 & -\sin{\theta_L} & \cos{\theta_L}
\end{pmatrix}.
\label{LDrotation}
} 
Through these flavor mixings, we will see significant contributions to flavor phenomenologies.
{The following factors are useful,}
%%%%%%%%%%%%%%%%%%%%%%%%
\al{
{X_{dd}} &\equiv
D^\dagger \begin{pmatrix} 0 & 0 & 0 \\ 0 & 0 & 0 \\ 0 & 0 & 1 \end{pmatrix} D
=
\begin{pmatrix}
0 & 0 & 0 \\
0 &  \sin^2{\theta_D} & -\cos{\theta_D}\sin{\theta_D} \\
0 & -\cos{\theta_D}\sin{\theta_D} & \cos^2{\theta_D}
\end{pmatrix}, \label{eq:M_transform_1} \\
%%%%%%
{X_{ll}} &\equiv
L^\dagger \begin{pmatrix} 0 & 0 & 0 \\ 0 & 0 & 0 \\ 0 & 0 & 1 \end{pmatrix} L
=
\begin{pmatrix}
0 & 0 & 0 \\
0 &  \sin^2{\theta_L} & -\cos{\theta_L}\sin{\theta_L} \\
0 & -\cos{\theta_L}\sin{\theta_L} & \cos^2{\theta_L}
\end{pmatrix}, \label{eq:M_transform_2} \\
%%%%%%
{X_{uu}} &\equiv
U^\dagger \begin{pmatrix} 0 & 0 & 0 \\ 0 & 0 & 0 \\ 0 & 0 & 1 \end{pmatrix} U
=
V_{\text{CKM}}
\begin{pmatrix}
0 & 0 & 0 \\
0 &  \sin^2{\theta_D} & -\cos{\theta_D}\sin{\theta_D} \\
0 & -\cos{\theta_D}\sin{\theta_D} & \cos^2{\theta_D}
\end{pmatrix}
V_{\text{CKM}}^\dagger, \label{eq:M_transform_3} \\
%%%%%%%
{X_{ld}} &\equiv
L^\dagger \begin{pmatrix} 0 & 0 & 0 \\ 0 & 0 & 0 \\ 0 & 0 & 1 \end{pmatrix} D
=
\begin{pmatrix}
0 & 0 & 0 \\
0 &  \sin{\theta_L}\sin{\theta_D} & -\sin{\theta_L}\cos{\theta_D} \\
0 & -\cos{\theta_L}\sin{\theta_D} &  \cos{\theta_L}\cos{\theta_D}
\end{pmatrix}, \label{eq:M_transform_4} \\
%%%%%%%
{X_{lu}} &\equiv
L^\dagger \begin{pmatrix} 0 & 0 & 0 \\ 0 & 0 & 0 \\ 0 & 0 & 1 \end{pmatrix} U
=
\begin{pmatrix}
0 & 0 & 0 \\
0 &  \sin{\theta_L}\sin{\theta_D} & -\sin{\theta_L}\cos{\theta_D} \\
0 & -\cos{\theta_L}\sin{\theta_D} &  \cos{\theta_L}\cos{\theta_D}
\end{pmatrix} V_{\text{CKM}}^\dagger,
\label{eq:M_transform_5} \\
%%%%%%%
{X_{ud}} &\equiv
U^\dagger \begin{pmatrix} 0 & 0 & 0 \\ 0 & 0 & 0 \\ 0 & 0 & 1 \end{pmatrix} D
=
V_{\text{CKM}} D \begin{pmatrix} 0 & 0 & 0 \\ 0 & 0 & 0 \\ 0 & 0 & 1 \end{pmatrix} D
=
V_{\text{CKM}}
\begin{pmatrix}
0 & 0 & 0 \\
0 &  \sin^2{\theta_D} & -\cos{\theta_D}\sin{\theta_D} \\
0 & -\cos{\theta_D}\sin{\theta_D} & \cos^2{\theta_D}
\end{pmatrix},
\label{eq:M_transform_6}
}
{with their charge conjugations}
\al{
{X_{dl} \equiv (X_{ld})^\dagger,\quad
X_{ul} \equiv (X_{lu})^\dagger,\quad
X_{du} \equiv (X_{ud})^\dagger.}
}
%%%%%%%%%%%%%%%%%%%%%%%% 

On the other hand, the indirect couplings derived from Eq.(\ref{mass:mixing}) have no contribution to FCNC phenomena at a tree-level\footnote{One-loop corrections from charged $\rho$ mesons would give extra contributions to the FCNC processes, which are, however, 
negligibly small as far as the $\rho$ mass scale on the order of TeV or higher is concerned.
} 
since these flavor transformations do not produce FCNCs. 
Note that charged flavor transitions such as $b \to c$ can be affected and we will see this point soon later.

\subsection{Effective {four-fermion} operators from the HC $\rho$ {and $\pi$} meson contributions}
\label{Sec:rhoEFT}
%%%%%%%%%%%%%%%%%%%%%
%%%%%%%%%%%%%%%%%%%%%

We see from Eq.{\eqref{couplings:rhoff:II}} that the octet HC mesons $\rho_{(8)}^{{\alpha},0}$ {induce} four-quark operators; the triplet $\rho_{(3)}^{{\alpha},0}$ {contribute} to two-quark and two-lepton operators; 
and the singlet $\rho_{(1^{(\prime)})}^{{\alpha},0}$ to all possible operators including four-lepton operators. 
By integrating out the HC $\rho$ mesons, the {four-fermion} operators are obtained as summarized in Eqs.\eqref{Lagrho8}--\eqref{Lagrho3} of the Appendix~\ref{App:ForFermiOP}. 
Employing Fierz transformations, the operators are simplified into six types: 
\begin{align}
 \mathcal O_{4q(1)}^{ijkl} &= \left( {\bar{q}}_L^i \gamma_\mu {q}_L^j \right) \left( {\bar{q}}_L^k \gamma^\mu {q}_L^l \right) \,,~~~\,
 \mathcal O_{4q(3)}^{ijkl} = \left( {\bar{q}}_L^i \gamma_\mu \sigma^{\alpha} {q}_L^j \right) \left( {\bar{q}}_L^k \gamma^\mu \sigma^{\alpha} {q}_L^l \right) \,, \\ 
 \mathcal O_{4\ell (1)}^{ijkl} &= \left( {\bar{l}}_L^i \gamma_\mu {l}_L^j \right) \left( {\bar{l}}_L^k \gamma^\mu {l}_L^l \right) \,,~~~~~\,
 \mathcal O_{4\ell (3)}^{ijkl} = \left( {\bar{l}}_L^i \gamma_\mu \sigma^{\alpha} {l}_L^j \right) \left( {\bar{l}}_L^k \gamma^\mu \sigma^{\alpha} {l}_L^l \right) \,, \\ 
 \mathcal O_{2q2\ell (1)}^{ijkl} &= \left( {\bar{q}}_L^i \gamma_\mu {q}_L^j \right) \left( {\bar{l}}_L^k \gamma^\mu {l}_L^l \right) \,,~~\, 
 \mathcal O_{2q2\ell (3)}^{ijkl} = \left( {\bar{q}}_L^i \gamma_\mu \sigma^{\alpha} {q}_L^j \right) \left( {\bar{l}}_L^k \gamma^\mu \sigma^{\alpha} {l}_L^l \right) \,,    
\end{align}
{along with the Wilson coefficients having the form}   
\begin{align}
 \label{EQ:couplingrho4f}
 %G_{\rho_{(n)}^{m}}^{ijkl} \equiv 
 {g_\rho^2\, g_L^{ij}\, g_L^{kl} \over m_{\rho_{(n)}^{m}}^2 } \,, 
\end{align}
with an appropriate {factor, derived from} Eqs.\eqref{rho:assign} and \eqref{couplings:rhoff:II}, for every HC $\rho$ meson. 
When we recall the degenerated $\rho$ meson masses, we have 
\begin{align}
 -\mathcal L_\text{eff} = 
 {g_\rho^2\, g_L^{ij}\, g_L^{kl} \over m_{\rho}^2}
  \left [ \Delta^{ij;kl} 
  \left( {1 \over 4} \mathcal O_{4q(3)}^{ijkl} 
  +{1 \over 4} \mathcal O_{4\ell (3)}^{ijkl}   
  +{3 \over 16} \mathcal O_{4q(1)}^{ijkl} 
  +{3 \over 16} \mathcal O_{4\ell (1)}^{ijkl} \right)
  +{7 \over 16} \mathcal O_{2q2\ell (1)}^{ijkl} \right] \,, 
  \label{FFOeffective}
\end{align}
where $\Delta^{ij;kl} = 1/2$ for two identical currents ($i=k, j=l$) and $1$ for the others. 
Note that the $\mathcal O_{2q2\ell (3)}^{ijkl}$ term is canceled out in this model with the degenerated masses. 
This term can be nonzero only if {$M_{\rho_{(1)}^{{\alpha}}} \neq M_{\rho_{(1)'}^{{\alpha}}}$ or $M_{\rho_{(3)}^{{\alpha}}} \neq M_{\rho_{(3)}^0}$}. 
The HLS-gauge invariance in the present model, however, does not allow such a sizable mass-split between them.

{On the other hand, scalar-type contributions to four fermion processes are given by HC pion exchanges in principle. From the interaction of Eq.(\ref{h-coupling}), corresponding contributions involve fermion masses (derived from derivative couplings of HC pions to external fermions), which is the NG boson nature of HC pions. Therefore, a typical size of the contribution is written as 
\begin{align}
\frac{h_L^{ij} h_L^{kl}}{m_\pi^2} \frac{m_{f,\,{\rm max}}^2}{f_\pi^2},
\end{align}
where $h_L^{ij}$ are the coefficients of the operators shown in Eq.(\ref{h-coupling}), $m_\pi$ represents a typical scale of the HC pion mass, and $m_{f,\,{\rm max}}$ indicates the largest mass of the external fermions which describes a typical scale of the momentum in meson-associated flavor physics.

This is compared with the one from the HC rho, $g_\rho^2 g_L^{ij} g_L^{kl}/m_\rho^2$. The couplings could be comparable such as $h_L^{ij} \sim g_\rho g_L^{ij}$ whereas $m_\pi$ becomes comparable with $m_\rho$ as discussed in Sec.~\ref{sec:rho_coupling_to_SMfermion} and Sec.~\ref{sec:mass_rho_and_pi}, respectively. Thus, we expect that the scalar contribution is suppressed by the factor $m_{f,\text{max}}^2 / f_\pi^2$, which is maximally of ${\cal O}(m_b/{\cal O}(10^2\,\text{GeV}))^2 \lesssim (1/20)^{2}${.}
Thereby, it is totally natural to neglect HC pion contributions to flavor physics and then we focus only on the HC rho contributions.\footnote{{An exceptional case is top-quark-associated processes, as we see around Eq.(\ref{eq:amplitude_pi0toGG}).}}}

%{On the other hand, a typical size of coefficients of the dimension-six operators by the exchange of HC pions can be estimated as
%\begin{align}
%\frac{h_L^{ij} h_L^{kl}}{m_\pi^2} \frac{m_{f,\,{\rm max}}^2}{f_\pi^2},
%\end{align}
%where $h_L^{ij}$ are the coefficients of the operators shown in Eq.(\ref{h-coupling}), $m_\pi$ represents a typical scale of the HC pion mass, and $m_{f,\,{\rm max}}$ indicates the largest mass of the external fermions which describes a typical scale of the momentum in meson-associated flavor physics.
%When the HC pion coupling is comparable with the HC rho coupling, i.e., $g_\rho \, g_L^{ij} \sim h_L^{ij}$, even though $m_\pi$ becomes comparable with $m_\rho$ due to huge quantum corrections observed in a walking gauge theory [as in Eq.(\ref{pi:masses})], the ratio ${m_{f,\,{\rm max}}^2}/{f_\pi^2}$ takes an illustrative digit 
%{which is maximally of ${\cal O}(m_b/{\cal O}(10^2\,\text{GeV}))^2 \lesssim (1/20)^{2}$.}
%
%This suppression is essentially due to the NG boson nature coupled to fermions (i.e. derivative coupling, converted to Yukawa coupling proportional to the target fermion mass).
%
%Thereby, contributions of HC pions to flavor physics are highly suppressed, compared with those of HC rho mesons.
%In the following discussion, we neglect HC pion contributions to flavor physics, and focus only on the HC rho contributions.\footnote{{An exceptional case is top-quark-associated processes, as we see around Eq.(\ref{eq:amplitude_pi0toGG}).}}}

\subsection{Effect on flavor observables}
%%%%%%%%%%%%%%%%%%%%%
%%%%%%%%%%%%%%%%%%%%%

When we transform the left-handed SM fermion fields to those in the mass basis by $U$, $D$, and $L$ under the assumptions of Eqs.\eqref{gL33} and \eqref{LDrotation}, 
the Lagrangian of Eq.~\eqref{FFOeffective} generates nonzero FCNC contributions in the $b$ -- $s$ system and the lepton flavor violating system. 
To be specific, intriguing processes are summarized as follows: 
\begin{itemize}
 \item $\mathcal O_{2q2\ell (n)}$: $\bar B \to D^{(*)} \tau\bar\nu$ (only for $n=3$), $\bar B \to K^{(*)} \mu^+\mu^-$, $\bar B \to K^{(*)} \nu\bar\nu$, and $\tau\to\phi\mu$, 
 \item $\mathcal O_{4q(n)}$~~: $B_s$--$\bar B_s$ mixing, 
 \item $\mathcal O_{4\ell(n)}$~~:  $\tau\to 3\mu$. 
\end{itemize}
Then Eq.\eqref{FFOeffective} implies that 
{the net effect of the HC $\rho$ mesons (which arises from the direct $\rho$-$f$-$f$ term of Eq.{\eqref{couplings:rhoff:II}}) on $\bar B \to D^{(*)} \tau\bar\nu$ is exactly zero} at the tree level, 
even though each {color-triplet and color-singlet} HC $\rho$ meson does contribute.

{In the following main body of the manuscript, we skip to show the prime symbol for fermion fields to emphasize them being mass eigenstates.}

\subsubsection{$\bar B \to K^{(*)} \ell^+\ell^-$}
%%%%%%%%%%%%
The effective Hamiltonian for $b \to s \ell^+\ell^-$ in the present model is described by 
\begin{align}
 H_{\rm eff}(b \to s \ell_{I}\bar\ell_{J}) 
 = 
 & - {\alpha G_F \over \sqrt 2 \pi} V_{tb} V_{ts}^* \left( C_9^\text{SM} \delta^{IJ} + C_9^{IJ} \right) 
 \left(\bar s_L \gamma^\mu b_L \right) \left( \bar\ell_{I} \gamma_\mu \ell_{J} \right)  \notag \\
 & - {\alpha G_F \over \sqrt 2 \pi} V_{tb} V_{ts}^* \left( C_{10}^\text{SM} \delta^{IJ} + C_{10}^{IJ} \right) 
 \left(\bar s_L \gamma^\mu b_L \right) \left( \bar\ell_{I} \gamma_\mu \gamma^5 \ell_{J} \right)  \,,  
 \label{EQ:bsellellLag}
\end{align}
with 
\begin{align}
 C_9^{IJ} = - C_{10}^{IJ} = - \frac{\sqrt{2} \pi}{ \alpha G_F V_{tb} V^\ast_{ts}} \cdot {7 \over 32} {g_\rho^2\, (g_L^{33})^2 \over m_{\rho}^2 } X^{23}_{dd} X^{IJ}_{ll} \,, 
	\label{eq:C9C10}
\end{align}
where the conventional coefficients $C_9^\text{SM} \simeq 0.95$ and $C_{10}^\text{SM} \simeq -1.00$ indicate the loop functions from the SM contributions, at the $m_b$ scale, (see, {\it e.g.}, \cite{Bobeth:2013uxa}).

Experimental measurements regarding $b \to s \ell^+\ell^-$ have been developed in {recent} years to search for new physics and then the net results have suggested significant deviations from the SM predictions:  
\begin{itemize}
\item
Since the decay $\bar B \to K^{*} \mu^+\mu^-$ includes detectable final state particles ($\mu^+,\mu^-,K^*\to K\pi$), it has variety of angular observables.  
LHCb and Belle have then measured the optimized observable, so-called $P'_5$~\cite{DescotesGenon:2012zf}, and found a sizable discrepancy between their results and the SM prediction, 
as in Refs.~\cite{Aaij:2013qta,Aaij:2015oid,Abdesselam:2016llu,Wehle:2016yoi}.
Recently, ATLAS and CMS have also reported (preliminary) results for several observables including $P'_5$ in Refs.~\cite{ATLAS-CONF-2017-023,CMS-PAS-BPH-15-008}, which are in agreement with the LHCb and Belle results. 
\item
LHCb {reported} a deviation in observables of $B_s \to\phi\mu^+\mu^-$~\cite{Aaij:2013aln,Aaij:2015esa}. 
\item
The tests of lepton-flavor-universality violation (LFUV) have been examined by measuring the ratio $R_{K^{(*)}} = \mathcal B (\bar B \to K^{(*)} \mu^+\mu^-) / \mathcal B (\bar B \to K^{(*)} e^+e^-)$ at LHCb. 
The result, again, suggests a deviation from the SM prediction in $R_K$~\cite{Aaij:2014ora} and very recently in $R_{K^*}$~\cite{NewRKstarLHCb}. 
Besides $R_{K^{(*)}}$, Belle has recently shown their result on {a} new measurement of the LFUV effect, $P^{\prime\mu}_5 - P^{\prime e}_5$~\cite{Capdevila:2016ivx}, 
which points to a presence of LFUV although the result is not yet statistically significant~\cite{Aaij:2015esa}.  
\end{itemize}
Immediately after the very recent report on $R_{K^*}$~\cite{NewRKstarLHCb}, {global fit analyses for new physics} have been investigated by several theory groups, 
see Refs.~\cite{Capdevila:2017bsm,Altmannshofer:2017yso,DAmico:2017mtc,Geng:2017svp,Ciuchini:2017mik,Hiller:2017bzc,Celis:2017doq,Alok:2017jaf,Alok:2017sui,Wang:2017mrd}. 
The results are usually shown in terms of the conventional coefficients $C_X^{(\prime)}$ for $X=9,10$, (where the primed coefficients are those for the operators replacing $P_L \to P_R$ in Eq.\eqref{EQ:bsellellLag}.) 
According to the literatures, favored solutions to accommodate the present data are obtained in the following three cases (among single degree of freedom of {new physics} contributions); 
$C_{9}^{\mu\mu}\neq 0$, $C_{9}^{\mu\mu} =-C_{10}^{\mu\mu} \neq 0$, and $C_{9}^{\mu\mu} =-C_{9}^{\prime\mu\mu} \neq 0$~\footnote
{To be precise, fit to the LFUV data ($R_K$ and $R_{K^*}$) prefers $C_{9}^{\mu\mu} =-C_{10}^{\mu\mu}\neq 0$ while that to all the data favors  $C_{9}^{\mu\mu}\neq 0$. 
}. 
The present model corresponds to the second case $C_{9}^{\mu\mu} =-C_{10}^{\mu\mu}$ and then the favored region is given as 
\al{
-0.87 \leq C_9^{\mu\mu} = - C_{10}^{\mu\mu} \leq -0.36 \,, 
\label{EQlimit_bsmumu}
}
at the $2\sigma$ level, whereas the best fit point is $-0.61$. 
We have quoted the favored region given in Ref.~\cite{Capdevila:2017bsm}, where all available {associated} data from LHCb, Belle, ATLAS and CMS were combined.

\subsubsection{$\bar B \to K^{(*)} \nu\bar\nu$}
%%%%%%%%%%%%
The effective Hamiltonian for $\bar B \to K^{(*)} \nu\bar\nu$ is written by 
\begin{align}
 H_{\rm eff}(b \to s \nu_{{I}}\bar\nu_{{J}}) 
 = 
 \left( - {\alpha G_F \over \sqrt 2 \pi} V_{tb} V_{ts}^* C_L^\text{SM} \delta^{{IJ}} + {7 \over 32} {g_\rho^2\, (g_L^{33})^2 \over m_{\rho}^2 } {X^{23}_{dd} X^{IJ}_{ll}}  \right)\, 
 \left(\bar s_L \gamma^\mu b_L \right) \left( \bar\nu_{{I}} \gamma_\mu (1-\gamma^5) \ell_{{J}} \right)  \,, 
\end{align}
where the SM loop function is $C_L^\text{SM} \simeq -6.36$. 
The experimental upper bounds on the branching ratios of $\bar B \to K^{(*)} \nu\bar\nu$~\cite{Lees:2013kla,Lutz:2013ftz} put constraints on {new physics} contributions~\cite{Buras:2014fpa}. 
The most severe constraint is then obtained as~\cite{Bhattacharya:2016mcc} 
\al{
-13 \sum_{{I}=1}^{3} \text{Re}[C_L^{{II}}] + \sum_{{I,J}=1}^{3} |C_L^{{IJ}}|^2 \leq 473 \,, 
\label{EQlimit_bsnunu}
}
at $90\%$ C.L., where 
\begin{align}
 C_L^{{IJ}} = -\frac{\sqrt{2} \pi}{ \alpha G_F V_{tb} V^\ast_{ts}} \cdot {7 \over 32} {g_\rho^2\, (g_L^{33})^2 \over m_{\rho}^2 } {X^{23}_{dd} X^{IJ}_{ll}} \,. 
	\label{eq:CLIJ}
\end{align}

\subsubsection{$\tau\to\phi\mu$}
%%%%%%%%%%%%
Our model produces lepton flavor violating decays such as $\tau\to\phi\mu$ and $\tau\to3\mu$ at the tree level while they are much suppressed in the SM. 
The effective Hamiltonian for $\tau\to\phi\mu$ is written by 
\begin{align}
 H_{\rm eff}(\tau\to\mu s \bar s) 
 = 
 {7 \over 16} {g_\rho^2\, (g_L^{33})^2 \over m_{\rho}^2 } {X^{22}_{dd} X^{23}_{ll}} \, ({\bar s}_L \gamma^\mu s_L)\, ({\bar \tau}_L \gamma_\mu \mu_L)~. 
\end{align}
We take the constraint obtained in~\cite{Bhattacharya:2016mcc} from the $90\%$ C.L. upper limit of $\mathcal{B}(\tau \to \mu \phi ) < 8.4 \times 10^{-8}$~\cite{Miyazaki:2011xe}, which leads 
\al{
\left| {7 \over 16} g_\rho^2\, (g_L^{33})^2 {X^{22}_{dd} X^{23}_{ll}} \right| < 0.019 \times \left( { m_{\rho} \over 1\,\text{TeV} } \right)^2~. 
\label{EQlimit_tauphimu}
}

\subsubsection{$\tau\to3\mu$}
%%%%%%%%%%%%
The lepton flavor violating decay $\tau\to3\mu$ plays an important role in this model. 
The effective Hamiltonian is 
\begin{align}
 H_{\rm eff}(\tau^- \to \mu^- \mu^+ \mu^-) 
 = 
 {7 \over 32} {g_\rho^2\, (g_L^{33})^2 \over m_{\rho}^2 } {X^{22}_{ll} X^{23}_{ll}} 
 ({\bar \mu}_L \gamma^\mu \mu_L)\, ({\bar \tau}_L \gamma_\mu \mu_L)~, 
\end{align}
and then the branching ratio is obtained by 
\begin{align}
 \mathcal B (\tau^- \to \mu^- \mu^+ \mu^-)
 = 
 {\left[{7 \over 32} {g_\rho^2\, (g_L^{33})^2 \over m_{\rho}^2 } X^{22}_{ll} X^{23}_{ll} \right]^2} \times \frac{0.94}{4} \frac{m_\tau^5 \tau_\tau}{192\pi^3} \,,  
	\label{eq:BR_tauto3mu}
\end{align}
{where {the factor $0.94$ came from the phase space suppression for the decay}~\cite{Bhattacharya:2016mcc}.}
The following experimental upper bound at $90\%$ C.L. is available~\cite{Hayasaka:2010np}:
\al{
\mathcal{B}(\tau^- \to \mu^- \mu^+ \mu^-) < 2.1 \times 10^{-8}. 
\label{EQlimit_tau3mu}
}

\subsubsection{$B^0_s$-$\bar{B}^0_s$ mixing}
%%%%%%%%%%%%
The effective Hamiltonian is 
\begin{align}
 H_{\rm eff}(bs \leftrightarrow bs) 
 =
 \left( \frac{G_F^2 m_W^2}{16\pi^2} (V_{tb} V_{ts}^*)^2 C_{VLL}^\text{SM} + {7 \over 32} {g_\rho^2\, (g_L^{33})^2 \over m_{\rho}^2 } {X^{23}_{dd} X^{23}_{dd}} \right)
 ({\bar s}_L \gamma^\mu b_L)\,({\bar s}_L \gamma_\mu b_L)~, 
\end{align}
with $C_{VLL}^\text{SM} \simeq 4.95$. 
The mass difference in the $B_s$ system is provided by
\al{
\Delta M_{B_s} = \frac{2}{3} M_{B_s} f_{B_s}^2 \hat{B}_{B_s} \left| \frac{G_F^2 m_W^2}{16\pi^2} (V_{tb} V_{ts}^*)^2 C_{VLL}^\text{SM} + {7 \over 32} {g_\rho^2\, (g_L^{33})^2 \over m_{\rho}^2 } {X^{23}_{dd} X^{23}_{dd}} \right|.
	\label{eq:DeltaMBs_Bsubs}
}
This is compared with the experimental measurement~\cite{Amhis:2014hma}
\al{
\Delta M_{B_s}^{\text{exp.}} = (17.757 \pm 0.021) \, \text{ps}^{-1}.
\label{EQlimit_bsbs}
}
Note that a theoretical uncertainty comes from the input parameters of $V_{tb} V_{ts}^*$ and $f_{B_s}^2 \hat{B}_{B_s}$, which are much more dominant than the above experimental {uncertainty} 
{\it e.g.}, $\Delta M_{B_s}^\text{SM} = (17.4 \pm 2.6) \, \text{ps}^{-1}$.
For a conservative choice, we take $\pm 1\sigma$ range for the theoretical uncertainty.

\subsubsection{$\bar B \to D^{(*)} \tau\bar\nu$}
%%%%%%%%%%%%
The semi-tauonic $B$ meson decays of $\bar B \to D^{(*)} \tau\bar\nu$ were measured~\cite{Lees:2013uzd,Huschle:2015rga,Aaij:2015yra} and then it has turned out that the experimental data deviate from the SM predictions. 
To be specific, with respect to the ratios 
\begin{align}
 R_{D} = \frac{ {\cal B}(\bar B \to D \tau\bar\nu) }{ {\cal B}(\bar B \to D \ell\bar\nu) } \,,~~~~~
 R_{D^*} = \frac{ {\cal B}(\bar B \to D^* \tau\bar\nu) }{ {\cal B}(\bar B \to D^* \ell\bar\nu) } \,,
\end{align}
(for $\ell = e$ or $\mu$), we see the deviations of the combined experimental data~\cite{Lees:2013uzd,Huschle:2015rga,Aaij:2015yra} from the SM of $1.7\sigma$ and $3.3\sigma$, respectively. 
On the other hand, the recent update of the analysis for $R_{D^*}$ at Belle reported in Ref.~\cite{Hirose:2016wfn} shows {consistency} with the SM. 
Combining this new result with the others\footnote{The recent Belle analysis in Ref.~\cite{Hirose:2016wfn} was performed using different decay modes of the tag-side $B$ mesons from the previous one in Ref.~\cite{Huschle:2015rga} 
so that the results in Refs.~\cite{Huschle:2015rga,Hirose:2016wfn} are independent with each other. 
Hence the two results can be statistically combined. 
}, one finds 
\begin{align}
 \frac{R_D^\text{exp}}{R_D^\text{SM}} = 1.29 \pm 0.17 \,,~~~~~
 \frac{R_{D^*}^\text{exp}}{R_{D^*}^\text{SM}} = 1.24 \pm 0.08 \,, 
\end{align}
where the deviations are now $1.7\sigma$ and $3.0\sigma$, respectively. 
Although they are still significant, the SM predictions are consistent within $3\sigma$ range.

The corresponding effective Hamiltonian that contributes to $\bar B \to D^{(*)} \ell\bar\nu$ can be written as 
\al{
H_\text{eff}(b \to c \ell_I \overline{\nu}_J) &= \frac{4 G_F}{\sqrt{2}} V_{cb}
	C_V^{IJ} (\overline{c}_L \gamma_\mu b_L) (\overline{e}_{I} \gamma^\mu \nu_{J}).
}
As we pointed out in Sec.~\ref{Sec:rhoEFT}, the net effect of the charged HC $\rho$ mesons [from Eq.\eqref{couplings:rhoff:II}] on $d \to u \ell\nu$ is exactly zero, namely, 
$$
{C_V^{IJ} =0}{,}
$$
for the degenerated HC $\rho$ masses.
%%%
{This is actually the consequence of the global $SU(8)$
invariance for HC rhos: the cancellation turns out to 
take place between $\rho_{(1)}^\alpha$-$\rho_{(1)'}^\alpha$ 
and $\rho_{(3)}^\alpha$-$\rho_{(3)}^0$ contributions, separately, 
with the common HC rho coupling $g_{\rho} g_L^{ij}$ in the gauge eigenbasis
{[See Eq.(\ref{eq:C3qqll})]}. 
Possible {non-zero} contributions arise from the mass difference in the charged 
HC mesons and the $V_{\rm SM}$-$\rho$ mixing effect 
which break the global $SU(8)$ symmetry. 
{However, both of those effects are suppressed}
%%%
%possible contributions arise from the mass difference in the charged HC $\rho$ mesons and the $V_\text{SM}$-$\rho$ mixing effect, both of which are suppressed 
(controlled) by the factor $g_W^2/g_\rho^2$.} 
For $g_\rho = 6$, {a reference point} {[Eq.(\ref{grho:value})]},
these contributions on $R_{D^{(*)}}$ do not exceed $5\%$ and thus it is not sufficient to account for ${\cal O}(10\%)$ deviations between the present data and the SM predictions in the ratios. 
{At Belle~II}, we may have potential to examine $R_{D^{(*)}}$ with a few $\%$ accuracy. 
Thus, if the discrepancy is reduced to a few $\%$ in future observation at Belle~II, it may point to the contributions from such small effects. 
See appendix~\ref{App:ForFermiOP} for explicit expression on $C_V^{IJ}(\text{NP})$.

\subsection{Allowed region in parameter space}
%%%%%%%%%%%%%%%%%%%%%
%%%%%%%%%%%%%%%%%%%%%

%%%%%%%%%%%%%%% [Table] %%%%%%%%%%%%%%%
\begin{table}[t] 
\begin{tabular}{c}
\hline\hline 
  $V_{tb} V_{ts}^* = -0.0405 \pm 0.0012$~\cite{Olive:2016xmw,Charles:2015gya}, 
  $f_{B_s}\sqrt{\hat B_{B_s}} = (266 \pm 18)$ MeV~\cite{Aoki:2013ldr,Aoki:2016frl}  \\
  $M_{B_s} = 5366.82\,\text{MeV}$~\cite{Olive:2016xmw}, $m_W = 80.4\,\text{GeV}$~\cite{Olive:2016xmw}, $G_F=1.1663787\times 10^{-5}\,\text{GeV}^{-2}$~\cite{Olive:2016xmw} \\
\hline\hline 
\end{tabular}
\caption{ 
Theoretical input to evaluate flavor observables described in the main text. 
  }
\label{tab:input}
\end{table}
%%%%%%%%%%%%%%% [Table] %%%%%%%%%%%%%%%
%
%%%%%%%%%%%%beginFIG1%%%%%%%%%%%%
\begin{figure}[t!!]
\begin{center}
\includegraphics[viewport=0 0 360 398, width=22em]{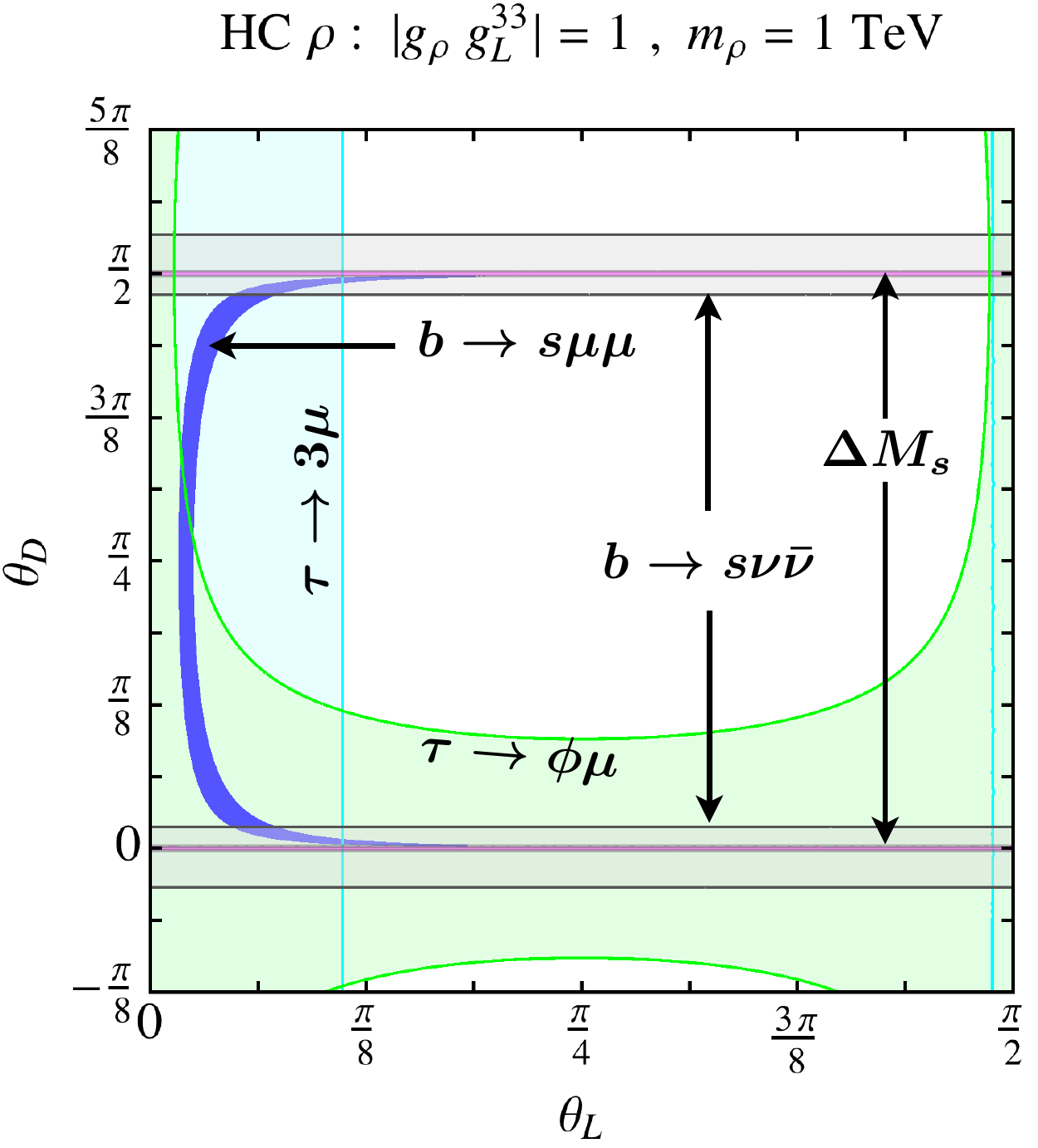}
\caption{
Allowed regions in the $(\theta_L, \theta_D)$ plane in the HC $\rho$ model for $m_\rho = 1\,\text{TeV}$ and $g_\rho\, g_L^{33} = 1$. 
The $b \to s \mu^+\mu^-$ anomaly can be explained in the blue region 
while the constraints from $\Delta M_s$, $\mathcal B(\tau\to3\mu)$, $\mathcal B(\tau\to\phi\mu)$, and $\mathcal B(B\to K^{(*)}\nu\bar\nu)$ are satisfied in the magenta, cyan, green, and gray regions, respectively. 
}
\label{Fig:flavorconstraint}
\end{center}
\end{figure}
%%%%%%%%%%%%endFIG1%%%%%%%%%%%%
%
Now we investigate the parameter space of the model consistent with the flavor measurements. 
Theoretical input for our evaluation is summarized in Table~\ref{tab:input}. 
In Fig.~\ref{Fig:flavorconstraint}, we show {a plane view of} allowed regions in the $(\theta_L, \theta_D)$ plane for $m_\rho = 1\,\text{TeV}$ and $g_\rho\, g_L^{33} = 1$ {constrained from} each observable: 
the $b\to s \ell^+\ell^-$ global fit in blue [Eq.\eqref{EQlimit_bsmumu}], $\Delta M_s$ in magenta [Eq.\eqref{EQlimit_bsbs}], 
$\mathcal B(\tau\to3\mu)$ in cyan, $\mathcal B(\tau\to\phi\mu)$ in green [Eq.\eqref{EQlimit_tauphimu}], and $\mathcal B(B\to K^{(*)}\nu\bar\nu)$ in gray [Eq.\eqref{EQlimit_bsnunu}], as denoted in the figure. 
{Note that $\Delta M_s$ is precisely measured at {experiments} and thus {new physics contributions are} allowed only within the theoretical uncertainties in Table~\ref{tab:input}.} 
One easily sees that the constraint from the $B_s$ mixing (magenta region) is much {stringent $|\theta_D| \ll 1${,}}
while the other constraints are consistent with the $b \to s \mu^+\mu^-$ anomaly (blue region) in {some limited regions.}

%%%%%%%%%%%%beginFIG2%%%%%%%%%%%%
\begin{figure}[t]
\begin{center}
\includegraphics[viewport=0 0 360 384, width=0.33\columnwidth]{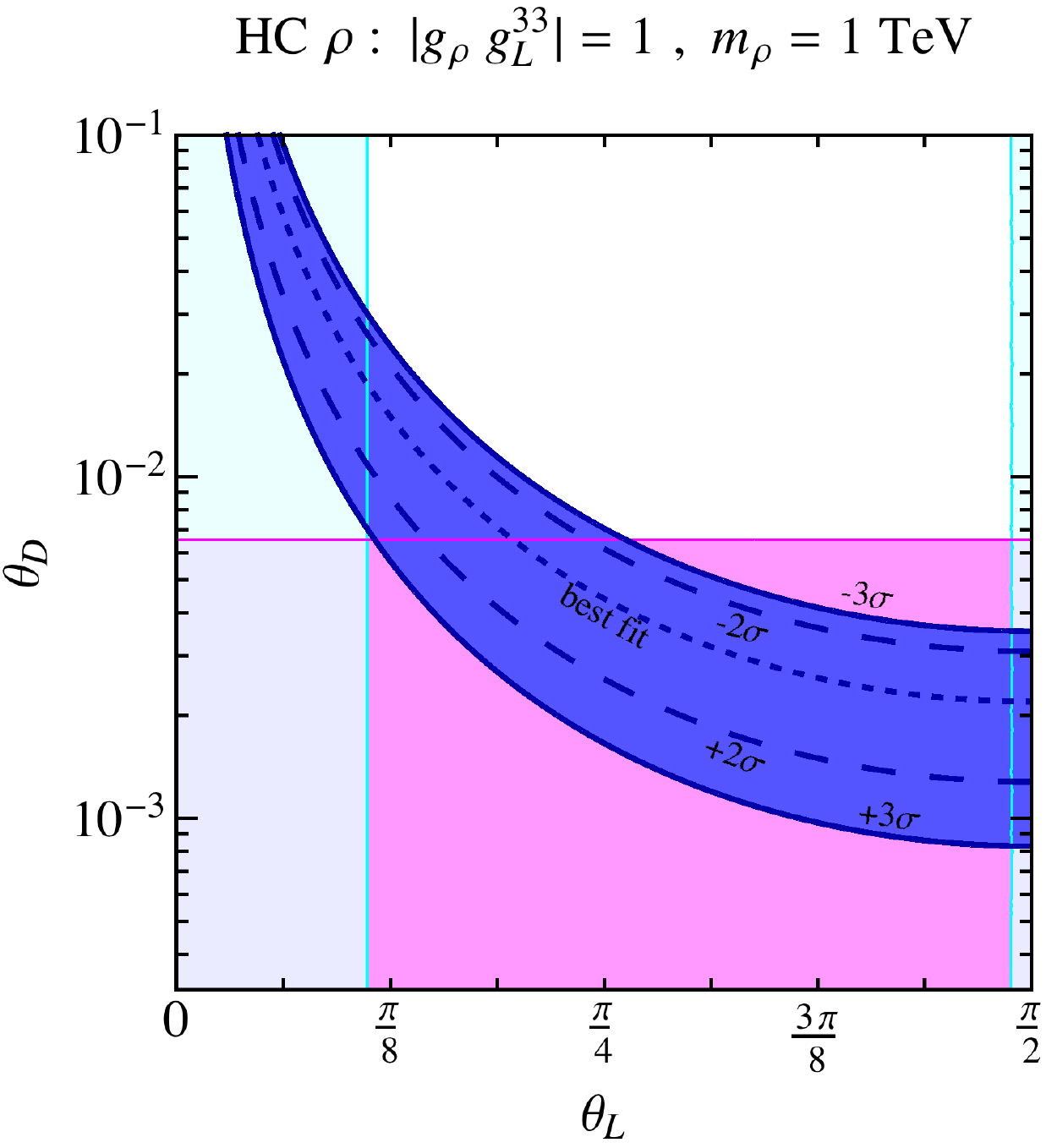}~~
\includegraphics[viewport=0 0 360 384, width=0.33\columnwidth]{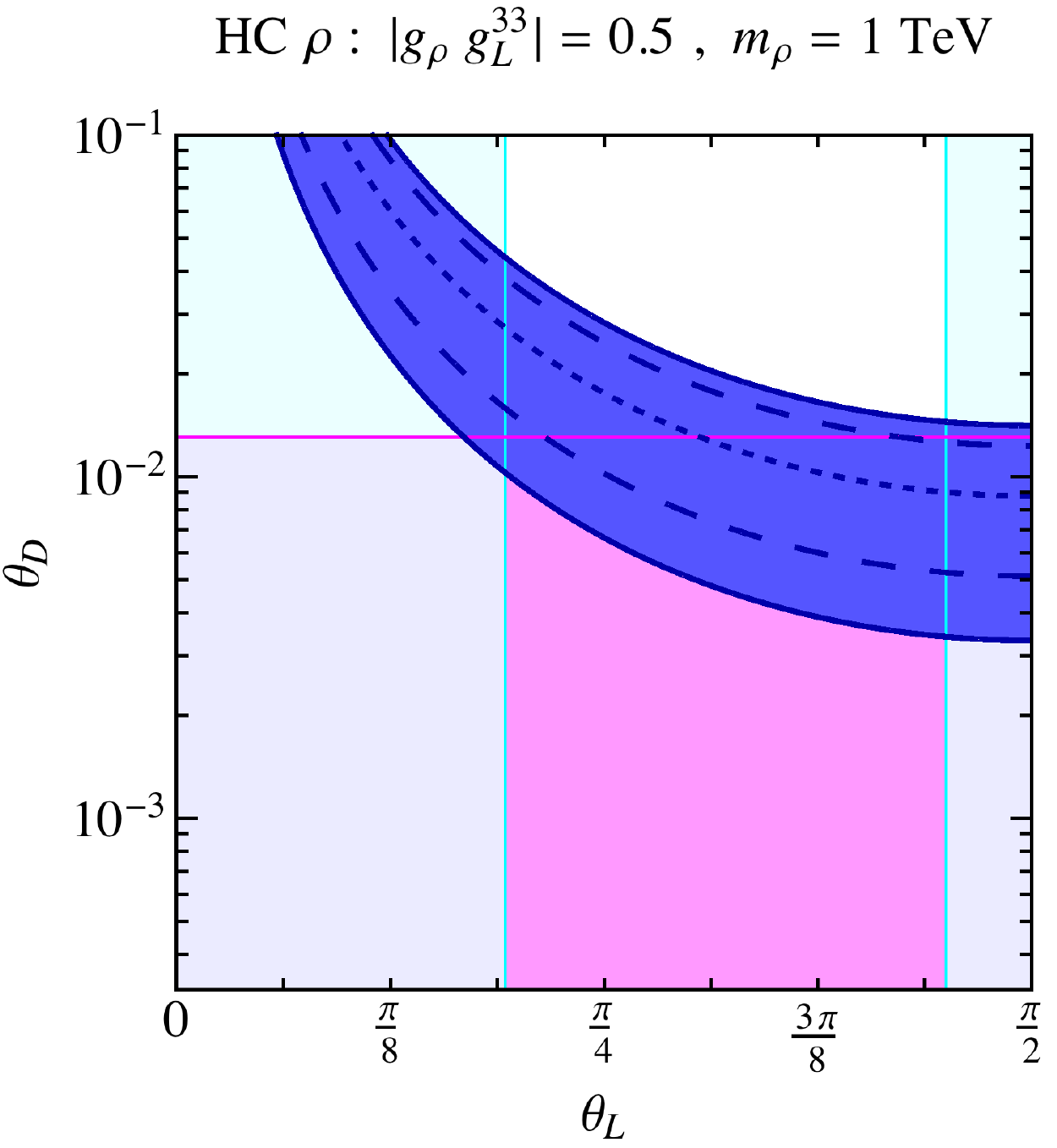}~~
\includegraphics[viewport=0 0 360 384, width=0.33\columnwidth]{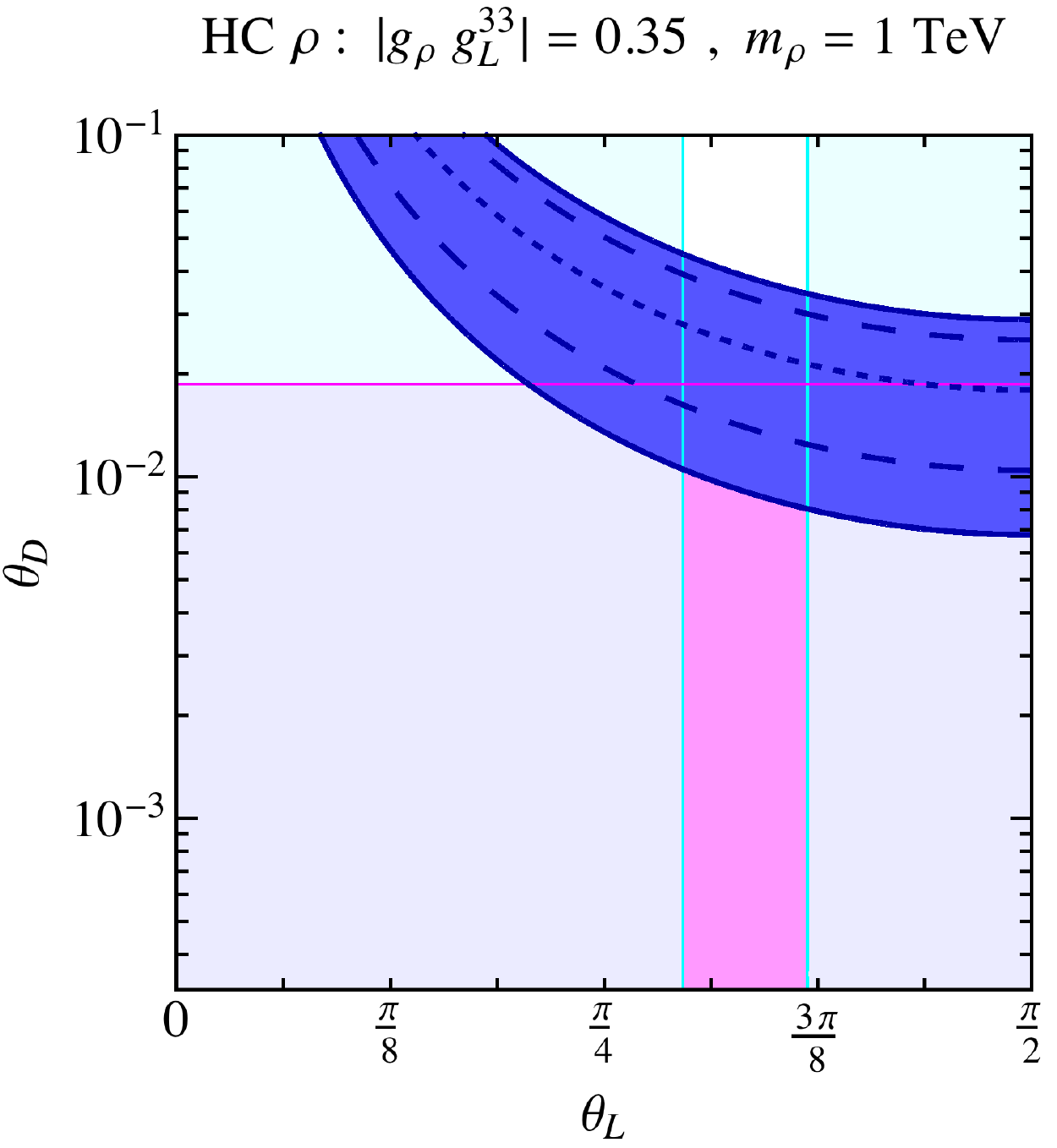}~~
\caption{{
Allowed regions of the flavor constraints for various choice of $|g_\rho \,g_L^{33}|$ with $m_\rho = 1\,\text{TeV}$ near $\theta_D \sim 0$. 
The significant constraints from $\Delta M_s$ (magenta), $\mathcal B(\tau\to3\mu)$ (cyan), and the $b \to s \mu^+\mu^-$ global fit (blue) are shown whereas the others are simply omitted. 
The best-fit point, $\pm 2\sigma$, and $\pm 3\sigma$ ranges to explain the $b \to s \mu^+\mu^-$ anomaly are presented with dotted, dashed, and solid curves, respectively. 
}}
\label{Fig:allowedregions}
\end{center}
\end{figure}
\begin{figure}[t]
\begin{center}
\includegraphics[viewport=0 0 360 407, width=22em]{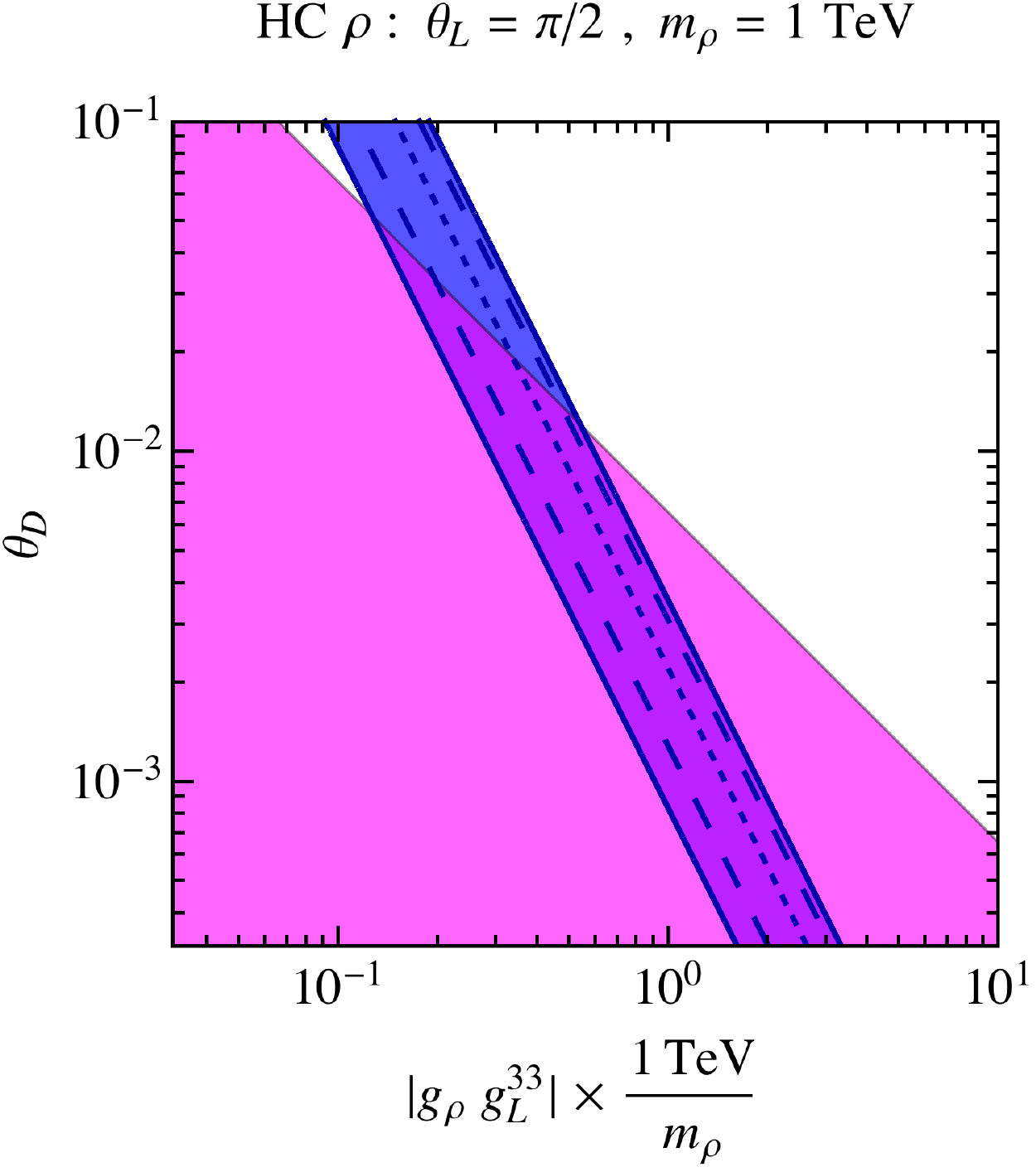}
\caption{{
The allowed range in terms of $|g_\rho\, g_L^{33}|$ and $\theta_D$ for the fixed value $\theta_L = \pi/2$ (``right-side'' spot). 
Conventions of the plots are the same as in Fig.~\ref{Fig:allowedregions}. 
}}
\label{Fig:allowedcoupling}
\end{center}
\end{figure}

%%%%%%%%%%%%endFIG2%%%%%%%%%%%%
%
To {more precisely see the allowed region} near $\theta_D \sim 0$,
in Fig.~\ref{Fig:allowedregions} we show the close-up version focused on the $\theta_D \sim 0$ region, for various values of $g_\rho g_L^{33}$ with $m_\rho = 1\,\text{TeV}$ fixed, where we have taken
the significant constraints, namely from $\Delta M_s$, $\mathcal B(\tau\to3\mu)$, and the $b \to s \mu^+\mu^-$ global fit. 
In the {close-up} plot, we have taken {the parameter range favored up to $\pm 3\sigma$ level} for the $b \to s \mu^+\mu^-$ anomaly. 
For $|g_\rho\, g_L^{33}| \gtrsim 1$, it turns out that the $b \to s \mu^+\mu^-$ anomaly is not consistent with the constraints from $\Delta M_s$ and $\mathcal B(\tau\to3\mu)$ in the present model. 
As for the range $0.35 \lesssim |g_\rho\, g_L^{33}| \lesssim 1$, several comments are in order: 
\begin{itemize}
\item
We found that there are two isolated regions, $\theta_L \lesssim \pi/4$ (``left-side'') and $\theta_L\sim\pi/2$ (``right-side''), where all the constraints are (marginally) satisfied. 
\item
The left-side spot is barely viable when the $b \to s \mu^+\mu^-$ anomaly is
{$3\sigma$ above in terms of the coefficient $C_9^{\mu\mu}$ from the best-fit point ($C_9^{\mu\mu}|_{\rm best} = -0.61$~\cite{Capdevila:2017bsm}), where $C_9^{\mu\mu}|_{+3\sigma} = -0.23$,}
which means that the deviation from the SM has to be rather small. 
\item
On the other hand, the right-side spot fairly satisfies all the constraints. 
In particular, the point of $\theta_L = \pi/2$ can accommodate the best fit point for the $b \to s \mu^+\mu^-$ anomaly. 
Note that $\theta_L = \pi/2$ is the point in which $\tau$ in the gauge basis is exactly equivalent to $\mu$ in the mass basis. 
In this case, the lepton sector only has the connection to the HC $\rho$ from the $\mu$-$\mu$-$\rho$ term. 
\end{itemize}
For $|g_\rho\, g_L^{33}| \lesssim 0.35$ we also found that the two allowed spots {(which were divided by the $\tau\to 3\mu$ constraint) are merged into a single spot and then the region near $\theta_L \sim \pi/2$ is only allowed.} 
In Fig.~\ref{Fig:allowedcoupling}, we survey the allowed range of $|g_\rho\, g_L^{33}|$ for the case $\theta_L = \pi/2$. 
The result implies that the present model for $\theta_L = \pi/2$ requires $|g_\rho\, g_L^{33}| \gtrsim 0.1$ in order to explain the $b \to s \mu^+\mu^-$ anomaly {consistently} with the bound from $\Delta M_s$.

%%%%%%%%%%%%beginFIG3%%%%%%%%%%%%
%%%%%%%%%%%%endFIG3%%%%%%%%%%%%

To summarize, we investigated the allowed regions in the parameter space of $\theta_L$, $\theta_D$, and $|g_\rho\, g_L^{33}|$, which satisfy all the flavor constraints. 
The situation is then divided by two cases; $\theta_L \lesssim \pi/4$ (``left-side'') and $\theta_L\sim\pi/2$ (``right-side''). 
As a result, the allowed region exists in 
\begin{align}
 \label{Eq:allowedgL}
 & ~~0.35 \lesssim \left| g_\rho\, g_L^{33} \right| \times \left( \frac{1\,\text{TeV}}{m_\rho} \right)  \lesssim 1 & & \quad\text{(for left-side spot)} \,, & \\[0.5em]
 & ~~~\,0.1 \lesssim \left| g_\rho\, g_L^{33} \right| \times \left( \frac{1\,\text{TeV}}{m_\rho} \right) & & \quad\text{(for right-side spot)} \label{Eq:allowedgL_2} \,.  &
\end{align}
%$0.35 \lesssim |g_\rho\, g_L^{33}| (\lesssim 1)$ 
The left-side spot is considered as the $\tau$-dominant case where the $\tau$-$\tau$-$\rho$ coupling is relatively larger than the other lepton couplings to HC $\rho$ (including LFV.) 
On the other hand, the right-side spot only involves the $\mu$-$\mu$-$\rho$ coupling. 
Indeed, we have to pay attention to this difference when we consider collider limit at LHC as will be discussed in the next section.

%%%%%%%%%%%%beginFIG4%%%%%%%%%%%%
\begin{figure}[t]
\begin{center}
\includegraphics[viewport=0 0 360 316, width=0.49\columnwidth]{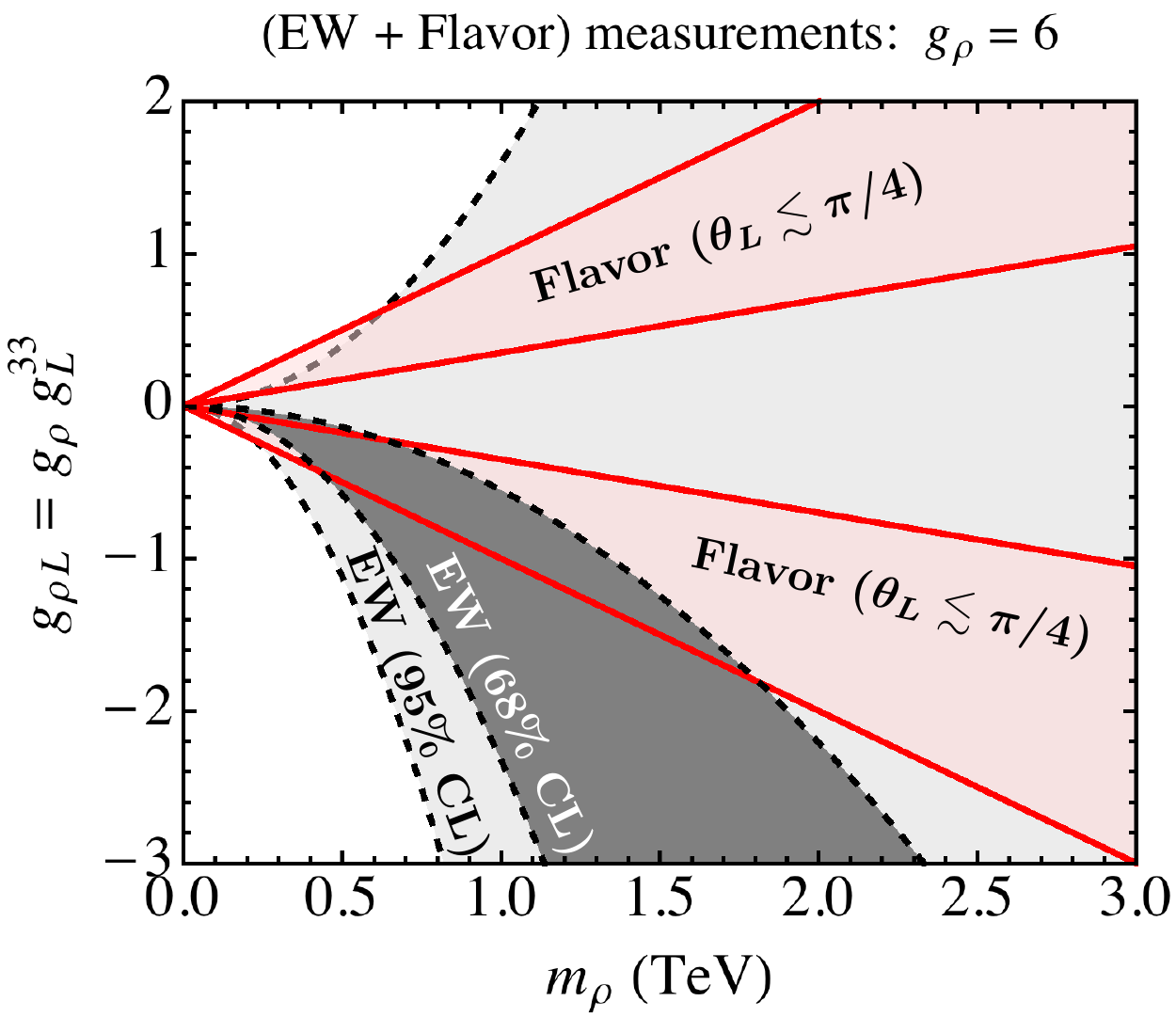}~~~
\includegraphics[viewport=0 0 360 316, width=0.49\columnwidth]{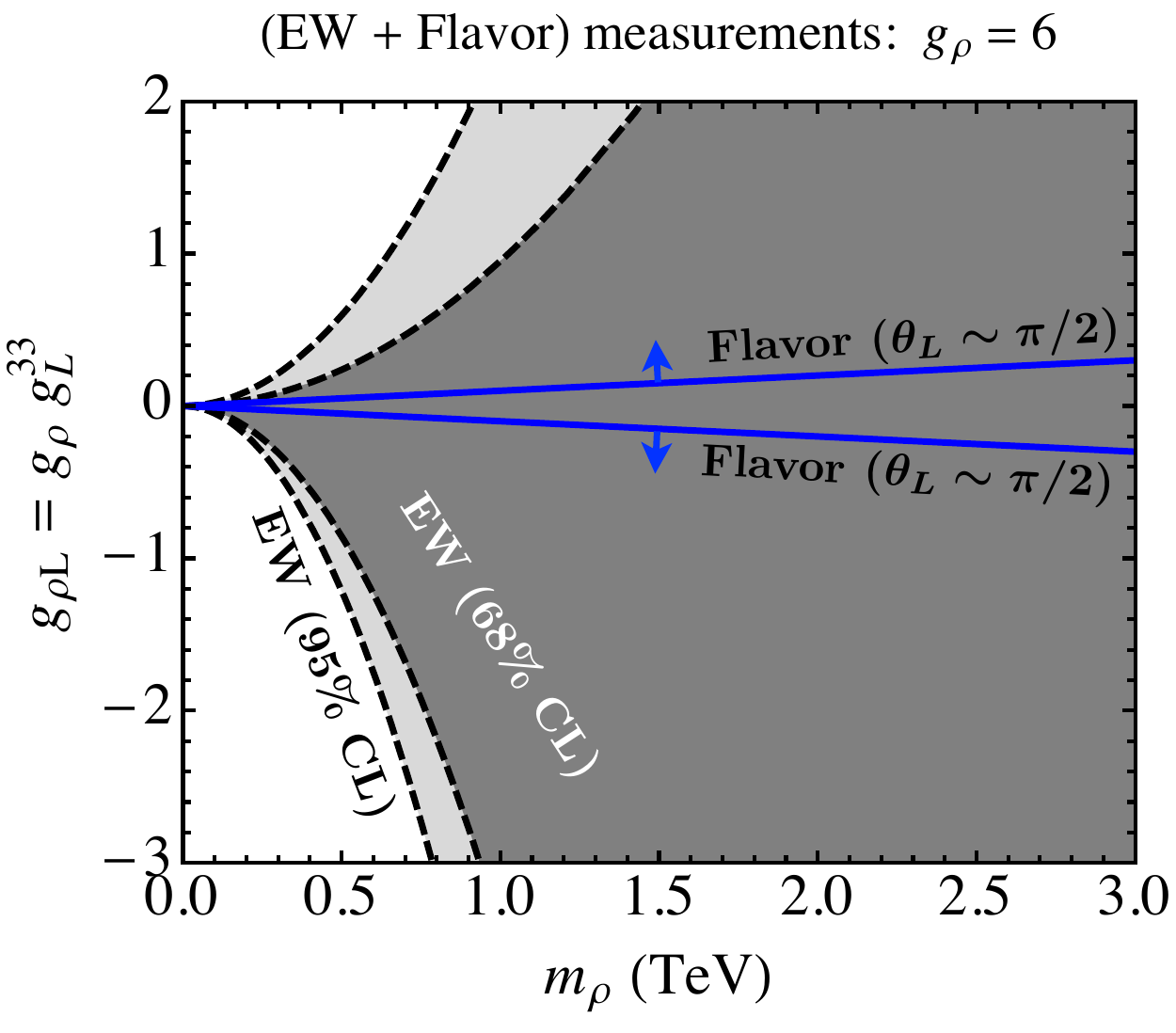}
\caption{{
Summary plot for constraints from the EW {precision} measurements and from the flavor observables on the $(g_\rho g_L^{33}, m_\rho)$ plane fixing $g_\rho =6$ for $\theta_L \lesssim \pi/4$ (left) and $\theta_L \sim \pi/2$ (right). 
The (darker) gray region is allowed by the EW {precision} measurements at $95\%$ ($68\%$) C.L. 
The regions in red (blue) line boundaries are favored by the $b \to s \mu^+\mu^-$ data that satisfies all the other flavor constraints, for $\theta_L \lesssim \pi/4$ ($\theta_L \sim \pi/2$).
}}
\label{Fig:EWandFlavor}
\end{center}
\end{figure}
%%%%%%%%%%%%endFIG4%%%%%%%%%%%%

The combined plots with the constraint from the {EW} precision measurements are shown in Fig.~\ref{Fig:EWandFlavor} on the $(g_\rho g_L^{33},\, m_\rho)$ plane 
for the $\theta_L \lesssim \pi/4$ and $\theta_L \sim \pi/2$ cases. 
The regions in red (blue) line boundaries are favored by the $b \to s \mu^+\mu^-$ data, along with the other constraints, for the {left-side ($\theta_L \lesssim \pi/4$) and right-side ($\theta_L \sim \pi/2$) spots}.  
The shaded regions show $68\%$ and $95\%$ C.L. constraints from the EW measurements as obtained in Sec.~\ref{sec:Fundamental_requirements}. 
(Note that, for the $\theta_L \sim \pi/2$ case, the EW limit is obtained by combining the $R_b$ and $A_{\rm FB}^{(0,\mu)}$ constraints.) 
In the figure, the reference number $g_\rho =6$ is taken. 
We can see that the favored regions from the $b \to s \mu^+\mu^-$ data are consistent with the $95\%$ C.L. EW {precision} measurements. 
One also finds that, in the $1\sigma$ range ($68\%$ C.L.), the EW {precision measurements} {exclude the HC rho mass} $m_\rho \gtrsim 1.8\,\text{TeV}$ and 
the positive value of $g_\rho\, g_L^{33}$ for the {left-side spot case with $\theta_L \lesssim \pi/4$}. 
The allowed parameter space can be examined by direct searches at the LHC. 
It will be discussed below.

%%%%%%%%%%%%%%%%%%%%%%%%%%%%%%%%%%%%%%%%%%%%%%%%%%%%%%%%%%%%%%%%%%%%%%%%%%%%%%%%%%%%%%%%%%
%%%%%%%%%%%%%%%%%%%%%%%%%%%%%%%%%%%%%%%%%%%%%%%%%%%%%%%%%%%%%%%%%%%%%%%%%%%%%%%%%%%%%%%%%%
%%%%%%%%%%%%%%%%%%%%%%%%%%%%%%%%%%%%%%%%%%%%%%%%%%%%%%%%%%%%%%%%%%%%%
\section{Collider-related issues}
\label{collider}
%%%%%%%%%%%%%%%%%%%%%%%%%%%%%%%%%%%%%%%%%%%%%%%%%%%%%%%%%%%%%%%%%%%%%
%%%%%%%%%%%%%%%%%%%%%%%%%%%%%%%%%%%%%%%%%%%%%%%%%%%%%%%%%%%%%%%%%%%%%%%%%%%%%%%%%%%%%%%%%%
%%%%%%%%%%%%%%%%%%%%%%%%%%%%%%%%%%%%%%%%%%%%%%%%%%%%%%%%%%%%%%%%%%%%%%%%%%%%%%%%%%%%%%%%%%

In this section, we discuss constraints from the latest null results in the new physics searches {and future prospects} at the $13\,\text{TeV}$ LHC. 
{The HC $\rho$'s as well as the HC $\pi$'s will be resonantly, or non-resonantly produced at the hadron collision machinery, to be constrained by the present experimental data.}

%%%%%%%%%%%%%%%%%%%%%%%%%%%%%%%%%%%%%%%%%%%%%%%%%%%%%%%%%%%%%%%%%%%%%
%%%%%%%%%%%%%%%%%%%%%%%%%%%%%%%%%%%%%%%%%%%%%%%%%%%%%%%%%%%%%%%%%%%%%
\subsection{Typical constraints on HC $\pi$}
%%%%%%%%%%%%%%%%%%%%%%%%%%%%%%%%%%%%%%%%%%%%%%%%%%%%%%%%%%%%%%%%%%%%%
%%%%%%%%%%%%%%%%%%%%%%%%%%%%%%%%%%%%%%%%%%%%%%%%%%%%%%%%%%%%%%%%%%%%%

Even though the details of the HC pion sector is out of our major interests, we briefly comment on possible constraints on this part.
Here, we focus on the color-singlet isospin-singlet HC pion $\pi^{0}_{(1)'}$ {as a typical signature}.
Two types of interactions can be derived {for the $\pi_{(1)'}^0$}.
The first one is from the global chiral anomalies of hypercolor fermions, which are {represented} by the covariantized WZW terms in the present non-Abelian {$SU(8)_{F_L} \times SU(8)_{F_R}$} case~\cite{Jia:2012kd,Matsuzaki:2015che} (based on the discussion on Ref.~\cite{Kaymakcalan:1983qq}),
\al{
S_{\text{WZW}} &\ni -\frac{N_{\rm HC}}{48 \pi^2} \int_{M^4}
\Bigg\{
	{\rm tr}
	\left[ (d{\cal L} {\cal L} + {\cal L} d{\cal L})\alpha + 
	   (d{\cal R} {\cal R} + {\cal R} d{\cal R})\beta \right] + i{\rm tr}
	\left[ d{\cal L} dU {\cal R} U^\dagger - d{\cal R} dU^\dagger {\cal L} U \right]
\Bigg\} \notag \\
&=
-\frac{N_{\rm HC}}{12 \pi^2 f_\pi} \int_{M^4}
	{\rm tr} \left[ (3d{\cal V} d{\cal V} + d{\cal A} d{\cal A}) \pi + {\cal O}(\pi^2) \right],
}
where we focus on {${\cal V}$-${\cal V}$}-$\pi$ interactions.
$M^4$ means the four-dimensional Minkowski manifold, $U$ denotes the chiral nonlinear basis $U \equiv e^{2i\pi/f_\pi} = \xi_{L}^\dagger \xi_{R}$, and the differential one-forms are defined as
\al{
\alpha = -i dU U^\dagger,\quad
\beta  = -i U^\dagger dU.
}
One-form of the external vector gauge boson includes the gluon as
\al{
{\cal V} = \frac{{\cal R + L}}{2} \ni g_s G^a (\sqrt{2} T_{(8)a}).
}
{Then we encounter the trace, ${\rm tr}\left[(\sqrt{2})^2 T_{(8)a} T_{(8)b} T_{(1)'} \right] = \delta^{ab}/4\sqrt{3}$, which leads to}
\al{
S_{\text{WZW}}(\pi^{0}_{(1)'}GG) &= - \frac{N_{\rm HC}}{16\sqrt{3} \pi^2} \frac{g_s^2}{f_\pi}
	\int_{M^4} \pi^{0}_{(1)'} \, dG^a dG^a \notag \\
&= - \frac{N_{\rm HC}}{16\sqrt{3} \pi^2} \frac{g_s^2}{f_\pi} \frac{1}{4} \int_{M^4} \!\!\!\!\! d^4 x \
	\pi^{0}_{(1)'} \, G_{\mu\nu}^a G_{\rho\sigma}^a \varepsilon^{\mu\nu\rho\sigma},
}
where we {used} the relation $d G^a d G^a = d^4x \, G_{\mu\nu}^a G_{\rho\sigma}^a \varepsilon^{\mu\nu\rho\sigma}/4$.
This effective action describes the anomaly-induced $G$-$G$-$\pi^0_{(1)'}$ interaction. 
{(Note that $G$ indicates gluon in this article.)}

Another possible origin of the $G$-$G$-$\pi^0_{(1)'}$ coupling  
comes from 
a top-quark loop contribution 
constructed from the $t$-$\bar{t}$-$\pi^0_{(1)'}$ interaction, 
which could be, in the present model, 
induced through an extended HC as given in Eq.(\ref{h-coupling}).  
As discussed in Ref.~\cite{Kauffman:1993nv} 
the amplitude corresponding to such a top-loop contribution 
is generically 
written down
\al{
{\cal M}\left( \pi^0_{(1)'} \to G^a(p_{1\mu}) G^b(p_{2\nu}) \right)
=
-i g_A r_y \, \delta^{ab} \varepsilon^{\mu\nu\rho\sigma} p_{1\rho} p_{2\sigma} \tau f(\tau) 
	\epsilon^{\ast}_{\mu} \epsilon^{\ast}_{\nu},
	\label{eq:amplitude_pi0toGG}
} 
with the factor $g_A = \alpha_s/(2 \pi v_{\rm VEV})$, the $t$-$\bar{t}$-$\pi^0_{(1)'}$ 
coupling strength 
{$r_y$ being real 
and defined} 
as a ratio to the SM like top-Higgs case ($m_t \gamma_5/v_{\rm VEV}$), and the gluon polarization vectors $\epsilon_{\mu,\nu}$.
{Here, we adopt the Feynman rule of the fundamental pseudoscalar to parametrize the coupling.}
The loop function $f(\tau)$ is a function of the parameter $\tau \equiv 4m_t^2/M_{\pi^{0}_{(1)'}}^2$ as
\al{
f(\tau) =
\begin{cases}
\displaystyle \left[ \sin^{-1} \left( \sqrt{\frac{1}{\tau}} \right) \right]^2 & \text{for } \tau \geq 1, \\
\displaystyle -\frac{1}{4} \left[ \ln\left( \frac{\eta_+}{\eta_-} \right) -i\,\pi \right]^2 
	& \text{for } \tau < 1,
\end{cases}
	\label{eq:function_ftau}
}
with $\eta_{\pm} = 1 \pm \sqrt{1 - \tau}$.

Combined with the above two sources, the squared amplitudes are {computed as} %After the summations on spin and color indices
\al{
\left|\overline{{\cal M}}\right|^2_{\rm WZW}
&=
1024 A^2  (p_1 \cdot p_2)^2, \\
%%%
\left|\overline{{\cal M}}\right|^2_{\rm top}
&=
16 \, g_A^2 |r_y \tau f(\tau)|^2 (p_1 \cdot p_2)^2, \\
%%%
\left|\overline{{\cal M}}\right|^2_{\rm int}
&= 
128 A g_A \left[ r_y \tau f(\tau) + \left( r_y \tau f(\tau) \right)^\ast \right] (p_1 \cdot p_2)^2, \\
%%%
{\left|\overline{{\cal M}}\right|^2}
&=
{\left|\overline{{\cal M}}\right|^2_{\rm WZW} +
\left|\overline{{\cal M}}\right|^2_{\rm top} +
\left|\overline{{\cal M}}\right|^2_{\rm int},}
}
with the factor
\al{
A = - \frac{N_{\rm HC}}{16\sqrt{3} \pi^2 \cdot 4} \frac{g_s^2}{f_\pi}.
}
Through the relations $2 \, p_1 \cdot p_2 = M_{\pi^{0}_{(1)'}}^2$ and $\Gamma({\pi^{0}_{(1)'}} \to GG) = |\overline{{\cal M}}|^2/(2 \times 16 \pi M_{\pi^{0}_{(1)'}})$, we obtain
\al{
\Gamma({\pi^{0}_{(1)'}} \to GG)
&=
\frac{\left(M_{\pi^{0}_{(1)'}}\right)^3}{\pi}
\left\{
8A^2 + \frac{1}{8} g_A^2 |r_y \tau f(\tau)|^2 + A g_A {r_y \left[ \tau f(\tau) + \left( \tau f(\tau) \right)^\ast \right]}
\right\} \notag \\
&=
\frac{\left(M_{\pi^{0}_{(1)'}}\right)^3}{\pi}
\left\{
8A^2 + \frac{1}{8} g_A^2 |r_y \tau f(\tau)|^2 + {2 A g_A r_y \text{Re}\left[ \tau f(\tau) \right]}
\right\}. 
\label{eq:decay_width_pizero_fullform}
}
Through the well known formula for {cross section with a spin-$J$ resonance that arises from a proton-proton collision with gluonic initial state}~\cite{Franceschini:2015kwy}
\al{
\sigma(GG \to \pi_{(1)'}^{0} \to \gamma \gamma) = \frac{2J+1}{s} C_{GG} \frac{\Gamma(\pi_{(1)'}^{0} \to GG)}{M_{\pi_{(1)'}^{0}}}
\frac{\Gamma({\pi_{(1)'}^{0}} \to \gamma \gamma)}{\Gamma_{{\pi_{(1)'}^{0}}}},
}
{(where $s$ is the center of mass energy and $C_{GG}|_{13\,\text{TeV}} = 2137$~\cite{Franceschini:2015kwy} denotes the luminosity coefficient for a pair of gluons as initial partons,)} 
{we can immediately calculate the diphoton cross section.}

First, we shall consider the simplest case with $r_y=0$, (namely, no coupling to top quark pair.) 
In this case, we estimate the diphoton cross section 
\al{
{
\sigma(GG \to \pi_{(1)'}^{0} \to \gamma\gamma)
\Bigg|_{r_y=0} 
} 
 \sim  {0.1\,\text{fb}} \times
\left[ \frac{N_{\rm HC}}{3} \right]^2
\left[ \frac{\alpha_s}{0.1} \right]^2
\left[ \frac{{\mathcal B} (\pi_{(1)'}^{0} \to \gamma \gamma)}{{10^{-3}}} \right]
\left( \frac{M_{\pi_{(1)'}^{0}}}{f_\pi} \right)^2.
	\label{eq:bound_pp_to_pi_to_diphoton}
}
Note that, in the case of $r_y=0$,
$\mathcal B(\pi_{(1)'}^{0} \to \gamma \gamma)$ 
is completely free from the $\pi_{(1)'}^0$ mass dependence 
because $\pi_{(1)'}$ decays only to the massless final states, $2\gamma$ and $2G$. 
Hence the diphoton 
cross section in Eq.(\ref{eq:bound_pp_to_pi_to_diphoton}) 
is controlled only by the ratio $(M_{\pi_{(1)'}^0}/f_\pi)$. 
To survey a generic parameter space in the present model, 
we shall momentarily take the value of $M_{\pi_{(1)'}^0}$ 
in a range from   
${\cal O}(100)$ GeV [low mass] 
up to ${\cal O}({\rm TeV})$ [high mass]~\footnote{{As listed in Eq.(\ref{pi:masses}), a typical size of the $M_{\pi_{(1)'}^0}$ 
is expected to be ${\cal O}(100)$ GeV. 
However, the TeV mass range might be achieved when one 
consider possible effects from extended HC sector, which could be 
enhanced in the case of many flavor QCD (nearly conformal/walking gauge 
theory), in a way similar to extended technicolor scenarios.}},  
and discuss the phenomenological constraints from the diphoton 
cross section of Eq.(\ref{eq:bound_pp_to_pi_to_diphoton}).

{In the high mass case ($M_{\pi_{(1)'}^0} \sim \text{TeV}$),}  
{the estimated diphoton cross section in Eq.(\ref{eq:bound_pp_to_pi_to_diphoton})} 
is compared with the $95\%$ C.L. upper bound on {fiducial cross sections} as $\sigma_{\rm ggF}^{\rm spin\,0}(\gamma\gamma) \equiv\sigma(GG \to [\text{spin-0 resonance}] \to \gamma \gamma) \lesssim 1\,\text{fb}$ { {for narrow-width resonance with mass} 
$\sim 1\,\text{TeV}$} 
reported in Refs.~\cite{ATLAS:2016eeo,Khachatryan:2016yec}. 
{Thus we can naively say} that the prediction in the HC theory has a tight tension with the present experimental data unless {$M_{\pi_{(1)'}^{0}}/f_\pi \lesssim 3$}. 
To be consistent with the above bound for $M_{\pi_{(1)'}^{0}} \sim 1\,\text{TeV}$  
implies that the HC pion decay constant $f_\pi$ should be {(at least) as large as around} $v_{\rm VEV}$. 
However, several nonperturbative estimates  
with some approximation~\cite{Harada:2003dc,Kurachi:2006ej}, 
as well as recent lattice simulations 
in  QCD with many flavors~\cite{Appelquist:2014zsa,Aoki:2016wnc},    
suggest the mass relation in magnitude of  
$m_\rho \sim {\cal O}(10) f_\pi$ 
(for the $m_\rho/f_\pi$ ratio in QCD with 8 flavors, 
see also Ref.~\cite{Matsuzaki:2015sya}.)  
{Therefore, the diphoton constraint ($M_{\pi_{(1)'}^{0}}/f_\pi \lesssim 3$) along with the estimated mass relation ($m_\rho \sim {\cal O}(10) f_\pi$) indicates the bound on the HC rho mass scale as 
$m_\rho \gtrsim 3\,\text{TeV}$.}

{As for the low mass case, the ATLAS $8\,\text{TeV}$ bound is available in the range $65$ -- $200\,\text{GeV}$.
In particular, the fluctuating $95\%$ C.L. bounds around $65$ -- $100\,\text{GeV}$ is $\sim 50\,\text{fb}$ as a crude average~\cite{Aad:2014ioa}, 
which corresponds to $\sim 250\,\text{fb}$ at $13\,\text{TeV}$~\cite{Franceschini:2015kwy}.}
Thereby the diphoton bound looks not serious 
for $M_{\pi_{(1)'}^{0}} \lesssim 200\,\text{GeV}$.

{For an intermediate-mass case ($200\,{\rm GeV} \lesssim M_{\pi_{(1)'}^{0}} \lesssim 1\,\text{TeV}$) -- referred to as the EW-mass case hereafter --}  
the (fluctuating) $13\,\text{TeV}$ $95\%$ C.L. upper bound 
for $\sigma_{\rm ggF}^{\rm spin\,0}(\gamma\gamma)$  
is around $\sim 1\,\text{fb}$ in the range $500\,\text{GeV}$ -- $1\,\text{TeV}$~\cite{ATLAS:2016eeo,Khachatryan:2016yec}.
{Below $500\,\text{GeV}$, the exclusion limit gets reduced such as 
$\sigma_{\rm ggF}^{\rm spin\,0}(\gamma\gamma) \sim 10\,\text{fb}$ for $M_{\pi_{(1)'}^{0}} = 200\,\text{GeV}$~\cite{ATLAS:2016eeo}; 
$\sim 4\,\text{fb}$ for $M_{\pi_{(1)'}^{0}} = 300\,\text{GeV}$; 
and $\sim 2\,\text{fb}$ for $M_{\pi_{(1)'}^{0}} = 400\,\text{GeV}$,} 
(here we do not take care of the rapid fluctuations in the cross section 
curves depicting $95\%$ C.L. upper bounds.)  
Those diphoton limits can safely be evaded  
if the decay constant $f_\pi$ 
is set to be as large as, {\it e.g.,} $\sim 500\,\text{GeV}$  
and the HC rho mass scale is $m_\rho\gtrsim 5\,\text{TeV}$, which is consistent with 
{the aforementioned relation, $m_\rho \sim {\cal O}(10)\, f_\pi \sim {\cal O}(10)\, M_{\pi_{(1)'}^{0}}$, based on the nonperturbative observation.}

%%%%%%%%%%%%%%%%%%%%%%%%%%%%%%%%%%%%%%%%%%%%%%%%%%%%%%%%%%%%%%%%%%%%%%%%%%%%
\begin{figure}[t]
\centering
\includegraphics[width=0.49\columnwidth]{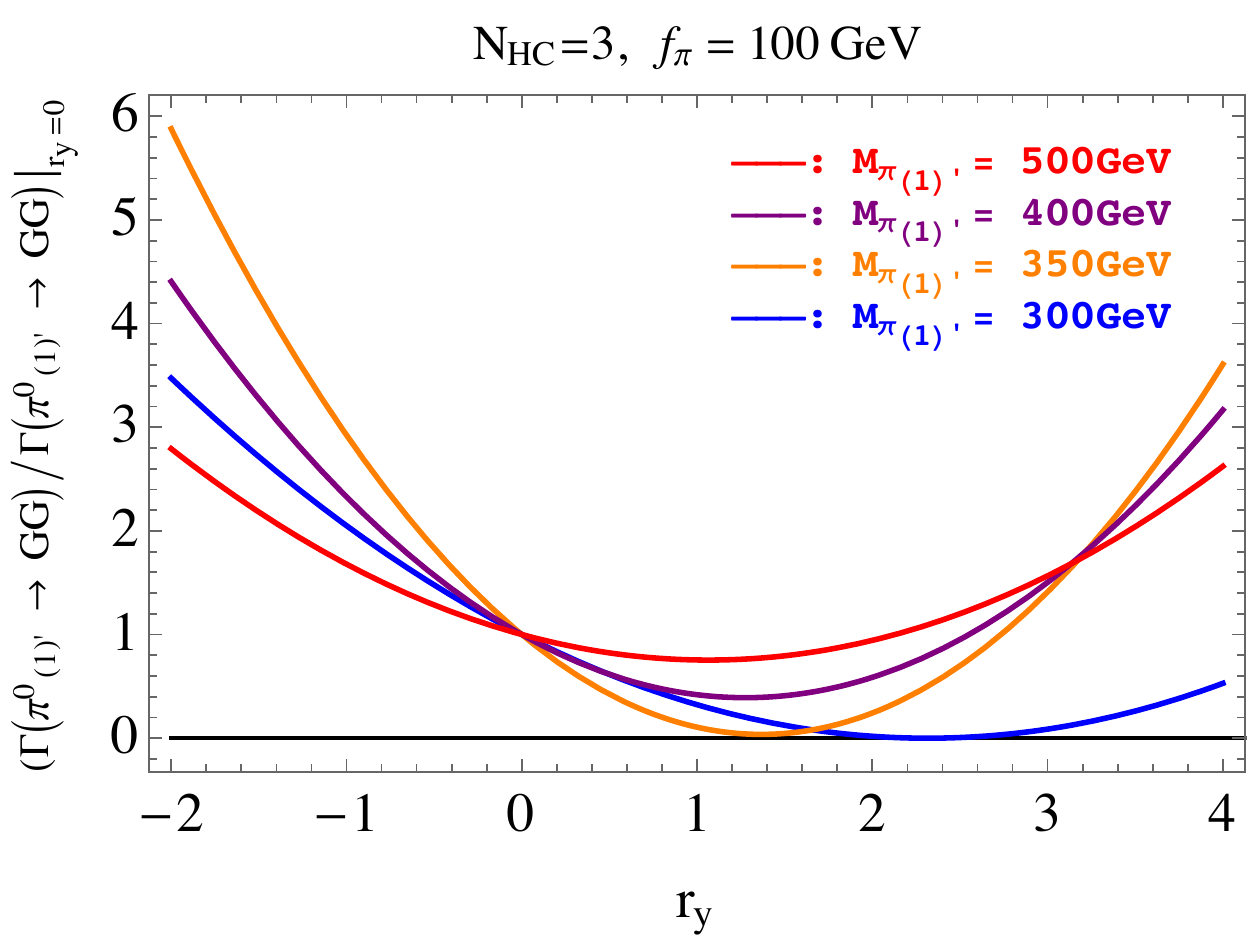}
%%%
\caption{
{Curves that show the ratio $\Gamma(\pi_{(1)'}^{0} \to GG)/\Gamma(\pi_{(1)'}^{0} \to GG)|_{r_y = 0}$ under the four different choices of $M_{\pi_{(1)'}^{0}}$ in $f_\pi = 10^2\,\text{GeV}$ and $N_{\rm HC} = 3$.}
}
%%%
\label{fig:tauftau}
\end{figure}
%%%%%%%%%%%%%%%%%%%%%%%%%%%%%%%%%%%%%%%%%%%%%%%%%%%%%%%%%%%%%%%%%%%%%%%%%%%%

The condition $M_{\pi_{(1)'}^{0}}/f_\pi \lesssim 3$ is violated, for example, if $f_\pi \sim 100\,\text{GeV}$ and $M_{\pi_{(1)'}^{0}} \gtrsim 300\,\text{GeV}$. 
In such a case, we may still evade the $95\%$ C.L. upper bound at $13\,\text{TeV}$ by taking the additional contribution 
from the coupling to top quark (i.e. $r_y \neq 0$) to
make the partial width $\Gamma(\pi_{(1)'}^{0} \to GG)$ reduced. 
Such a reduction can originate from the interference effect in Eq.(\ref{eq:decay_width_pizero_fullform}).
{We illustrate this case in the following.}
In Fig.~\ref{fig:tauftau}, we show how $\Gamma(\pi_{(1)'}^{0} \to GG)$ is changed under the presence of nonzero $r_y$ in $f_\pi = 10^2\,\text{GeV}$ and $N_{\rm HC} = 3$~\footnote{{{For $N_{\rm HC} \ge 4$, 
the cancellation still works 
for greater $r_y$ values 
to realize the complete cancellation  
in $\Gamma(\pi_{(1)'}^{0} \to GG)$.}}
}.
For $M_{\pi_{(1)'}^{0}} = 300-350\,\text{GeV}$, a suitably tuned $r_y$ leads to {an (almost)} vanishing value of $\Gamma(\pi_{(1)'}^{0} \to GG)$, while for $M_{\pi_{(1)'}^{0}} \gtrsim 400\,\text{GeV}$, only $\lesssim 50\%$ cancellation is possible at most,
which may be enough to alleviate the tension for $M_{\pi_{(1)'}^{0}} \sim 400\,\text{GeV}$. 
Let us remind that the constraint on the cross section for $M_{\pi_{(1)'}^{0}} = 300\,\text{GeV}$ {
($\sigma_{\rm ggF}^{\rm spin\,0}(\gamma\gamma) \lesssim 4\,\text{fb}$)} 
is looser than that for $M_{\pi_{(1)'}^{0}} = 400\,\text{GeV}$ 
($\sigma_{\rm ggF}^{\rm spin\,0}(\gamma\gamma) \lesssim 2\,\text{fb}$).
{Thus, the interference contribution helps us to revive the possibility of $\pi_{(1)'}^{0}$ with the mass around $400\,\text{GeV}$} 
even when $f_\pi \sim 100\,\text{GeV}$, while no cancellation may be required for $300\,\text{GeV} \lesssim M_{\pi_{(1)'}^{0}} \lesssim 400\,\text{GeV}$. 
{This consequence is followed by discussion for a LHC bound on the HC $\rho$ meson mass.}

Another constraint comes from the scalar leptoquark search derived from a non-resonant pair production of $\pi_{(3)}^{0}$ in our model, 
as has been reported by ATLAS in Ref.~\cite{Aaboud:2016qeg} for first and second generation leptoquarks. 
The result says that the corresponding mass scale of $\pi_{(3)}^{0}$ should be $\gtrsim 1050 - 1100\,\text{GeV}$ at $95\%$ C.L. assuming $100\%$ branching fraction for a $\pi_{(3)}^{0}$ decay. 
We note that final states are less ambiguous since possible decay branches are limited, but still there is a parameter dependence on $a$ in general as shown in Eq.(\ref{eq:V-pi-pi_form}).
Here, we provide a simple conclusion such that the bound from the pair production is harmless if a typical mass scale of HC pions is sufficiently greater than $1\,\text{TeV}$. 
As shown in Eq.(\ref{pi:masses}), vector leptoquarks indeed obtain $\sim 3\,\text{TeV}$ masses for $\Lambda_{\rm HC} \sim 1\,\text{TeV}$, 
which come most dominantly through QCD gluon exchange corrections amplified due to 
the characteristic feature of many flavor QCD. 
Thus our heavy HC pion scenario can still avoid the current scalar leptoquark bound.

%%%%%%%%%%%%%%%%%%%%%%%%%%%%%%%%%%%%%%%%%%%%%%%%%%%%%%%%%%%%%%%%%%%%%
%%%%%%%%%%%%%%%%%%%%%%%%%%%%%%%%%%%%%%%%%%%%%%%%%%%%%%%%%%%%%%%%%%%%%
\subsection{Resonant productions of HC $\rho$}
%%%%%%%%%%%%%%%%%%%%%%%%%%%%%%%%%%%%%%%%%%%%%%%%%%%%%%%%%%%%%%%%%%%%%
%%%%%%%%%%%%%%%%%%%%%%%%%%%%%%%%%%%%%%%%%%%%%%%%%%%%%%%%%%%%%%%%%%%%%

In this part, we discuss constraints from resonance searches at the $13\,\text{TeV}$ LHC, which provide stringent bounds on mass scales of HC rho mesons directly.

%%%%%%%%%%%%%%%%%%%%%%%%%%%%%%%%%%%%%%%%%%%%%%%%%%%%%%%%
\subsubsection{Basic backgrounds}
%%%%%%%%%%%%%%%%%%%%%%%%%%%%%%%%%%%%%%%%%%%%%%%%%%%%%%%%

As we have pointed out, although the {$V_{\rm SM}$}-$\rho$ mixing (in the covariantized HLS formulation of this model) generates mass splittings among HC $\rho$ mesons in physical eigenstates,  
only a few-percent splittings are allowed because of the large coupling $g_\rho \sim 6$ 
as in Eq.(\ref{grho:value}), which would be
 supported from the QCD-like vector dominance. 
Thereby, 
it is a good approximation {to} consider 
all of the HC $\rho$ components 
to be degenerated in the common mass scale $m_\rho$,  
and interactions induced through such mass mixings to be negligible.

{We should also recall that} 
the mixing angles $\theta_D$ and $\theta_L$ defined in Eq.(\ref{LDrotation}) were already restricted 
from the {EW} (in Sec.~\ref{sec:Fundamental_requirements}) and flavor (in Sec.~\ref{sec:flavor_issues}) observables. 
{To address the $b \to s\mu^+\mu^-$ anomaly consistently with the other constraints, we found that $\theta_D$ needs to be much tiny such as $\theta_D \sim 10^{-2}$ -- $10^{-3}$. 
Thus} we can take the limit of $\theta_D \to 0$ for {all the calculations} on collider phenomena (though the minuscule values are mandatory for addressing the $b \to s\mu^+\mu^-$ anomaly.)
On the other hand, {it has turned out that the favored regions {for} $\theta_L$ are categorized by two spots, $\theta_L \lesssim \pi/4$ and $\theta_L\sim\pi/2$ 
depending on the magnitude of the coupling $g_{\rho L} \equiv g_\rho g_L^{33}$.}
This angle {$\theta_L$} determines the relative {coupling strength} of HC $\rho$ mesons to {$2\tau$ and $2\mu$}, which are $\cos^2{\theta_L}$ and $\sin^2{\theta_L}$, respectively,
{hence the coupling to $2\tau$ becomes dominant for $\theta_L \lesssim \pi/4$, while the coupling to $2 \mu$ does for $\theta_L \sim \pi/2$.
Thus, significant collider bounds would be derived from the $2\tau$ and $2\mu$ channel searches, depending on the $\theta_L$.
To be concrete and conservative, we shall hereafter take $\theta_L = 0$ (for the $2\tau$ channel) and $\theta_L = \pi/2$ (for the $2\mu$ channel) as the reference values corresponding to the two cases.}

To calculate resonant processes, values of total decay widths are important.
Possible decay branches are {shared by} a pair of the SM fermions and a pair of HC pions {[see Appendix~\ref{appendix:rho-pi-pi}]}.
Here, we assume that the latter case is kinematically blocked ($m_\rho < 2 m_\pi$), which 
{turns out to be reasonable.  
To see how it works, first recall that  
some} HC pion masses should be 
sufficiently as heavy as {${\cal O}(1)\,\text{TeV}$}, 
as listed in Eq.(\ref{pi:masses}).   
{More precisely, 
we see from Eq.(\ref{pi:masses}) that} 
when $f_\pi \sim {\cal O}(100)\,\text{GeV}$ and $\Lambda_{\rm HC} \sim 1\,\text{TeV}$, 
{the mass of the color-singlet isospin-triplet mesons is $\sim 2\,\text{TeV}$},
while those of the colored mesons are $\sim 3\,\text{TeV}$ ($\sim 4\,\text{TeV}$) for color triplets (color octets). 
To make sure how the decay channels to HC pion pairs open, 
we list up
the total values of 
{the final-state particle masses ($m_{\pi\pi}$):}   
({\it c.f.,} Appendix~\ref{appendix:rho-pi-pi}),
\begin{itemize}
\item $\rho^0_{(3)}\to \bar{\pi}^0_{(3)}\pi^0_{(1)'}$\,: $m_{\pi\pi}\sim (3 + {\cal O}(0.1))\,\text{TeV}$,
\item $\rho^\alpha_{(3)} \to \bar{\pi}^\alpha_{(3)} \pi^0_{(1)'}$\,: $m_{\pi\pi}\sim (3 + {\cal O}(0.1))\,\text{TeV}$,
\item $\rho^0_{(8)}\to \bar{\pi}^0_{(3)} \pi^0_{(3)}$\,: $m_{\pi\pi}\sim (3 + 3)\,\text{TeV} = 6\,\text{TeV}$,
\item $\rho^\alpha_{(8)}\to \bar{\pi}^0_{(3)}\pi^\alpha_{(3)}$\,: $m_{\pi\pi}\sim (3 + 3)\,\text{TeV} = 6\,\text{TeV}$,
\item $\rho^0_{(1)'} \to \bar{\pi}^0_{(3)} \pi^0_{(3)}$\,: $m_{\pi\pi}\sim (3 + 3)\,\text{TeV} = 6\,\text{TeV}$,
\item $\rho^\alpha_{(1)'}\to \bar{\pi}^\beta_{(1)} \pi^\gamma_{(1)'}$\,: $m_{\pi\pi}\sim (1 + 2)\,\text{TeV} = 3\,\text{TeV}$,
\item $\rho^\alpha_{(1)}\to \bar{\pi}^\beta_{(1)} \pi^\gamma_{(1)}$\,: $m_{\pi\pi}\sim (1 + {2})\,\text{TeV} = {3}\,\text{TeV}${.}
\end{itemize}
\noindent
Then, additional contributions to the HC rho's decay branches {appear} when {$m_\rho \gtrsim 3\,\text{TeV}$} at the present benchmark point, $f_\pi \sim {\cal O}(100)\,\text{GeV}$ and $\Lambda_{\rm HC} \sim 1\,\text{TeV}$.
On the other hand, when $f_\pi$ is {somewhat} greater than $\sim 100\,\text{GeV}$ (with a sizable explicit breaking scale $m^0_F$), the HC pions becomes heavier and we may block the HC rho's decays to the HC pions consistently, keeping the relation $m_\rho \sim {\cal O}(10) f_\pi$ intact.

\begin{figure}[t]
\centering
\includegraphics[width=0.49\columnwidth]{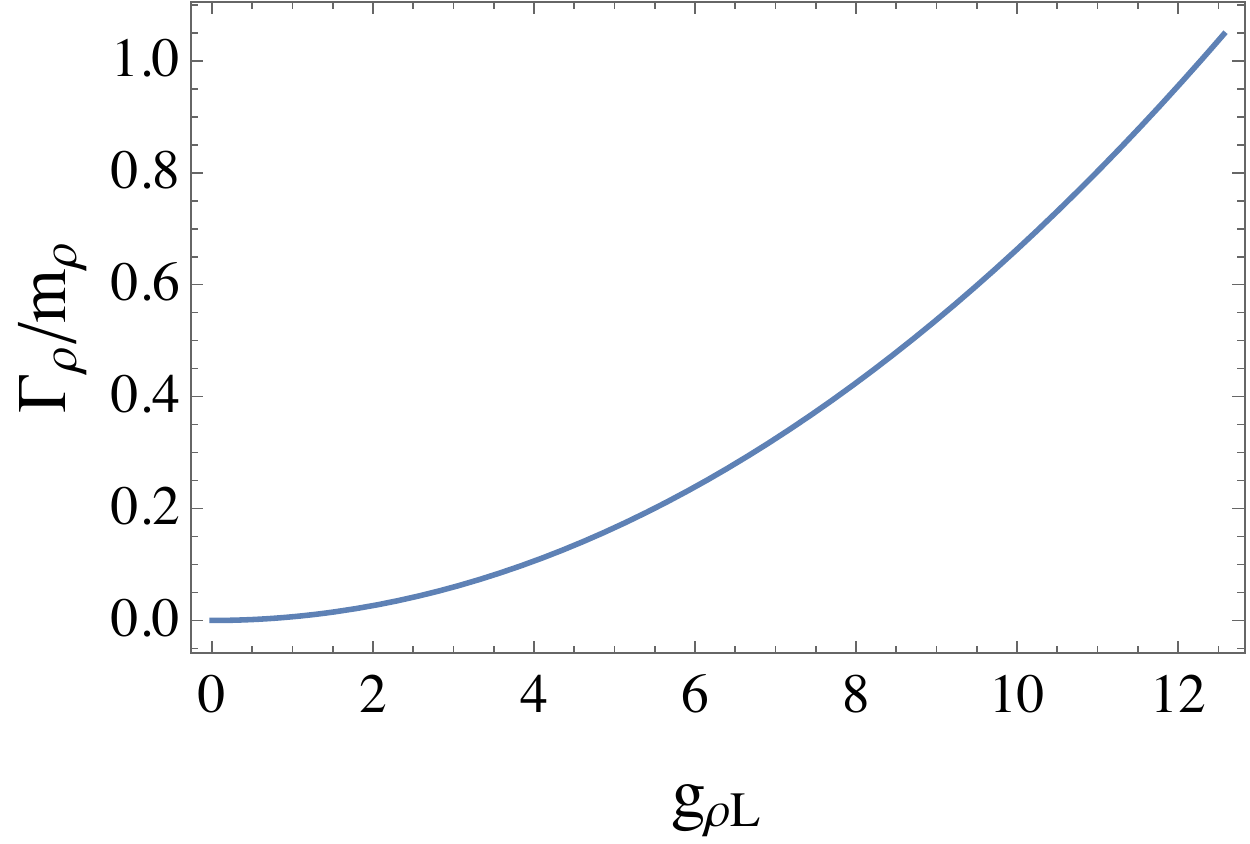}
%%%
\caption{The ratio $\Gamma_\rho/m_\rho$ in Eq.(\ref{eq:form_of_totalwidth_rho}) as a function of $g_{\rho L}$.
}
%%%
\label{fig:totalwidth_rho}
\end{figure}

{Thus our assumption $m_\rho < 2 m_\pi$ may be justified even in the range {$m_\rho \gtrsim 3\,\text{TeV}$}, {so that we may be able to ignore decays to HC pion pairs.}}
{Of interest enough is then that} 
all of the physical HC $\rho$ components have the common value in the total width as
\al{
\Gamma_\rho = \frac{g_{\rho L}^2 m_\rho}{48\pi},
	\label{eq:form_of_totalwidth_rho}
}
where we simply ignored tiny contributions through mixing effects.
Details of partial widths are provided in appendix~\ref{appendix:rho_decay_width}.
The curve of the ratio $\Gamma_\rho/m_\rho$ as a function of $g_{\rho L}$ is illustrated in Fig.~\ref{fig:totalwidth_rho}.

%%%%%%%%%%%%%%%%%%%%%%%%%%%%%%%%%%%%%%%%%%%%%%%%%%%%%%%%%%%%%%%%%%%%%%%%%%%%
%%%%%%%%%%%%%%%%%%%%%%%%%%%%%%%%%%%%%%%%%%%%%%%%%%%%%%%%%%%%%%%%%%%%%%%%%%%%

%%%%%%%%%%%%%%%%%%%%%%%%%%%%%%%%%%%%%%%%%%%%%%%%%%%%%%%%
\subsubsection{Forms of resonant cross sections}
%%%%%%%%%%%%%%%%%%%%%%%%%%%%%%%%%%%%%%%%%%%%%%%%%%%%%%%%

We summarize the forms of differential production (on the solid angle $\Omega$ in the center-of-mass flame) cross section at the LHC.
As we pointed out, we set the mixing angles 
$\theta_D=0$ and $\theta_L=0$ or $\pi/2$ 
and consider that all of the HC $\rho$ mesons are completely degenerated.
We note that in the limit of $\theta_D =0$, the possible initial state is $b \bar{b}$ only.
{Due to this mass degeneracy, we should take all of the HC $\rho$} contributions simultaneously.

First, we look at the dijet final state {that} originates from $b$ (or anti-$b$) quark {where} $\rho_{(8)}^{0}$, $\rho_{(8)}^{3}$, $\rho_{(1)}^{3}$, $\rho_{(1)'}^{3}$, and $\rho_{(1)'}^{0}$ contribute as intermediate states ($b \bar{b} \to \text{$\rho$'s} \to b \bar{b}$).
The {differential cross section forms} for color-singlets and color-octets are summarized as
\al{
\left( \frac{d \sigma_{jj}}{d \Omega} \right)_{\text{singlet}}
	&= \frac{1}{2\hats} \frac{1}{32\pi^2}
	   \frac{1}{2^2} \frac{1}{3^2}
	   \frac{4 \, g_{\rho L}^4 f^{bb}_{\rm singlet}}{(\hats-m_\rho^2)^2 + (m_\rho \Gamma_\rho)^2}
	   \left( \frac{\hats}{2} (1+\cos{\theta}) \right)^2, \\[0.5em]
%%%
\left( \frac{d \sigma_{jj}}{d \Omega} \right)_{\text{octet}}
	&=  \frac{1}{2\hats} \frac{1}{32\pi^2} \frac{1}{2^2} \frac{1}{3^2}
	    \frac{4 \, g_{\rho L}^4 f^{bb}_{\rm octet}}{(\hats-m_\rho^2)^2 + (m_\rho \Gamma_\rho)^2}
	    \left( \frac{\hats}{2} (1+\cos{\theta}) \right)^2,
}
where we take all of the quarks are massless.
Now, both of color-octets and color-singlets contribute in $s$-channels, and no interference term appears between the singlet amplitudes and those of octets.
Here, the following relations hold in the hatted Mandelstam variables in the parton system
\al{
\hats + \hatt + \hatu \simeq 0,\quad
\hatt \simeq  -\frac{\hats(1-\cos{\theta})}{2},\quad
\hatu \simeq  -\frac{\hats(1+\cos{\theta})}{2},
}
where the angle $\theta$ is defined in the center-of-mass frame.
The factors $f^{bb}_{\rm singlet}$ and $f^{bb}_{\rm octet}$ represent summations of all possible combinations of the couplings plus the overall color factor,
\al{
f^{bb}_{\rm singlet}
	&= {3^2} \times \left[ \left( -\frac{1}{4} \right)^2 + \left( -\frac{1}{4\sqrt{3}} \right)^2 + \left( \frac{1}{4\sqrt{3}} \right)^2 \right]^2 \simeq 0.098, \\
%%%
f^{bb}_{\rm octet}
	&= {2} \times \left[ \left( -\frac{1}{\sqrt{2}} \right)^2 + \left( \frac{1}{\sqrt{2}} \right)^2 \right]^2 = 2.
}

Next, we go for the {ditau and dimuon final states}, where $\rho_{(3)}^{3}$, $\rho_{(3)}^{0}$, $\rho_{(1)}^{3}$, $\rho_{(1)'}^{3}$, $\rho_{(1)'}^{0}$ are possible intermediate states ({$b \bar{b} \to \text{$\rho$'s} \to \tau \bar{\tau}/\mu\bar{\mu}$}).
{We note that in the present limit ($\theta_D = 0$ and $\theta_L = 0$ or $\pi/2$), {the cross section formula for the ditau channel is exactly the same as that for the dimuon channel.}}
In the present case, color-triplets/-singlets contribute in $t$/$s$-channels, and interference terms are observed between the color-triplets and color-singlets.
The {differential cross section forms} for color-singlets, color-triplets, and interferences are summarized as
\al{
\left( \frac{d \sigma_{{\tau\tau/\mu\mu}}}{d \Omega} \right)_{\text{singlet}}
	&=  \frac{1}{2\hats} \frac{1}{32\pi^2} \frac{1}{2^2} \frac{1}{3^2}
	    \frac{4 \, g_{\rho L}^4 f^{{\tau\tau/\mu\mu}}_{\rm singlet}}{(\hats-m_\rho^2)^2 + (m_\rho \Gamma_\rho)^2}
	    \left( \frac{\hats}{2} (1+\cos{\theta}) \right)^2, \\
%%%
\left( \frac{d \sigma_{{\tau\tau/\mu\mu}}}{d \Omega} \right)_{\text{triplet}}
	&=  \frac{1}{2\hats} \frac{1}{32\pi^2} \frac{1}{2^2} \frac{1}{3^2}
	    \frac{4 \, g_{\rho L}^4 f^{{\tau\tau/\mu\mu}}_{\rm triplet}}{(\hatt-m_\rho^2)^2 + (m_\rho \Gamma_\rho)^2}
	    \left( \frac{\hats}{2} (1+\cos{\theta}) \right)^2, \\
%%%
\left( \frac{d \sigma_{{\tau\tau/\mu\mu}}}{d \Omega} \right)_{\text{int}}
	&=  \frac{1}{2\hats} \frac{1}{32\pi^2} \frac{1}{2^2} \frac{1}{3^2}
	    \frac{g_{\rho L}^4 f^{{\tau\tau/\mu\mu}}_{\rm int} \left[  (\hats-m_\rho^2)(\hatt-m_\rho^2) + (m_\rho \Gamma_\rho)^2  \right] \times {2}}
	         {\left[(\hats-m_\rho^2)(\hatt-m_\rho^2) + (m_\rho \Gamma_\rho)^2\right]^2 + \left[ m_\rho \Gamma_\rho (\hatt-\hats) \right]^2}
	    (-1) \left( \frac{\hats}{2} (1+\cos{\theta}) \right)^2,
}
with
\al{
f^{{\tau\tau/\mu\mu}}_{\rm singlet}
	&= {3} \times \left[ \left( -\frac{1}{4} \right)^2 + 
	                   \left( - \frac{1}{4\sqrt{3}} \right)\left( \frac{\sqrt{3}}{4} \right) +
	                   \left(  \frac{1}{4\sqrt{3}} \right)\left( - \frac{\sqrt{3}}{4} \right) \right]^2
	                   \simeq 0.012, 
	                   \label{eq:ftautau_1}\\
%%%
f^{{\tau\tau/\mu\mu}}_{\rm triplet}
	&= 3 \times \left[ \left( -\frac{1}{2} \right)^2 + \left( \frac{1}{2} \right)^2 \right]^2 = 0.75,
	\label{eq:ftautau_2} \\
%%%
f^{{\tau\tau/\mu\mu}}_{\rm int}
	&= 3 \times \left[ \left( -\frac{1}{4} \right)^2 + 
	                   \left( - \frac{1}{4\sqrt{3}} \right)\left( \frac{\sqrt{3}}{4} \right) +
	                   \left(  \frac{1}{4\sqrt{3}} \right)\left( - \frac{\sqrt{3}}{4} \right) \right]
	            \left[ \left( -\frac{1}{2} \right)^2 + \left( \frac{1}{2} \right)^2 \right] \simeq -0.094.
	            \label{eq:ftautau_3}
}

%%%%%%%%%%%%%%%%%%%%%%%%%%%%%%%%%%%%%%%%%%
\subsubsection{Results}
%%%%%%%%%%%%%%%%%%%%%%%%%%%%%%%%%%%%%%%%%%

%%%%%%%%%%%%%%%%%%%%%%%%%%%%%%%%%%%%%%%%%%%%%%%%%%%%%%%%%%%%%%%%%%%%%%%%%%%%
\begin{figure}[t]
\centering
\includegraphics[width=0.49\columnwidth]{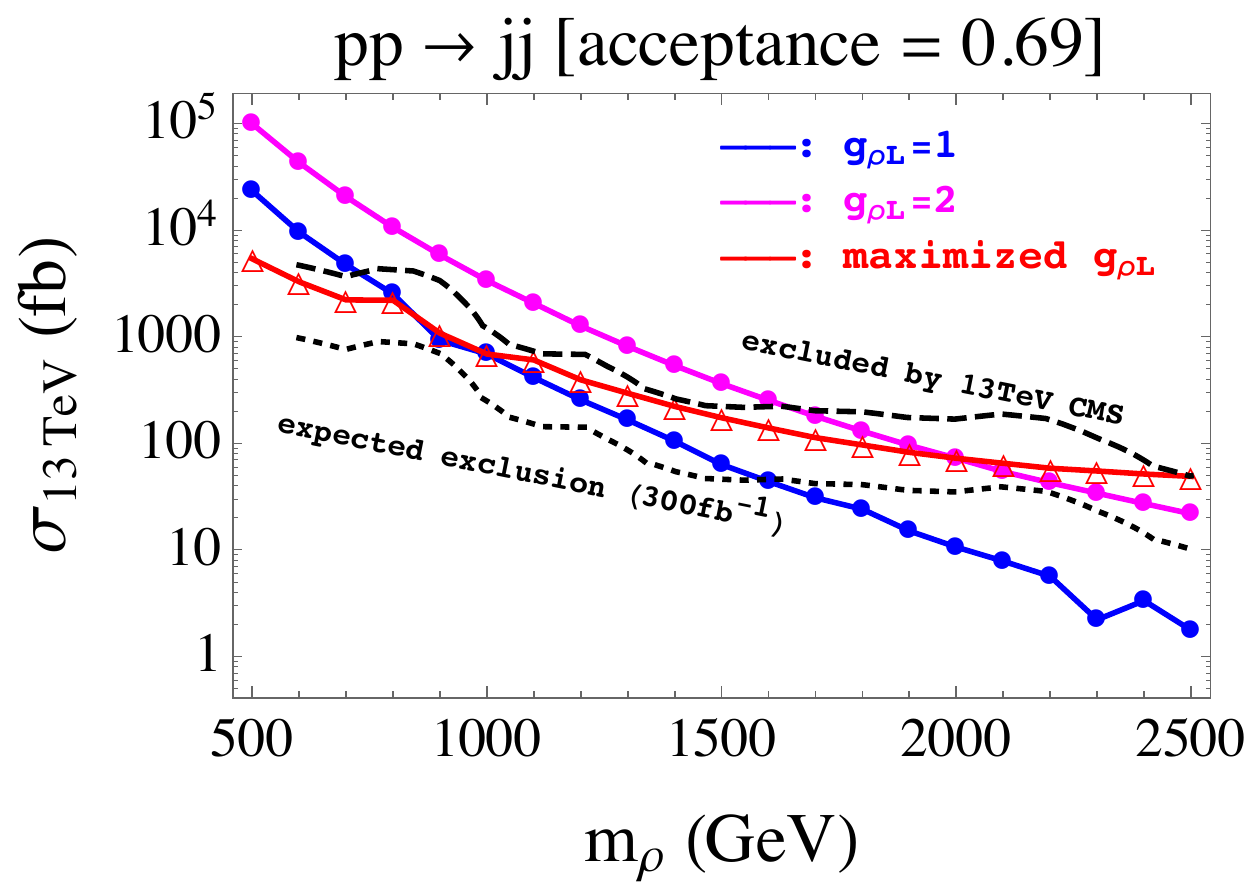}
\includegraphics[width=0.49\columnwidth]{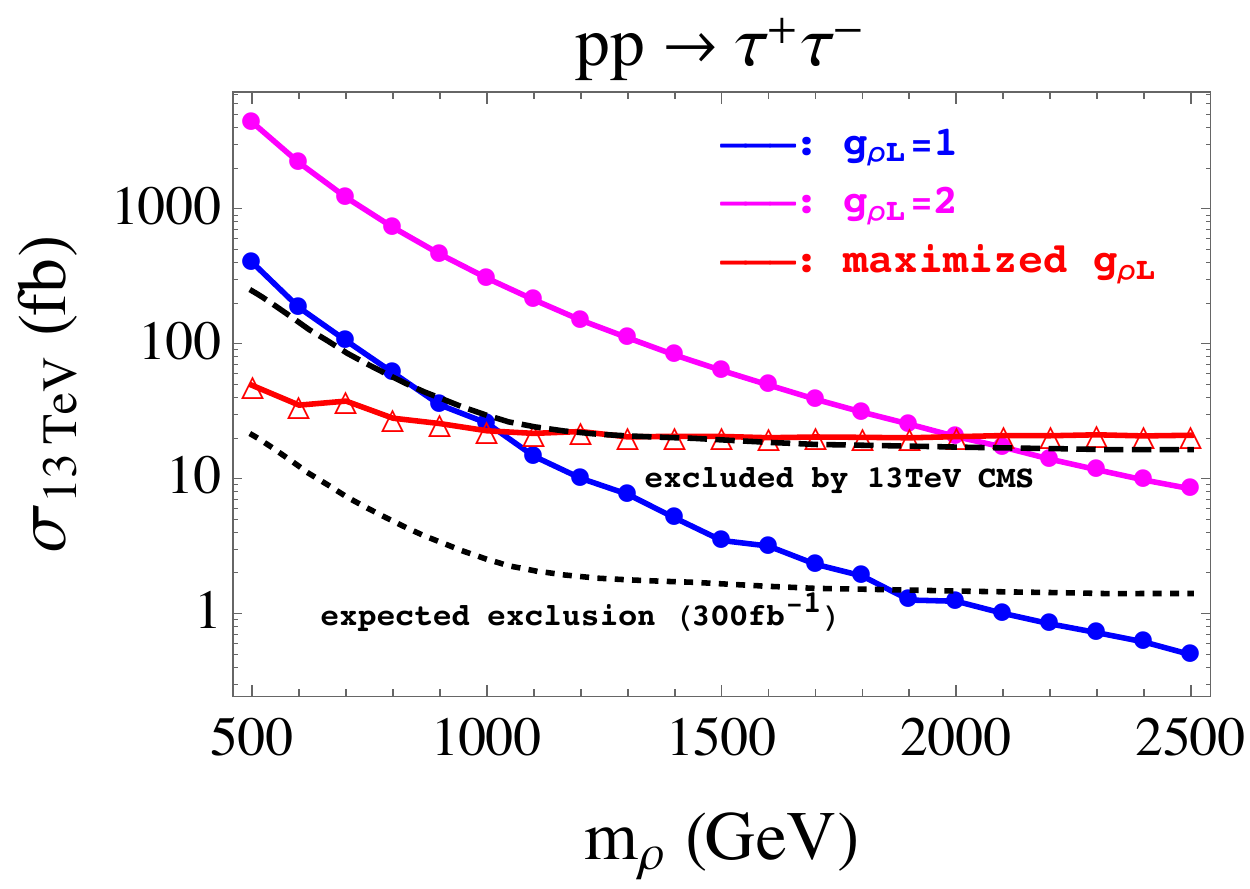}
%%%
\caption{Naive constraints on $\sigma(pp \to jj)$ [left panel] and $\sigma(pp \to \tau \bar{\tau})$ [right panel] in $\theta_D = \theta_L = 0$, where the CMS experimental results are from Refs.~\cite{Sirunyan:2016iap,Khachatryan:2016qkc}.
For the dijet resonance, we adopt the calculated value of acceptance ${\cal A} = 0.69$ being different from the isotropic decays (${\cal A} \approx 0.6$)~\cite{Sirunyan:2016iap}.
For the red curves showing `maximized $g_{\rho L}$' being consistent with the requisites from flavor issues in Eq.(\ref{Eq:allowedgL}), the value of $g_{\rho L}$ is tuned as $|g_{\rho L}| = {1.0} \times (m_{\rho}/1\,\text{TeV})$.
If $g_{\rho L}$ is greater {than the tuned value} (in each of $m_{\rho}$), we cannot address the flavor issues in the present model correctly.
The regions above the black dashed (black dotted) lines are excluded (expected to be excluded after a $300\,\text{fb}^{-1}$ integrated luminosity accumulation) {at $95\%$ C.L.s}.
}
%%%
\label{fig:cross_section_13TeV}
\end{figure}
%%%%%%%%%%%%%%%%%%%%%%%%%%%%%%%%%%%%%%%%%%%%%%%%%%%%%%%%%%%%%%%%%%%%%%%%%%%%

The convolution with the (anti-)bottom quark parton distribution function~(PDF) inside the proton $f_{b/p}(x,\mu_F)$ ($f_{\bar{b}/p}(x,\mu_F)$) with the Bjorken $x$ and the PDF factorization scale $\mu_F$ is formulated as
\al{
\int_{\tau_0}^{1} dx_1 \int_{\tau_0/x_1}^{1} dx_2  \,\sigma 
\left[ f_{b/p}(x_1,\mu_F) f_{\bar{b}/p}(x_2,\mu_F) + f_{b/p}(x_2,\mu_F) f_{\bar{b}/p}(x_1,\mu_F) \right].
}
Here, the total energy squared of the present LHC $s$($=13^2\,\text{TeV}^2$) is related to the total energy squared of the focused parton system $\widehat{s}$ as $\widehat{s} = x_1 x_2 s \equiv \tau \, s$.
Kinematically, the minimal of the fraction $\tau$ is estimated as~\cite{Han:2005mu}
\al{
\tau_0 = {\frac{\{4 m_b^2,\ 4 m_\tau^2,\ 4 m_\mu^2\}}{s} \sim 
         \{ 4 \times 10^{-7},\ 8 \times 10^{-8},\ 3 \times 10^{-10} \}},
	\label{eq:tau_zero_kinematical}
}
where $\mu_F$ is set as $m_\rho$.
We adopt the {\tt CTEQ6L1} PDF~\cite{Pumplin:2002vw} in calculations in {\tt Mathematica} with the help of a PDF parser package, {\tt ManeParse\_2.0}~\cite{Clark:2016jgm}, and set $\tau_0$ as $10^{-6}$ of the minimal value of the Bjorken $x$ in the {\tt CTEQ6L1} PDF set, which {is} not far from the values shown in Eq.(\ref{eq:tau_zero_kinematical}).

In Fig.~\ref{fig:cross_section_13TeV}, we summarize the cross sections of $p(b)p(\bar{b}) \to \text{$\rho$'s} \to j(b)j(\bar{b})$ (left panel) and $p(b)p(\bar{b}) \to \text{$\rho$'s} \to \tau \bar{\tau}$ (right panel).
To calculate the numerical integration including the PDF convolution in the {\tt Divonne} method~\cite{Friedman:1978ed,Friedman:1981ak}, we use the {\tt CUBA} package~\cite{Hahn:2004fe} with the {\tt Mathlink} protocol in {\tt Mathematica}.
The values of cross sections were cross-checked with {\tt MadGraph5\_aMC@NLO}~\cite{Alwall:2011uj,Alwall:2014hca}, where the {\tt UFO}-style model file~\cite{Degrande:2011ua} was generated by the {\tt FeynRules} package~\cite{Christensen:2008py,Alloul:2013bka}.
For estimating the acceptance ${\cal A}$ of dijet events, we generated parton-level events in {\tt MadGraph5\_aMC@NLO} and analyzed them in the {\tt ROOT} framework~\cite{Brun:1997pa} with the help of {\tt ExRootAnalysis}, which is a part of the integrated package of {\tt MadGraph5\_aMC@NLO}.
We obtained ${\cal A} \simeq 0.69$ in our case, which a bit deviates from the isotropic case (${\cal A} \approx 0.6$) shown in Ref.~\cite{Sirunyan:2016iap}.
The $95\%$ C.L. upper bounds at $\sqrt{s} = 13\,\text{TeV}$ were extracted from Refs.~\cite{Sirunyan:2016iap} (dijet, based on $12.9\,\text{fb}^{-1}$ CMS data),\footnote{{The latest ATLAS result was reported in Ref.~\cite{Aaboud:2017yvp} after~\cite{Sirunyan:2016iap}, where the constraints on $W'$ and $W^\ast$ scenarios do not overwhelm the bound of~\cite{Sirunyan:2016iap} in the range of the invariant mass, $q^2 \lesssim 2.5\,\text{TeV}$.}
}
\cite{Khachatryan:2016qkc} (ditau, based on $2.2\,\text{fb}^{-1}$ CMS data), and \cite{ATLAS-CONF-2017-027} (dimuon, based on $36.1\,\text{fb}^{-1}$ ATLAS data).
The expectations for the limits after $300\,\text{fb}^{-1}$ data accumulation were simply calculated by rescaling from the present bounds~\cite{Sirunyan:2016iap,Khachatryan:2016qkc,ATLAS-CONF-2017-027}.

From Fig.~\ref{fig:cross_section_13TeV}, we see the constraints on $g_{\rho L}$ and $m_\rho$ for the $\theta_L =0$ case. 
The dijet bound {in} the left panel shows that no constraint is imposed on our flavor-specific HC $\rho$ mesons if the value of $g_{\rho L}$ is maximized so as to be consistent with Eq.(\ref{Eq:allowedgL}), 
which is the combined constraints from all the appreciable flavor observables, (the corresponding maximal value is taken for each of $m_\rho$ from Eq.(\ref{Eq:allowedgL}).) 
On the other hand, the ditau channel excludes {a part of possibilities to have the maximized $g_{\rho L}$}, {where $m_{\rho}$ is greater than $\sim 1500 \,\text{GeV}$,} 
whereas it excludes {the HC rho mass scale as} $m_{\rho} \gtrsim 900 \ (2100) \,\text{GeV}$ for the fixed {coupling} value $g_{\rho L} = 1 \ (2)$, as shown in the right panel {at $95\%$ C.L.s}. 
Therefore, the ditau channel plays a significant role in probing this scenario at the LHC~\footnote
{We observed that the largeness of the $t$-channel effective coupling shown in Eqs.(\ref{eq:ftautau_1})--(\ref{eq:ftautau_3}) results in the situation that 
the $t$-channel contribution becomes a major part to the cross section in the {ditau and dimuon} production. 
Thus some deviations from the present acceptance times efficiency may be expected when a dedicated collider simulation is performed.  
We do not take into account of this point in our ballpark estimations of current constraints and future prospects.
}.

%%%%%%%%%%%%%%%%%%%%%%%%%%%%%%%%%%%%%%%%%%%%%%%%%%%%%%%%%%%%%%%%%%%%%%%%%%%%
\begin{figure}[t]
\centering
\includegraphics[width=0.49\columnwidth]{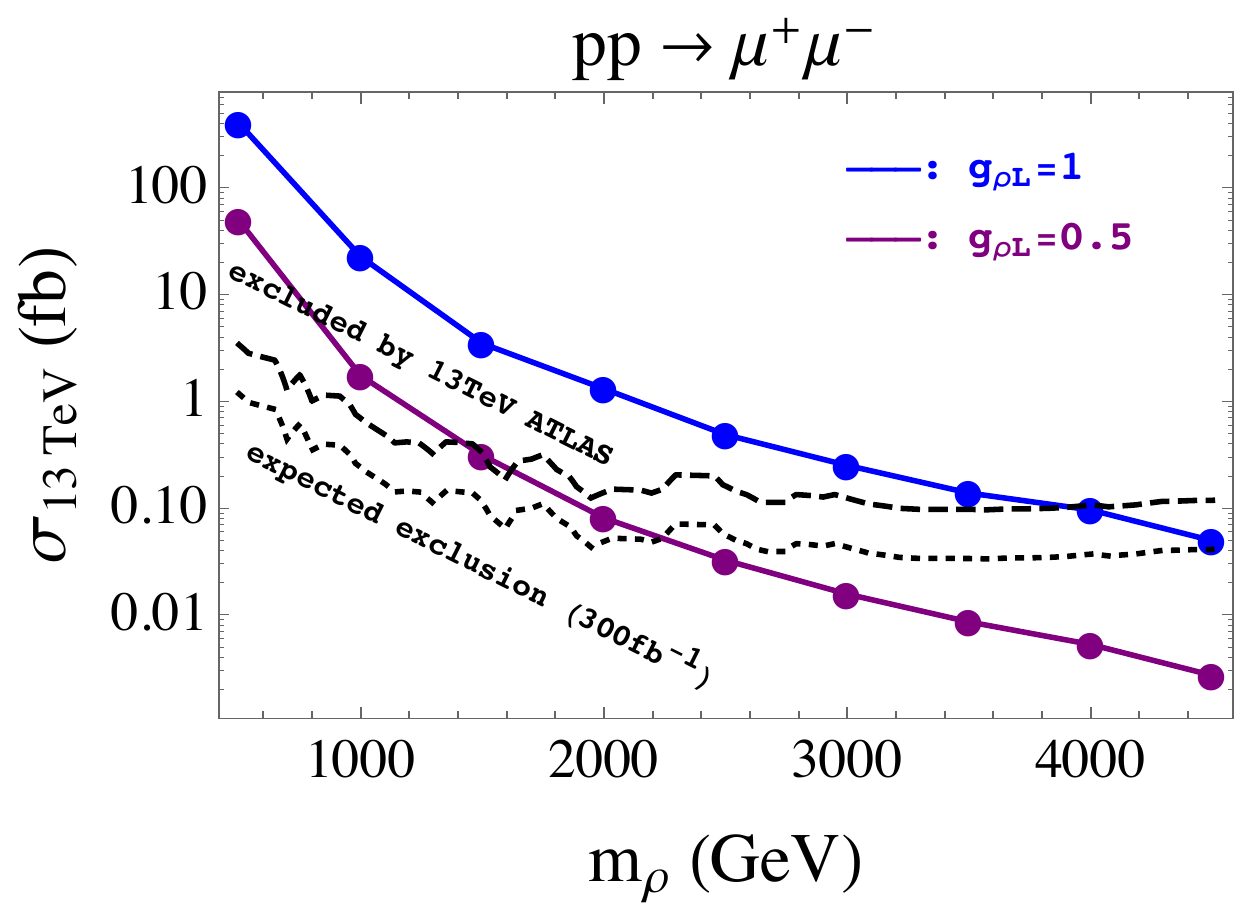}
%%%
\caption{{{Naive} constraint on $\sigma(pp \to \mu^+ \mu^-)$ for $\theta_D = 0$, $\theta_L = \pi/2$ and $g_{\rho L} = 0.5, 1$, 
where the latest experimental result provided by the ATLAS group is obtained from Ref.~\cite{ATLAS-CONF-2017-027}.}
{The regions above the black dashed (black dotted) lines are excluded (expected to be excluded after a $300\,\text{fb}^{-1}$ integrated luminosity accumulation) {at $95\%$ C.L.s}.}
}
%%%
\label{fig:cross_section_13TeV_dimuon}
\end{figure}
%%%%%%%%%%%%%%%%%%%%%%%%%%%%%%%%%%%%%%%%%%%%%%%%%%%%%%%%%%%%%%%%%%%%%%%%%%%%

In Fig.~\ref{fig:cross_section_13TeV_dimuon}, we show the collider bound for the $\theta_L =\pi/2$ case from the dimuon searches. 
It indicates that our scenario was already tested and {excluded} up to $m_\rho = 4 \, (1.5)\,\text{TeV}$ at $95\%$ C.L. when $g_{\rho L} = 1 \, (0.5)$. 
We should keep in mind, however, that the present scenario with $g_{\rho L} =1$ {for the mass range} $4\,\text{TeV} \lesssim m_\rho \lesssim 10\,\text{TeV}$ can still accommodate the $b \to s\mu^+\mu^-$ anomaly as seen in Eq.\eqref{Eq:allowedgL_2}. 
Note that the dijet bound for the $\theta_L =\pi/2$ case is not significant {when} $g_{\rho L} < 1$.

%%%%%%%%%%%%%%%%%%%%%%%%%%%%%%%%%%%%%%%%%%%%%%%%%%%%%%%%%%%%%%%%%%%%%
%%%%%%%%%%%%%%%%%%%%%%%%%%%%%%%%%%%%%%%%%%%%%%%%%%%%%%%%%%%%%%%%%%%%%
\section{Summary} 
\label{summary}
%%%%%%%%%%%%%%%%%%%%%%%%%%%%%%%%%%%%%%%%%%%%%%%%%%%%%%%%%%%%%%%%%%%%%
%%%%%%%%%%%%%%%%%%%%%%%%%%%%%%%%%%%%%%%%%%%%%%%%%%%%%%%%%%%%%%%%%%%%%

In this paper, we analyzed the flavorful composite physics of composite vector bosons arising from the one-family model of the vectorlike HC theory, 
which we dubbed the HC rho model formulated based on the HLS construction. 
The model structure has been unambiguously fixed by the ``chiral'' symmetry for the HC fermions and the HLS gauge invariance. 
The model involves the 63 HC rho mesons and {63} HC pions with phenomenologically rich structure, which couple to
the SM gauge bosons as the essential consequence of the HLS gauge invariance. 
Coupling properties to the SM fermions are restricted due to the HLS gauge invariance of the vectorlike theory.
The flavor-dependent couplings are required to be the third-generation {fermion-philic} by the flavor-dependent EW precision measurements. 
The flavor-universal couplings, on the other hand, 
hardly get limited by the EW sector constraints (oblique corrections), 
which is indeed due to the vectorlike model construction. 
To be specific, we have found that the forward-backward asymmetry of tau lepton $A_{\rm FB}^{(0,\tau)}$ and the {$Z$} boson decay to bottom quark pair $(R_b)$ are fairly sensitive to the flavor-dependent couplings of the HC rho mesons.

In turn, we surveyed the allowed coupling parameter space of the HC rho mesons from the relevant flavor observables 
by taking into account the flavor mixing structures between the second and third generations involved in the left-handed down-quarks and leptons, 
parametrized by the angle $\theta_D$ and $\theta_L$ as in Eq.\eqref{LDrotation}.
The most stringent bound come from the $B^0_s$-$\bar{B}^0_s$ mixing.
In conjunction with other flavor observables, it has turned out that the down-quark mixing $\theta_D$ has to be much tiny (but non-zero) 
while the lepton mixing $\theta_L$ has wider allowed range depending on the mass scale of HC rho mesons and the coupling strength as seen in Fig.~\ref{Fig:flavorconstraint}. 
In particular, the viable scenarios are classified into two spots with respect to $\theta_L$:
one class is the case that the HC rho mesons predominantly couple to third-generation leptons $(\theta_L \lesssim \pi/4)$, and 
another is to the second-generation ($\theta_L \sim \pi/2$), as obtained in Figs.~\ref{Fig:flavorconstraint} -- \ref{Fig:allowedcoupling}. 
The exclusion plots combined with the flavor-dependent EW precision tests for the above two cases are shown in Fig.~\ref{Fig:EWandFlavor}.
Of interest for both two scenarios is that the HC rho mesons hardly contribute to $B \to D^{(*)}\ell\bar\nu$, which in turn implies that the ratios $R_{D^{(*)}}$ 
do not significantly deviate from the SM predictions. 
This is essentially because of the almost complete degeneracy in the HC rho mass spectra due to the large flavor-universal coupling ($g_\rho\sim 6$). 
In contrast, the HC rho mesons with mass of TeV scale can give significant contributions to $b \to s\mu^+\mu^-$. 
Hence the present scenario can achieve the large values of the Wilson coefficients for the effective operators of $b \to s\mu^+\mu^-$ with the $V-A$ form ($C_9 = -C_{10}$), 
which can account for the present anomalies in the experimental data from LHCb, Belle, ATLAS, and CMS.

We then discussed the implications to the collider physics at LHC 
and found that the HC rho mesons with mass of TeV scale 
can be consistent with the current 13 TeV LHC data:  
In the case with $\theta_L \lesssim \pi/4$, 
the most stringent limit comes from dijet and ditau channels, 
which turns out to exclude the mass up to $\sim 1 - 2$ TeV 
for the flavored HC rho coupling $g_{\rho L}=1-2$ [Fig.~\ref{fig:cross_section_13TeV}]. 
The HC rhos in the other case with $\theta \sim \pi/2$ 
is, on the other hand, 
more severely constrained by the dimuon channel, 
and the mass has already been excluded up to $\sim 4\, (1.5)$ TeV for $g_{\rho L}=1\, (0.5)$.

Thus, our HC rho model has interesting correlations 
sensitive to the EW precision measurements, the $B$-decay anomalies 
and the LHC collider signatures. 
Through the present anatomy of our model,  
we can reach a definite conclusion: 
if the current $R_{D^{(*)}}$ anomaly goes away, but the $R_{K^{(*)}}$ 
deficit further grows to be explained by the flavorful HC rhos on 
TeV mass scale,  
then those HC rhos will show up also in 
the future LHC data on the ditau, dijet and dimuon channels 
with higher luminosity. 
In particular, 
the case with $\theta_L \sim \pi/2$ is the most intriguing 
even when viewed from any point of EW precision, flavor and collider physics;    
(i) the experimental bound for {$b\to s \mu^+ \mu^-$} anomaly can easily be satisfied 
if and only if a tiny $\theta_D$ is at hand, 
while the case with $\theta_L \lesssim \pi/4$ is barely allowed 
within the {$3\sigma$} range even for the tiny $\theta_D$ {if $g_{\rho} g_{L}^{33} \, (= g_{\rho L})$ is close to unity} [Fig.~\ref{Fig:allowedregions}];   
(ii) the EW precision tests give {the {severer} constraint on the HC rho mass when the $68\%$ C.L. bound is taken into account}
%the constraints at 68\% C.L. more severely for the HC rho mass scale 
in the case with $\theta_L \lesssim \pi/4$, 
which is much milder in the case with $\theta_L\sim\pi/2$ 
[Fig.~\ref{Fig:EWandFlavor}]; 
(iii) the dimuon channel search at the LHC generically has higher sensitivity 
than the ditau channel, so does the case with $\theta_L \sim \pi/2$ {[Figs.~\ref{fig:cross_section_13TeV} and \ref{fig:cross_section_13TeV_dimuon}]}.

It is also interesting to note that
the net effects on the flavor observables in the present model are similar to those in a low-energy effective $U(1)'$ model, in spite of the fact that our model includes 63 {components of} new vector bosons. 
This can be checked by looking at the analytic formulae for the observables presented in this paper. 
To be specific, the $U(1)'$ model can be described, in terms of the HC rho model, by replacing 
the coefficient of the four-fermion operators
$(\bar{q}_L \gamma^\mu q_L)(\bar{q}_L \gamma_\mu q_L)$,
$(\bar{q}_L \gamma^\mu q_L)(\bar{\ell}_L \gamma_\mu \ell_L)$,
$(\bar{\ell}_L \gamma^\mu \ell_L)(\bar{\ell}_L \gamma_\mu \ell_L)$: 
$\frac{7}{16} \frac{g_\rho (g_L^{33})^2}{m_\rho^2} \to \frac{g_a^{33} g_b^{33}}{m_{Z'}^2}$ {[for $(a,b)= (q,q),(q,\ell),(\ell,\ell)$]}, see {\it e.g.}, Ref.~\cite{Cline:2017lvv} {and {\it c.f.,}~Eqs.(\ref{eq:C9C10}), (\ref{eq:CLIJ}), (\ref{EQlimit_tauphimu}), (\ref{eq:BR_tauto3mu}), (\ref{eq:DeltaMBs_Bsubs})}. 
Therefore, the HC rho model can be considered as one of the UV completed models that realize the extended $U(1)'$ gauge symmetry to the SM at the low energy scale. 
Unlike the $U(1)'$ model, on the other hand, vector leptoquark bosons are involved in the present model. 
Therefore, individual searches for signals of the vector leptoquark bosons would be of great importance to distinguish our scenario with the low energy $U(1)'$ model that only contains $Z'$~\cite{KSRfuture}.

Besides the flavorful HC rhos and HC pions, 
the one-family model of the HC 
predicts a number of other HC hadrons like 
composite scalars and baryons. 
All those HC hadrons are expected to have the same order of the mass 
as the HC rhos (on the order of TeV scale), 
and thus are potentially sensitive enough to be detected 
at the LHC. 
The lightest HC baryon might be a candidate of dark matter, 
because of the stability by the HC baryon number conservation. 
Those interesting issues are to be discussed in another publication.

In closing, we briefly sketch how to discriminate our scenario based on the vector-like confinement and other proposals based on composite Higgs scenarios~\cite{Gripaios:2014tna,Niehoff:2015bfa,Niehoff:2015iaa,Carmona:2015ena,Barbieri:2016las,DAmico:2017mtc}.
Two general guiding principles can be proposed.
One is to measure the properties of the observed $125\,\text{GeV}$ Higgs boson precisely.
At the leading order no deviation is expected in Higgs couplings in our scenario with the $SU(2)_W$ doublet fundamental Higgs boson, while deviations are expected to be observed in Higgs couplings in composite Higgs scenarios (see e.g.,~\cite{Contino:2010rs,Bellazzini:2014yua,Panico:2015jxa}{.})
The other is to clarify the species of vector-$\rho$ mesons, 
which depends on how global symmetries break down, at the LHC.
For example in the composite Higgs scenario discussed in Ref.~\cite{Barbieri:2016las} (which follows discussions in Refs.~\cite{Barbieri:2012uh,Barbieri:2015yvd}),
the original global symmetry is $SU(4) \times SU(2)_L \times SU(2)_R \times U(1)_X$ while that of ours is {$SU(8)$}.
A possible difficulty is that expected spectra of these vector-$\rho$ mesons may be fairly degenerated to evade the bounds from the electroweak precision measurements as discussed in section~\ref{sec:Fundamental_requirements}.
Thereby, more dedicated discussions are required to declare how relevant this way is for discriminating models with (hidden) strong dynamics.

\vspace{2em}

{\bf Note added:}  %\noindent

The new LHCb measurement of $R_{D^*}$ has been reported as in Ref.~\cite{NewRDstarLHCb} after when we submitted the first version of this work on the arXiv.org. 
Its result is consistent with the SM prediction within $\sim 1 \sigma$, but a naive private combination with the other results leads $R_{D^*}^\text{exp}/R_{D^*}^\text{SM} \approx 1.21 \pm 0.07$. 
Although the deviation is unchanged, the central value becomes smaller, which is good direction for the HC $\rho$ model.

\acknowledgments 
This work was supported in part by the JSPS Grant-in-Aid for Young Scientists (B) \#15K17645 (S.M.). 
K.N. thanks Dong Woo Kang, Pyungwon Ko, Takaaki Nomura, Tomoki Nosaka, Seong Chan Park, Yasuhiro Yamamoto {and Naoki Yamatsu for useful discussions, and especially Takuya Morozumi for various fruitful comments and pointing out a typo.}
K.N. is also grateful to David London and the Group of Particle Physics of Universit\'e de Montr\'eal for the kind hospitality at the middle stage of this work.
R.W. thanks Ryuichiro Kitano for a discussion at very early stage of this work.

\appendix

%%%%%%%%%%%%%%%%%%%%%%%%%%%%%%%%%%%%%%%
%%%%%%%%%%%%%%%%%%%%%%%%%%%%%%%%%%%%%%%
%%%%%%%%%%%%%%%%%%%%%%%%%%%%%%%%%%%%%%%
\section{The $\rho-\pi-\pi$ coupling terms \label{appendix:rho-pi-pi}} 
%%%%%%%%%%%%%%%%%%%%%%%%%%%%%%%%%%%%%%%
%%%%%%%%%%%%%%%%%%%%%%%%%%%%%%%%%%%%%%%
%%%%%%%%%%%%%%%%%%%%%%%%%%%%%%%%%%%%%%%

In this Appendix we give the explicit formulae for the $\rho$-$\pi$-$\pi$ 
coupling terms derived from Eq.(\ref{Lag:HLS}){.} 
The rho meson couplings to two pions  
arise from the rho mass term $m_\rho^2/g_\rho^2 {\rm tr}[\hat{\alpha}_{|| \mu}^2]$ 
in Eq.(\ref{Lag:HLS}):    
\begin{eqnarray} 
\frac{m_\rho^2}{g_\rho^2} {\rm tr}[\hat{\alpha}_{\mu ||}^2] 
\Bigg|_{\textrm{SM gauges =0}}^{\textrm{unitary gauge for HLS}} 
&=& 
\frac{m_\rho^2}{g_\rho^2} {\rm tr}\left[\left( \frac{\partial_\mu \xi_R \cdot \xi_R^\dag + 
\partial_\mu \xi_L \cdot \xi_L^\dag}{2i} \right)^2 \right]  
\nonumber \\
&=& \frac{m_\rho^2}{g_\rho^2} {\rm tr}\left[ 
\left(- \frac{i}{2 f_\pi^2} [\partial_\mu \pi, \pi] - g_\rho \rho_\mu + {\cal O}(\pi^4) 
\right)^2 \right] 
\nonumber \\ 
&=& 
m_\rho^2 {\rm tr}[\rho_\mu^2] 
+ 
\frac{m_\rho^2}{g_\rho f_\pi^2} i \, 
{\rm tr}\left[ 
 [\partial_\mu \pi, \pi] \rho^\mu 
\right] 
+ {\cal O}(\pi^4) 
\,, 
\label{rho-pi-pi}
\end{eqnarray}
where we have used $\xi_{R,L} = e^{{\pm} i \pi/f_\pi}$ in the unitary gauge of the HLS. 
%%%
{From Eq.(\ref{rho-pi-pi}) [or Eq.(\ref{eq:rho-pi-pi_form}) equivalently], we derive concrete forms of the $\rho$-$\pi$-$\pi$ couplings.
\al{
\mathcal{L}_{\rho_{(3)}^0-\pi-\pi}
&=
+
\frac{i\,m_\rho^2}{g_\rho f_\pi^2}
\Bigg[
\frac{1}{2\sqrt{2}} 
\left( \stackrel{\leftrightarrow}{\partial}_\mu \left[ \overline{\pi}^{0}_{(3)} \pi^{0}_{(8)a} \right] \right)
\left( \frac{\lambda^a}{2} \right) \rho^{0\mu}_{(3)}
+
\frac{1}{2\sqrt{2}} 
\left( \stackrel{\leftrightarrow}{\partial}_\mu \left[ \overline{\pi}^{\alpha}_{(3)} \pi^{\alpha}_{(8)a} \right] \right)
\left( \frac{\lambda^a}{2} \right) \rho^{0\mu}_{(3)} \notag \\
&\quad
+
\frac{1}{2\sqrt{2}} 
\left( \stackrel{\leftrightarrow}{\partial}_\mu \left[ \overline{\pi}^{0}_{(3)} \pi^{0}_{(1)'} \right] \right)
\left( \frac{\lambda^a}{2} \right) \rho^{0\mu}_{(3)}
+
\frac{1}{2\sqrt{2}} 
\left( \stackrel{\leftrightarrow}{\partial}_\mu \left[ \overline{\pi}^{\alpha}_{(3)} \pi^{\alpha}_{(1)'} \right] \right)
\left( \frac{\lambda^a}{2} \right) \rho^{0\mu}_{(3)}
\Bigg], \\[0.5em]
%%%%%%%%%%%%
%%%%%%%%%%%%
\mathcal{L}_{\rho_{(3)}^\alpha-\pi-\pi}
&=
+
\frac{i\,m_\rho^2}{g_\rho f_\pi^2}
\Bigg[
\frac{1}{2\sqrt{2}} 
\left( \stackrel{\leftrightarrow}{\partial}_\mu \left[ \overline{\pi}^{\alpha}_{(3)} \pi^{0}_{(8)a} \right] \right)
\left( \frac{\lambda^a}{2} \right) \rho^{\alpha\mu}_{(3)}
+
\frac{1}{2\sqrt{2}} 
\left( \stackrel{\leftrightarrow}{\partial}_\mu \left[ \overline{\pi}^{0}_{(3)} \pi^{\alpha}_{(8)a} \right] \right)
\left( \frac{\lambda^a}{2} \right) \rho^{\alpha\mu}_{(3)} \notag \\
&\quad
+
\frac{i}{2\sqrt{2}} \epsilon^{\beta\gamma\alpha}
\left( \stackrel{\leftrightarrow}{\partial}_\mu \left[ \overline{\pi}^{\beta}_{(3)} \pi^{\gamma}_{(8)a} \right] \right)
\left( \frac{\lambda^a}{2} \right) \rho^{\alpha\mu}_{(3)}
+
\frac{1}{2\sqrt{3}} 
\left( \stackrel{\leftrightarrow}{\partial}_\mu \left[ \overline{\pi}^{\alpha}_{(3)} \pi^{0}_{(1)'} \right] \right)
\rho^{\alpha\mu}_{(3)} \notag \\
&\quad
+
\frac{1}{2\sqrt{3}} 
\left( \stackrel{\leftrightarrow}{\partial}_\mu \left[ \overline{\pi}^{0}_{(3)} \pi^{\alpha}_{(1)'} \right] \right)
\rho^{\alpha\mu}_{(3)}
-
\frac{i}{4\sqrt{3}} \epsilon^{\beta\gamma\alpha}
\left( \stackrel{\leftrightarrow}{\partial}_\mu \left[ \overline{\pi}^{\beta}_{(3)} \pi^{\gamma}_{(1)'} \right] \right)
\rho^{\alpha\mu}_{(3)} \notag \\
&\quad
+
\frac{i}{4} \epsilon^{\beta\gamma\alpha}
\left( \stackrel{\leftrightarrow}{\partial}_\mu \left[ \overline{\pi}^{\beta}_{(3)} \pi^{\gamma}_{(1)} \right] \right)
\rho^{\alpha\mu}_{(3)}
\Bigg], \\[0.5em]
%%%%%%%%%%%%%%%
%%%%%%%%%%%%%%%
\mathcal{L}_{\rho_{(\overline{3})}^{0}-\pi-\pi} &= \left(\mathcal{L}_{\rho_{(3)}^{0}-\pi-\pi}\right)^\dagger,
\qquad
\mathcal{L}_{\rho_{(\overline{3})}^\alpha-\pi-\pi}  = \left(\mathcal{L}_{\rho_{(3)}^\alpha-\pi-\pi}\right)^\dagger, \\[0.5em]
%%%%%%%%%%%%%%%
%%%%%%%%%%%%%%%
\mathcal{L}_{\rho_{(8)}^{0}-\pi-\pi}
&=
+
\frac{i\,m_\rho^2}{g_\rho f_\pi^2}
\Bigg[
\frac{i}{2\sqrt{2}} f^{bca}
\left( {\partial}_\mu {\pi}^{0}_{(8)b} \cdot \pi^{0}_{(8)c} \right)
\rho^{0\mu}_{(8)a}
-
\frac{1}{2\sqrt{2}}
\left( \stackrel{\leftrightarrow}{\partial}_\mu 
\left[ \overline{\pi}^{0}_{(3)} \left( \frac{\lambda^a}{2} \right) \pi^{0}_{(3)} \right] \right)
\rho^{0\mu}_{(8)a} \notag \\
&\quad
-
\frac{1}{2\sqrt{2}}
\left( \stackrel{\leftrightarrow}{\partial}_\mu 
\left[ \overline{\pi}^{\alpha}_{(3)} \left( \frac{\lambda^a}{2} \right) \pi^{\alpha}_{(3)} \right] \right)
\rho^{0\mu}_{(8)a}
+
\frac{i}{2\sqrt{2}} f^{bca}
\left( {\partial}_\mu {\pi}^{\alpha}_{(8)b} \cdot \pi^{\alpha}_{(8)c} \right)
\rho^{0\mu}_{(8)a}
\Bigg], \\[0.5em]
%%%%%%%%%%%%%%%
%%%%%%%%%%%%%%%
\mathcal{L}_{\rho_{(8)}^{\alpha}-\pi-\pi}
&=
+
\frac{i\,m_\rho^2}{g_\rho f_\pi^2}
\Bigg[
\frac{i}{2\sqrt{2}} f^{bca}
\left( \stackrel{\leftrightarrow}{\partial}_\mu 
\left[ {\pi}^{\alpha}_{(8)b} \pi^{0}_{(8)c} \right] \right)
\rho^{\alpha\mu}_{(8)a}
+
\frac{i}{2\sqrt{2}} \epsilon^{\beta\gamma\alpha} d^{bca}
\left( {\partial}_\mu {\pi}^{\beta}_{(8)b} \cdot \pi^{\gamma}_{(8)c} \right)
\rho^{\alpha\mu}_{(8)a} \notag \\
&\quad
-
\frac{1}{2\sqrt{2}}
\left( \stackrel{\leftrightarrow}{\partial}_\mu 
\left[ \overline{\pi}^{0}_{(3)} \left( \frac{\lambda^a}{2} \right) \pi^{\alpha}_{(3)} \right] \right)
\rho^{\alpha\mu}_{(8)a}
-
\frac{1}{2\sqrt{2}}
\left( \stackrel{\leftrightarrow}{\partial}_\mu 
\left[ \overline{\pi}^{\alpha}_{(3)} \left( \frac{\lambda^a}{2} \right) \pi^{0}_{(3)} \right] \right)
\rho^{\alpha\mu}_{(8)a} \notag \\
&\quad
-
\frac{i}{2\sqrt{2}} \epsilon^{\beta\gamma\alpha}
\left( \stackrel{\leftrightarrow}{\partial}_\mu 
\left[ \overline{\pi}^{\gamma}_{(3)} \left( \frac{\lambda^a}{2} \right) \pi^{\beta}_{(3)} \right] \right)
\rho^{\alpha\mu}_{(8)a}
\Bigg], \\[0.5em]
%%%%%%%%%%%%%%%
%%%%%%%%%%%%%%%
\mathcal{L}_{\rho_{(1)'}^{0}-\pi-\pi}
&=
+
\frac{i\,m_\rho^2}{g_\rho f_\pi^2}
\Bigg[
\frac{1}{4\sqrt{3}}
\left( \stackrel{\leftrightarrow}{\partial}_\mu 
\left[ \overline{\pi}^{0}_{(3)} \pi^{0}_{(3)} \right] \right)
\rho^{0\mu}_{(1)'}
+
\frac{1}{4\sqrt{3}}
\left( \stackrel{\leftrightarrow}{\partial}_\mu 
\left[ \overline{\pi}^{\alpha}_{(3)} \pi^{\alpha}_{(3)} \right] \right)
\rho^{0\mu}_{(1)'}
\Bigg], \\[0.5em]
%%%%%%%%%%%%%%%
%%%%%%%%%%%%%%%
\mathcal{L}_{\rho_{(1)'}^{\alpha}-\pi-\pi}
&=
+
\frac{i\,m_\rho^2}{g_\rho f_\pi^2}
\Bigg[
\frac{i}{4\sqrt{3}} \epsilon^{\beta\gamma\alpha}
\left( {\partial}_\mu {\pi}^{\beta}_{(8)a} \cdot \pi^{\gamma}_{(8)a} \right)
\rho^{\alpha\mu}_{(1)'}
-
\frac{1}{2\sqrt{3}}
\left( \stackrel{\leftrightarrow}{\partial}_\mu 
\left[ \overline{\pi}^{0}_{(3)} \pi^{\alpha}_{(3)} \right] \right)
\rho^{\alpha\mu}_{(1)'} \notag \\
&\quad
-
\frac{1}{2\sqrt{3}}
\left( \stackrel{\leftrightarrow}{\partial}_\mu 
\left[ \overline{\pi}^{\alpha}_{(3)} \pi^{0}_{(3)} \right] \right)
\rho^{\alpha\mu}_{(1)'}
+
\frac{i}{4\sqrt{3}} \epsilon^{\beta\gamma\alpha}
\left( \stackrel{\leftrightarrow}{\partial}_\mu 
\left[ \overline{\pi}^{\gamma}_{(3)} \pi^{\beta}_{(3)} \right] \right)
\rho^{\alpha\mu}_{(1)'} \notag \\
&\quad
-
\frac{i}{2\sqrt{3}} \epsilon^{\beta\gamma\alpha}
\left( {\partial}_\mu 
\overline{\pi}^{\beta}_{(1)'} \cdot \pi^{\gamma}_{(1)'} \right)
\rho^{\alpha\mu}_{(1)'}
+
\frac{i}{4} \epsilon^{\beta\gamma\alpha}
\left( \stackrel{\leftrightarrow}{\partial}_\mu 
\left[ \overline{\pi}^{\beta}_{(1)} \pi^{\gamma}_{(1)'} \right] \right)
\rho^{\alpha\mu}_{(1)'} 
\Bigg], \\[0.5em]
%%%%%%%%%%%%%%%
%%%%%%%%%%%%%%%
\mathcal{L}_{\rho_{(1)}^{\alpha}-\pi-\pi}
&=
+
\frac{i\,m_\rho^2}{g_\rho f_\pi^2}
\Bigg[
\frac{i}{4} \epsilon^{\beta\gamma\alpha}
\left( {\partial}_\mu {\pi}^{\beta}_{(8)a} \cdot \pi^{\gamma}_{(8)a} \right)
\rho^{\alpha\mu}_{(1)}
+
\frac{i}{4} \epsilon^{\beta\gamma\alpha}
\left( {\partial}_\mu {\pi}^{\beta}_{(1)'} \cdot \pi^{\gamma}_{(1)'} \right)
\rho^{\alpha\mu}_{(1)} \notag \\
&\quad
+
\frac{i}{4} \epsilon^{\beta\gamma\alpha}
\left( {\partial}_\mu {\pi}^{\beta}_{(1)} \cdot \pi^{\gamma}_{(1)} \right)
\rho^{\alpha\mu}_{(1)}
-
\frac{i}{4} \epsilon^{\beta\gamma\alpha}
\left( \stackrel{\leftrightarrow}{\partial}_\mu 
\left[ \overline{\pi}^{\gamma}_{(3)} \pi^{\beta}_{(3)} \right] \right)
\rho^{\alpha\mu}_{(1)}
\Bigg],
}
with
\al{
\stackrel{\leftrightarrow}{\partial}_\mu \left[A B\right]
&\equiv 
\partial_\mu A \cdot B - A \cdot  \partial_\mu B
\,, \\
\left[ \tau^\alpha, \tau^\beta \right] &= i \epsilon^{\alpha\beta\gamma} \tau^{\gamma}\,, \\
\left[ \frac{\lambda^a}{2}, \frac{\lambda^b}{2} \right] &= i f^{abc} \left( \frac{\lambda^c}{2} \right)\,, \\
\left\{ \frac{\lambda^a}{2}, \frac{\lambda^b}{2} \right\} &= \frac{1}{3} \delta^{ab} + d^{abc} \left( \frac{\lambda^c}{2} \right)\,.
}
Here, $\epsilon^{\alpha\beta\gamma}$ is the {$SU(2)_W$} antisymmetric tensor, $f^{abc}$ {(antisymmetric)} and $d^{abc}$ {(symmetric)} describe the $SU(3)_c$ group structure.}

%%%%%%%%%%%%%%%%%%%%%%%%%%%%%%%%%%%%%%%
%%%%%%%%%%%%%%%%%%%%%%%%%%%%%%%%%%%%%%%
%%%%%%%%%%%%%%%%%%%%%%%%%%%%%%%%%%%%%%%
\section{Decomposition of four-fermion currents}
\label{App:ForFermiOP}
%%%%%%%%%%%%%%%%%%%%%%%%%%%%%%%%%%%%%%%
%%%%%%%%%%%%%%%%%%%%%%%%%%%%%%%%%%%%%%%
%%%%%%%%%%%%%%%%%%%%%%%%%%%%%%%%%%%%%%%

{In this Appendix we derive the four-fermion operators 
induced from the HC rho meson exchanges, which are all relevant to 
flavor physics study in the text.}

By integrating out the HC rho mesons coupled to the SM fermions with the $g_L^{{ij}}$ in Eq.{(\ref{couplings:rhoff:II})} and using Eq.(\ref{rho:assign}), the following effective Lagrangian forms are obtained: 
\al{
-\mathcal{L}^{\text{(8)}}_\text{eff} &=
	\left(\sqrt{2}\right)^2 \frac{g_\rho^2 g^{{ij}}_L g^{{kl}}_L}{({M_{\rho^{{\alpha}}_{(8)}}})^2}
	{\Delta^{ik;jl}}
	\left[ (\overline{q}^{{i}}_L \gamma_\mu \tau^{\alpha} T^a q^{{j}}_L)
	       (\overline{q}^{{k}}_L \gamma^\mu \tau^{\alpha} T^a q^{{l}}_L) \right] \notag \\
&\, +
	\left(\frac{1}{\sqrt{2}}\right)^2 \frac{g_\rho^2 g^{{ij}}_L g^{{kl}}_L}{({M_{\rho^0_{(8)}}})^2}
	{\Delta^{ik;jl}}
	\left[ (\overline{q}^{{i}}_L \gamma_\mu T^a q^{{j}}_L)
	       (\overline{q}^{{k}}_L \gamma^\mu T^a q^{{l}}_L) \right], 
	\label{Lagrho8} \\[0.5em]
%%%%%%%%%%%%%%%%%%%%%%%%%%%%%%%%
-\mathcal{L}^{\text{(1)}}_\text{eff} &=
	\left(\frac{1}{2}\right)^2 \frac{g_\rho^2 g^{{ij}}_L g^{{kl}}_L}{({M_{\rho^{{\alpha}}_{(1)}}})^2}
	{\Delta^{ik;jl}}
	\left[ (\overline{q}^{{i}}_L \gamma_\mu \tau^{\alpha} q^{{j}}_L)
	       (\overline{q}^{{k}}_L \gamma^\mu \tau^{\alpha} q^{{l}}_L) \right] \notag \\ 
&\, +
	\left(\frac{1}{2}\right)^2 \frac{g_\rho^2 g^{{ij}}_L g^{{kl}}_L}{({M_{\rho^{{\alpha}}_{(1)}}})^2}
	{\Delta^{ik;jl}}
	\left[ (\overline{l}^{{i}}_L \gamma_\mu \tau^{\alpha} l^{{j}}_L)
	       (\overline{l}^{{k}}_L \gamma^\mu \tau^{\alpha} l^{{l}}_L) \right] \notag \\
&\, +
	\left(\frac{1}{2}\right)^2 \frac{g_\rho^2 g^{{ij}}_L g^{{kl}}_L}{({M_{\rho^{{\alpha}}_{(1)}}})^2}
	\left[ (\overline{q}^{{i}}_L \gamma_\mu \tau^{\alpha} q^{{j}}_L)
	       (\overline{l}^{{k}}_L \gamma^\mu \tau^{\alpha} l^{{l}}_L) \right] \notag \\
%%%%%%%%%
&\, +
	\left(\frac{1}{2\sqrt{3}}\right)^2 \frac{g_\rho^2 g^{{ij}}_L g^{{kl}}_L}{({M_{\rho^{{\alpha}}_{{(1)'}}}})^2}
	{\Delta^{ik;jl}}
	\left[ (\overline{q}^{{i}}_L \gamma_\mu \tau^{\alpha} q^{{j}}_L)
	       (\overline{q}^{{k}}_L \gamma^\mu \tau^{\alpha} q^{{l}}_L) \right] \notag \\ 
&\, +
	\left(\frac{-\sqrt{3}}{2}\right)^2 \frac{g_\rho^2 g^{{ij}}_L g^{{kl}}_L}{({M_{\rho^{{\alpha}}_{{(1)'}}}})^2}
	{\Delta^{ik;jl}}
	\left[ (\overline{l}^{{i}}_L \gamma_\mu \tau^{\alpha} l^{{j}}_L)
	       (\overline{l}^{{k}}_L \gamma^\mu \tau^{\alpha} l^{{l}}_L) \right] \notag \\
&\, +
	\left(\frac{1}{2\sqrt{3}}\right)
	\left(\frac{-\sqrt{3}}{2}\right) \frac{g_\rho^2 g^{{ij}}_L g^{{kl}}_L}{({M_{\rho^{{\alpha}}_{{(1)'}}}})^2}
	\left[ (\overline{q}^{{i}}_L \gamma_\mu \tau^{\alpha} q^{{j}}_L)
	       (\overline{l}^{{k}}_L \gamma^\mu \tau^{\alpha} l^{{l}}_L) \right] \notag \\
%%%%%%%%%
%%%%%%%%%
&\, +
	\left(\frac{1}{4\sqrt{3}}\right)^2 \frac{g_\rho^2 g^{{ij}}_L g^{{kl}}_L}{({M_{\rho^0_{{(1)'}}}})^2}
	{\Delta^{ik;jl}}
	\left[ (\overline{q}^{{i}}_L \gamma_\mu q^{{j}}_L)
	       (\overline{q}^{{k}}_L \gamma^\mu q^{{l}}_L) \right] \notag \\
&\, +
	\left(\frac{-\sqrt{3}}{4}\right)^2 \frac{g_\rho^2 g^{{ij}}_L g^{{kl}}_L}{({M_{\rho^0_{{(1)'}}}})^2}
	{\Delta^{ik;jl}}
	\left[ (\overline{l}^{{i}}_L \gamma_\mu l^{{j}}_L)
	       (\overline{l}^{{k}}_L \gamma^\mu l^{{l}}_L) \right] \notag \\
&\, +
	\left(\frac{1}{4\sqrt{3}}\right)
	\left(\frac{-\sqrt{3}}{4}\right) \frac{g_\rho^2 g^{{ij}}_L g^{{kl}}_L}{({M_{\rho^0_{{(1)'}}}})^2}
	\left[ (\overline{q}^{{i}}_L \gamma_\mu q^{{j}}_L)
	       (\overline{l}^{{k}}_L \gamma^\mu l^{{l}}_L) \right], 
	\label{Lagrho1} \\[0.5em]
%%%%%%%%%%%%%%%%%%%%%%%%%%%%%%%%
-\mathcal{L}^{\text{(3)}}_\text{eff} &=
	\frac{g_\rho^2 g^{{ij}}_L g^{{kl}}_L}{({M_{\rho^{{\alpha}}_{(3)}}})^2}
	\left[ (\overline{q}^{{i}}_L \gamma_\mu \tau^{\alpha} l^{{j}}_L)
	       (\overline{l}^{{k}}_L \gamma^\mu \tau^{\alpha} q^{{l}}_L) \right] +
	\left(\frac{1}{2}\right)^2
	\frac{g_\rho^2 g^{{ij}}_L g^{{kl}}_L}{({M_{\rho^0_{(3)}}})^2}
	\left[ (\overline{q}^{i}_L \gamma_\mu l^{j}_L)(\overline{l}^{k}_L \gamma^\mu q^{l}_L) \right]{,}
	\label{Lagrho3}
}
with $T^a \equiv \lambda^a/2$.
%In the above forms, the interactions via the $V$-$\rho$ mixing are not taken into %account, where part of them inevitably couples with right-handed components.
%We note that in the region where $g_\rho$ is sufficiently greater than the other %gauge couplings, the induced couplings
{Here, the factor $\Delta^{IK;JL}$ describes combinatorics factor, which is defined as}\footnote{{In general, {the effective current-current interaction by the exchange of vector particle} should be defined as
\al{
\mathcal{L} = - \frac{1}{2 M_V^2} J_\mu J^\mu,
}
where $M_V$ means the mass of the vector boson and $J_\mu = \sum_{i,j} \varepsilon_{ij} \overline{f}^i_{R/L} \gamma_\mu f^j_{R/L}$ with charges $\varepsilon_{i,j}$ (ignoring the $SU(3)_C$ and $SU(2)_W$ generators to be shown).
Here, the overall minus sign originate from the vector-boson propagator, and the factor two is introduced to compensate the combinatoric factor of two from expanding the quadratic form of $J_\mu J^\mu$ or deriving corresponding Feynman rules.
In any case, no additional factor of two is found at the corresponding amplitudes.
The factor $\Delta^{{ik;jl}}$ should be introduced to derive correct combinatoric factor in the case that the two fermion bi-linear forms are the same, where no factor of two emerges at the stage of expanding the quadrature, while the factor appears at the stage of deriving Feynman rules.}
}
\al{
{
\Delta^{ik;jl} \left(= \Delta^{ki;jl} = \Delta^{ik;lj} = \Delta^{ki;lj}\right) = 
	\begin{cases}
	1/2 & \text{for } i=k \text{ and } j=l, \\
	1   & \text{for } \text{others}.
	\end{cases}}
	\label{eq:Delta_factor}
} 
With the help of the following relations in \cite{Bhattacharya:2016mcc,Barger:1987nn},
\al{
\delta_{{xy}} \delta_{{zw}} &= \frac{1}{2} \delta_{{xw}} \delta_{{zy}} + 
	\frac{1}{2} \sigma^{{\alpha}}_{{xw}} \sigma^{{\alpha}}_{{zy}}, \\
%%%
\sigma^{{\alpha}}_{{xy}} \sigma^{{\alpha}}_{{zw}} &= \frac{3}{2} \delta_{{xw}} \delta_{{zy}} - 
	\frac{1}{2} \sigma^{{\alpha}}_{{xw}} \sigma^{{\alpha}}_{{zy}}, \\
%%%
T^a_{{xy}} T^a_{{zw}} &= \frac{1}{2} \delta_{{xw}} \delta_{{zy}} - \frac{1}{6} \delta_{{xy}} \delta_{{zw}},
}
we can rewrite the interactions as 
\al{
-\mathcal{L}_\text{eff} &\supset 
	-\left(\mathcal{L}_\text{eff}^{(8)} + \mathcal{L}_\text{eff}^{(1)} + \mathcal{L}_\text{eff}^{(3)}\right) \notag \\
%%%%%%%%
&= 
	C^{[3]}_{q_{{i}} q_{{j}} q_{{k}} q_{{l}}}
	(\overline{q}^{{i}}_L \gamma_\mu \sigma^{{\alpha}} q^{{j}}_L)
	(\overline{q}^{{k}}_L \gamma^\mu \sigma^{{\alpha}} q^{{l}}_L) +
	C^{[3]}_{l_{{i}} l_{{j}} l_{{k}} l_{{l}}}
	(\overline{l}^{{i}}_L \gamma_\mu \sigma^{{\alpha}} l^{{j}}_L)
	(\overline{l}^{{k}}_L \gamma^\mu \sigma^{{\alpha}} l^{{l}}_L) \notag \\
&\, +
	C^{[3]}_{q_{{i}} q_{{j}} l_{{k}} l_{{l}}}
	(\overline{q}^{{i}}_L \gamma_\mu \sigma^{{\alpha}} q^{{j}}_L)
	(\overline{l}^{{k}}_L \gamma^\mu \sigma^{{\alpha}} l^{{l}}_L) +
%%%
	C^{[1]}_{q_{{i}} q_{{j}} q_{{k}} q_{{l}}}
	(\overline{q}^{{i}}_L \gamma_\mu q^{{j}}_L)
	(\overline{q}^{{k}}_L \gamma^\mu q^{{l}}_L) \notag \\
&\, +
	C^{[1]}_{l_{{i}} l_{{j}} l_{{k}} l_{{l}}}
	(\overline{l}^{{i}}_L \gamma_\mu l^{{j}}_L)
	(\overline{l}^{{k}}_L \gamma^\mu l^{{l}}_L) +
	C^{[1]}_{q_{{i}} q_{{j}} l_{{k}} l_{{l}}}
	(\overline{q}^{{i}}_L \gamma_\mu q^{{j}}_L)
	(\overline{l}^{{k}}_L \gamma^\mu l^{{l}}_L),
}
with
\al{
C^{[3]}_{q_{{i}} q_{{j}} q_{{k}} q_{{l}}} &= 
	{\Delta^{ik;jl}} \Bigg\{
	{\frac{1}{2}}
	\left[ \frac{1}{2} \alpha^{{il;kj}} - \frac{1}{6} \alpha^{{ij;kl}} \right] \frac{1}{({M_{\rho^{{\alpha}}_{(8)}}})^2} +
	{\frac{1}{16}} \frac{\alpha^{{ij;kl}}}{({M_{\rho^{{\alpha}}_{(1)}}})^2} +
	{\frac{1}{48}} \frac{\alpha^{{ij;kl}}}{({M_{\rho^{{\alpha}}_{{(1)'}}}})^2}
	\Bigg\}, \\[0.5em]
%%%
C^{[3]}_{l_{{i}} l_{{j}} l_{{k}} l_{{l}}} &=
	{\Delta^{ik;jl}} \Bigg\{
	{\frac{1}{16}} \frac{\alpha^{{ij;kl}}}{({M_{\rho^{{\alpha}}_{(1)}}})^2} +
	{\frac{3}{16}} \frac{\alpha^{{ij;kl}}}{({M_{\rho^{{\alpha}}_{{(1)'}}}})^2}
	\Bigg\}, \\[0.5em]
%%%
C^{[3]}_{q_{{i}} q_{{j}} l_{{k}} l_{{l}}} &=
	{\frac{1}{16}} \frac{\alpha^{{ij;kl}}}{({M_{\rho^{{\alpha}}_{(1)}}})^2} -
	{\frac{1}{16}} \frac{\alpha^{{ij;kl}}}{({M_{\rho^{{\alpha}}_{{(1)'}}}})^2} -
	{\frac{1}{8}} \frac{\beta^{{il;kj}}}{({M_{\rho^{{\alpha}}_{(3)}}})^2} +
	{\frac{1}{8}} \frac{\beta^{{il;kj}}}{({M_{\rho^0_{(3)}}})^2},
	\label{eq:C3qqll} \\[0.5em]
%%%%%%%%%%%%%%
C^{[1]}_{q_{{i}} q_{{j}} q_{{k}} q_{{l}}} &= 
	{\Delta^{ik;jl}} \Bigg\{
	{\frac{1}{2}}
	\left[ \frac{1}{2} \alpha^{{il;kj}} - \frac{1}{6} \alpha^{{ij;kl}} \right] \frac{1}{({M_{\rho^0_{(8)}}})^2} +
	{\frac{1}{48}} \frac{\alpha^{{ij;kl}}}{({M_{\rho^0_{{(1)'}}}})^2}
	\Bigg\}, \\[0.5em]
%%%
C^{[1]}_{l_{{i}} l_{{j}} l_{{k}} l_{{l}}} &=
	{\Delta^{ik;jl}} \Bigg\{
	{\frac{3}{16}} \frac{\alpha^{{ij;kl}}}{({M_{\rho^0_{{(1)'}}}})^2}
	\Bigg\}, \\[0.5em]
%%%
C^{[1]}_{q_{{i}} q_{{j}} l_{{k}} l_{{l}}} &=
	{- \frac{1}{16}}
	\frac{\alpha^{{ij;kl}}}{({M_{\rho^0_{{(1)'}}}})^2} +
	{\frac{3}{8}} \frac{\beta^{{ij;kl}}}{({M_{\rho^{{\alpha}}_{(3)}}})^2} +
	{\frac{1}{8}} \frac{\beta^{{ij;kl}}}{({M_{\rho^0_{(3)}}})^2},
}
where the two factors $\alpha^{ij;kl}$ and $\beta^{ij;kl}$ are defined as
\al{
\alpha^{{ij;kl}} &\equiv g_\rho^2 g_L^{{ij}} g_L^{{kl}}, \\
\beta^{{ij;kl}}  &\equiv g_\rho^2 g_L^{{ij}} (g_L^\dagger)^{{kl}}.
}
{Here, the overall minus sign has a clear nature of vector-boson exchanges, namely, originating from the vector propagator.
However, to avoid clumsy negative signs in the contexts of the Wilson coefficients, we do not include these signs in the definition of the $C$'s.}
Note that all of the coefficients have mass dimensions of two.
The definition,
\al{
q_L^{{i}} = \begin{pmatrix} u_L^{{i}} \\[5pt] d_L^{{i}} \end{pmatrix},\quad
l_L^{{i}} = \begin{pmatrix} \nu_L^{{i}} \\[5pt] e_L^{{i}} \end{pmatrix},
}
immediately lead to the following explicit decompositions,
\al{
(\overline{q}^{{i}}_L \gamma_\mu q^{{j}}_L)
(\overline{l}^{{k}}_L \gamma^\mu l^{{l}}_L)
&=
	(\overline{u}^{{i}}_L \gamma_\mu u^{{j}}_L)
	(\overline{\nu}^{{k}}_L \gamma^\mu \nu^{{l}}_L) +
	(\overline{u}^{{i}}_L \gamma_\mu u^{{j}}_L)
	(\overline{e}^{{k}}_L \gamma^\mu e^{{l}}_L) \notag \\
& +
	(\overline{d}^{{i}}_L \gamma_\mu d^{{j}}_L)
	(\overline{\nu}^{{k}}_L \gamma^\mu \nu^{{l}}_L) +
	(\overline{d}^{{i}}_L \gamma_\mu d^{{j}}_L)
	(\overline{e}^{{k}}_L \gamma^\mu e^{{l}}_L), \\[0.5em]
%%%%%%
(\overline{q}^{{i}}_L \gamma_\mu \sigma^{{\alpha}} q^{{j}}_L)
(\overline{l}^{{k}}_L \gamma^\mu \sigma^{{\alpha}} l^{{l}}_L)
&=
	2 (\overline{u}^{{i}}_L \gamma_\mu d^{{j}}_L)
	  (\overline{e}^{{k}}_L \gamma^\mu \nu^{{l}}_L) +
	2 (\overline{d}^{{i}}_L \gamma_\mu u^{{j}}_L)
	  (\overline{\nu}^{{k}}_L \gamma^\mu e^{{l}}_L) \notag \\
&+
	(\overline{u}^{{i}}_L \gamma_\mu u^{{j}}_L)
	(\overline{\nu}^{{k}}_L \gamma^\mu \nu^{{l}}_L) +
	(\overline{d}^{{i}}_L \gamma_\mu d^{{j}}_L)
	(\overline{e}^{{k}}_L \gamma^\mu e^{{l}}_L) \notag \\
&-
	(\overline{u}^{{i}}_L \gamma_\mu u^{{j}}_L)
	(\overline{e}^{{k}}_L \gamma^\mu e^{{l}}_L) -
	(\overline{d}^{{i}}_L \gamma_\mu d^{{j}}_L)
	(\overline{\nu}^{{k}}_L \gamma^\mu \nu^{{l}}_L), \\[0.5em]
%%%%%%%%%%%%%%%%%%%%%%%%%%%%%%%%%%%
(\overline{q}^{{i}}_L \gamma_\mu q^{{j}}_L)
(\overline{q}^{{k}}_L \gamma^\mu q^{{l}}_L)
&=
	(\overline{u}^{{i}}_L \gamma_\mu u^{{j}}_L)
	(\overline{u}^{{k}}_L \gamma^\mu u^{{l}}_L) +
	(\overline{u}^{{i}}_L \gamma_\mu u^{{j}}_L)
	(\overline{d}^{{k}}_L \gamma^\mu d^{{l}}_L) \notag \\
& +
	(\overline{d}^{{i}}_L \gamma_\mu d^{{j}}_L)
	(\overline{u}^{{k}}_L \gamma^\mu u^{{l}}_L) +
	(\overline{d}^{{i}}_L \gamma_\mu d^{{j}}_L)
	(\overline{d}^{{k}}_L \gamma^\mu d^{{l}}_L), \\[0.5em]
%%%%%%
(\overline{q}^{{i}}_L \gamma_\mu \sigma^{{\alpha}} q^{{j}}_L)
(\overline{q}^{{k}}_L \gamma^\mu \sigma^{{\alpha}} q^{{l}}_L)
&=
	2 (\overline{u}^{{i}}_L \gamma_\mu d^{{j}}_L)
	  (\overline{d}^{{k}}_L \gamma^\mu u^{{l}}_L) +
	2 (\overline{d}^{{i}}_L \gamma_\mu u^{{j}}_L)
	  (\overline{u}^{{k}}_L \gamma^\mu d^{{l}}_L) \notag \\
&+
	(\overline{u}^{{i}}_L \gamma_\mu u^{{j}}_L)
	(\overline{u}^{{k}}_L \gamma^\mu u^{{l}}_L) +
	(\overline{d}^{{i}}_L \gamma_\mu d^{{j}}_L)
	(\overline{d}^{{k}}_L \gamma^\mu d^{{l}}_L) \notag \\
&-
	(\overline{u}^{{i}}_L \gamma_\mu u^{{j}}_L)
	(\overline{d}^{{k}}_L \gamma^\mu d^{{l}}_L) -
	(\overline{d}^{{i}}_L \gamma_\mu d^{{j}}_L)
	(\overline{u}^{{k}}_L \gamma^\mu u^{{l}}_L), \\[0.5em]
%%%%%%%%%%%%%%%%%%%%%%%%%%%%%%%%%%%
(\overline{l}^{{i}}_L \gamma_\mu l^{{j}}_L)
(\overline{l}^{{k}}_L \gamma^\mu l^{{l}}_L)
&=
	(\overline{\nu}^{{i}}_L \gamma_\mu \nu^{{j}}_L)
	(\overline{\nu}^{{k}}_L \gamma^\mu \nu^{{l}}_L) +
	(\overline{\nu}^{{i}}_L \gamma_\mu \nu^{{j}}_L)
	(\overline{e}^{{k}}_L \gamma^\mu e^{{l}}_L) \notag \\
& +
	(\overline{e}^{{i}}_L \gamma_\mu e^{{j}}_L)
	(\overline{\nu}^{{k}}_L \gamma^\mu \nu^{{l}}_L) +
	(\overline{e}^{{i}}_L \gamma_\mu e^{{j}}_L)
	(\overline{e}^{{k}}_L \gamma^\mu e^{{l}}_L), \\[0.5em]
%%%%%%
(\overline{l}^{{i}}_L \gamma_\mu \sigma^{{\alpha}} l^{{j}}_L)
(\overline{l}^{{k}}_L \gamma^\mu \sigma^{{\alpha}} l^{{l}}_L)
&=
	2 (\overline{\nu}^{{i}}_L \gamma_\mu e^{{j}}_L)
	  (\overline{e}^{{k}}_L \gamma^\mu \nu^{{l}}_L) +
	2 (\overline{e}^{{i}}_L \gamma_\mu \nu^{{j}}_L)
	  (\overline{\nu}^{{k}}_L \gamma^\mu e^{{l}}_L) \notag \\
&+
	(\overline{\nu}^{{i}}_L \gamma_\mu \nu^{{j}}_L)
	(\overline{\nu}^{{k}}_L \gamma^\mu \nu^{{l}}_L) +
	(\overline{e}^{{i}}_L \gamma_\mu e^{{j}}_L)
	(\overline{e}^{{k}}_L \gamma^\mu e^{{l}}_L) \notag \\
&-
	(\overline{\nu}^{{i}}_L \gamma_\mu \nu^{{j}}_L)
	(\overline{e}^{{k}}_L \gamma^\mu e^{{l}}_L) -
	(\overline{e}^{{i}}_L \gamma_\mu e^{{j}}_L)
	(\overline{\nu}^{{k}}_L \gamma^\mu \nu^{{l}}_L),
}
where $u$, $d$, $\nu$, $e$ are up-type, down-type quarks, neutrinos, charged leptons, respectively. 
Here we shall extract operators relevant for our discussion in the next stage,
\al{
-\mathcal{L}_\text{eff} &\supset
	\left( C^{[1]}_{q_{{i}} q_{{j}} l_{{k}} l_{{l}}} +
	       C^{[3]}_{q_{{i}} q_{{j}} l_{{k}} l_{{l}}} \right)
	(\overline{d}^{{i}}_L \gamma_\mu d^{{j}}_L)
	(\overline{e}^{{k}}_L \gamma^\mu e^{{l}}_L) +
	\left( C^{[1]}_{q_{{i}} q_{{j}} l_{{k}} l_{{l}}} - 
	       C^{[3]}_{q_{{i}} q_{{j}} l_{{k}} l_{{l}}} \right)
	(\overline{d}^{{i}}_L \gamma_\mu d^{{j}}_L)
	(\overline{\nu}^{{k}}_L \gamma^\mu \nu^{{l}}_L) \notag \\
&+
	2 C^{[3]}_{q_{{i}} q_{{j}} l_{{k}} l_{{l}}} \left(
	(\overline{u}^{{i}}_L \gamma_\mu d^{{j}}_L)
	(\overline{e}^{{k}}_L \gamma^\mu \nu^{{l}}_L) +
	(\overline{d}^{{i}}_L \gamma_\mu u^{{j}}_L)
	(\overline{\nu}^{{k}}_L \gamma^\mu e^{{l}}_L) \right) \notag \\
&+
	\left( C^{[1]}_{q_{{i}} q_{{j}} q_{{k}} q_{{l}}} + 
	       C^{[3]}_{q_{{i}} q_{{j}} q_{{k}} q_{{l}}} \right)
	(\overline{d}^{{i}}_L \gamma_\mu d^{{j}}_L)
	(\overline{d}^{{k}}_L \gamma^\mu d^{{l}}_L) +
	\left( C^{[1]}_{l_{{i}} l_{{j}} l_{{k}} l_{{l}}} + 
	       C^{[3]}_{l_{{i}} l_{{j}} l_{{k}} l_{{l}}} \right)
	(\overline{e}^{{i}}_L \gamma_\mu e^{{j}}_L)
	(\overline{e}^{{k}}_L \gamma^\mu e^{{l}}_L).
	\label{eq:effective_operators}
}
When $(g_L)^{ij}$ takes unsuppressed generic form, as widely known, such possibilities are immediately discarded by severe constraints on flavor-changing neutral current~(FCNC) processes, especially the $K^0$-$\overline{K^0}$ mixing.
We adopt the safety setup in Eq.(\ref{gL33}) adopted in e.g.,~\cite{Bhattacharya:2016mcc},
which leads to the relations,
\al{
\alpha^{ij;kl} = {\beta^{ij;kl}} =
\begin{cases}
g_\rho^2 {(g_L^{33})^2}  &  \text{for } i=j=k=l=3 \\
0  &  \text{for others}
\end{cases}.
%%%%%
%\beta^{ij;kl} &=
%\begin{cases}
%g_\rho^2 |g_L|^2  &  \text{for } i=j=k=l=3 \\
%0  &  \text{for others}
%\end{cases}.
}
Note that the factor $\Delta^{ik;jl}$ corresponds to $1/2$ when $i=j=k=l\,(=3)$.
For clarity, we write down all of the Wilson coefficients in Eq.~(\ref{eq:effective_operators}) when all of the masses of the vector particles are completely degenerated [under the coupling texture of $g_L$ in Eq.(\ref{gL33})],
\al{
C^{[3]}_{qqqq} &= 
	{\frac{1}{2}} \frac{1}{4} \frac{g_\rho^2 {(g_L^{33})^2}}{m_\rho^2},&
%%%
C^{[3]}_{llll} &=
	{\frac{1}{2}} \frac{1}{4} \frac{g_\rho^2 {(g_L^{33})^2}}{m_\rho^2},&
%%%
C^{[3]}_{qqll} &= 0, \notag \\
%%%%%%%%%%%%%%
C^{[1]}_{qqqq} &= 
	{\frac{1}{2}} \frac{3}{16} \frac{g_\rho^2 {(g_L^{33})^2}}{m_\rho^2},&
%%%
C^{[1]}_{llll} &=
	{\frac{1}{2}} \frac{3}{16} \frac{g_\rho^2 {(g_L^{33})^2}}{m_\rho^2},&
%%%
C^{[1]}_{qqll} &=
	\frac{7}{16} \frac{g_\rho^2 {(g_L^{33})^2}}{m_\rho^2},
	\label{eq:coefficients_limit}
}
where $m_\rho$ represents the universal vector boson mass and {the coupling ${g_L^{33}}$ is a real quantity}.
It would be noted that the vanishing condition for $C^{[3]}_{qqll}$ is, concretely speaking, {$M_{\rho^{{\alpha}}_{(1)}} = M_{\rho^{{\alpha}}_{(1)'}}$ and $M_{\rho^{{\alpha}}_{(3)}} = M_{\rho^0_{(3)}}$}.

In the current setup where only $2 \leftrightarrow 3$ flavor transition exists, 
it suffices to consider the effective Hamiltonians for $b \to s \ell^{+} \ell^{-}$, $b \to s \nu \overline{\nu}$, $b \to c \tau^- \overline{\nu}$, $\tau \to \mu s \overline{s}$, $B^0_s \,(=s\overline{b}) \leftrightarrow \overline{B^0_s} \,(=b\overline{s})$, $\tau^- \to \mu^- \mu^+ \mu^-$. 
%\footnote{
%In the main body, we sometimes skip to show the prime symbol for denoting mass %eigenstates for simplicity.
%Hereafter, $I$, $J$, $K$, $L$ discriminate the mass basis.
%}
The ones for the first three categories are
\al{
H_\text{eff}(b \to s e_I \overline{e}_J) &= - \frac{\alpha G_F}{\sqrt{2}\pi} V_{tb} V^\ast_{ts}
	\left[
	C_9^{IJ} (\overline{s'}_L \gamma_\mu b'_L) (\overline{e'}_{I} \gamma^\mu e'_{J}) +
	C_{10}^{IJ} (\overline{s'}_L \gamma_\mu b'_L) (\overline{e'}_{I} \gamma^\mu \gamma_5 e'_{J})
	\right], \\[0.5em]
%%%
H_\text{eff}(b \to s \nu_I \overline{\nu}_J) &= - \frac{\alpha G_F}{\sqrt{2}\pi} V_{tb} V^\ast_{ts}
	C_L^{IJ} (\overline{s'}_L \gamma_\mu b'_L) (\overline{\nu'}_{I} \gamma^\mu (1-\gamma_5) \nu'_{J}), \\[0.5em]
%%%
H_\text{eff}(b \to c \tau_I \overline{\nu}_J) &= + \frac{4 G_F}{\sqrt{2}} V_{cb}
	C_V^{IJ} (\overline{c'}_L \gamma_\mu b'_L) (\overline{e'}_{I} \gamma^\mu \nu'_{J}),
}
where $\alpha$ and $G_F$ are the QED fine structure constant and 
the Fermi constant, respectively; 
the Wilson coefficients include both of the SM and new physics~(NP) contributions: $C_X = C_X(\text{SM}) + C_X(\text{NP})$;  
the coefficients $C_9^{IJ}$, $C_{10}^{IJ}$, $C_L^{IJ}$, $C_V^{IJ}$ have zero mass dimensions.
{Let us remind that, as defined in Eq.(\ref{eq:definition_masseigenstates}), the fermion fields with the prime symbol represent the mass eigenstates, where we skipped to show the symbol throughout almost all the part of the main text for clarity.}
We note that when only the left-type vector interactions exist 
as in the case of the present model, 
the following relation is realized, 
\al{
C_9^{IJ} = - C_{10}^{IJ}.
}
Eq.(\ref{eq:effective_operators}), 
%(\ref{eq:simplified_C}), 
%(\ref{eq:M_transform_1})--(\ref{eq:M_transform_6}) 
thus brings us to the concrete results,
\al{
C_9^{IJ}(\text{NP}) &= {-}
	\frac{\pi}{\sqrt{2} \alpha G_F V_{tb} V^\ast_{ts}}
	\left( C^{[1]}_{qqll} + C^{[3]}_{qqll} \right) {X_{dd}^{23} X_{ll}^{IJ}}
	\left( = - C_{10}^{IJ}(\text{NP}) \right), \\[0.5em]
%%%%
C_L^{IJ}(\text{NP}) &= {-}
	\frac{\pi}{\sqrt{2} \alpha G_F V_{tb} V^\ast_{ts}}
	\left( C^{[1]}_{qqll} - C^{[3]}_{qqll} \right) {X_{dd}^{23} X_{ll}^{IJ}}, \\[0.5em]
%%%%
C_V^{IJ}(\text{NP}) &= {+}
	\frac{1}{2\sqrt{2} G_F V_{cb}}
	\left( 2 C^{[3]}_{qqll} \right)
	\underbrace{\left[ V_{cs} {X_{dd}^{23}} + V_{cb} {X_{dd}^{33}} \right]}_{= {X_{ud}^{23}}} {X_{ll}^{IJ}},
}
{where the coefficients $C_V^{IJ}(\text{NP})$ become zero when the HC vector bosons are completely degenerated [{\it c.f.}~Eq.(\ref{eq:coefficients_limit})].}
The effective Lagrangians which describe the remaining three processes are also discussed in Ref.~\cite{Bhattacharya:2016mcc} as
\al{
\mathcal{L}_\text{eff}(\tau \to \mu s \overline{s})(\text{NP}) &= {-}
	\left( C^{[1]}_{qqll} + C^{[3]}_{qqll} \right) {X_{dd}^{22} X_{ll}^{23}}
	(\overline{\mu'}_L \gamma_\mu \tau'_L) (\overline{s'}_L \gamma^\mu s'_L), \\[5pt]
%%%%
\mathcal{L}_\text{eff}(B^0_s \leftrightarrow \overline{B^0_s})(\text{NP})
&=
	{-}
	\left( C^{[1]}_{qqqq} + C^{[3]}_{qqqq} \right) {(X_{dd}^{23})^2}
	(\overline{s'}_L \gamma_\mu b'_L) (\overline{s'}_L \gamma^\mu b'_L) \notag \\
&=
	{-} \frac{1}{4}
	\left( C^{[1]}_{qqqq} + C^{[3]}_{qqqq} \right) \sin^2{\theta_D} \cos^2{\theta_D}
	(\overline{s'} \gamma_\mu (1-\gamma_5) b') (\overline{s'} (1-\gamma_5) \gamma^\mu b'), \\[5pt]
%%%%
\mathcal{L}_\text{eff}(\tau^- \to \mu^- \mu^+ \mu^-)(\text{NP}) &=
	{-}
	\left( C^{[1]}_{llll} + C^{[3]}_{llll} \right) {X_{ll}^{23} X_{ll}^{22}}
	(\overline{\mu'}_L \gamma_\mu \tau'_L) (\overline{\mu'}_L \gamma^\mu \mu'_L) \notag \\
&=
	{+}
	\left( C^{[1]}_{llll} + C^{[3]}_{llll} \right) \sin^3{\theta_L} \cos{\theta_L}
	(\overline{\mu'}_L \gamma_\mu \tau'_L) (\overline{\mu'}_L \gamma^\mu \mu'_L).
}

%%%%%%%%%%%%%%%%%%%%%%%%%%%%%%%%%%%%%%%%%%%%%%%%%%%%%%%%%%%%%%%%%%%%%%%%%%%%%%%%%%%%%%%%%%
%%%%%%%%%%%%%%%%%%%%%%%%%%%%%%%%%%%%%%%%%%%%%%%%%%%%%%%%%%%%%%%%%%%%%%%%%%%%%%%%%%%%%%%%%%
%%%%%%%%%%%%%%%%%%%%%%%%%%%%%%%%%%%%%%%%%%%%%%%%%%%%%%%%%%%%%%%%%%%%%
\section{Decay widths of $\rho$ mesons \label{appendix:rho_decay_width}}
%%%%%%%%%%%%%%%%%%%%%%%%%%%%%%%%%%%%%%%%%%%%%%%%%%%%%%%%%%%%%%%%%%%%%
%%%%%%%%%%%%%%%%%%%%%%%%%%%%%%%%%%%%%%%%%%%%%%%%%%%%%%%%%%%%%%%%%%%%%%%%%%%%%%%%%%%%%%%%%%
%%%%%%%%%%%%%%%%%%%%%%%%%%%%%%%%%%%%%%%%%%%%%%%%%%%%%%%%%%%%%%%%%%%%%%%%%%%%%%%%%%%%%%%%%%

{In this Appendix}
we summarize the partial decay widths of the $\rho$ mesons.
{Here we ignore the mixing effect of the $\rho$ mesons and the SM vector bosons 
as have been done throughout the present paper {[see Eq.(\ref{eq:rho_mass_splitting_summary})]}.  
We also assume the kinematic relation of $m_\rho < 2 m_\pi$, 
where no decay branch into a pair of pions is possible. 
This situation could be realized in the case of HC with many flavors such as 
the present {one}-family model (See Eq.(\ref{pi:masses})).} 
We use the short-hand notation $\Gamma_0 = g_{\rho L}^2 m_\rho/8\pi$ and we claim that all of the $\rho$ mesons take the universal mass $m_\rho$ by ignoring small mass splittings due to the $\rho$-$V_{\text{SM}}$ mixing. 
We treat all of the final-state fermions as massless particles. 

%%%%%%%%%%%%%%%%%%%%%
\subsection{color-triplet $\rho$}
%%%%%%%%%%%%%%%%%%%%%
\al{
\Gamma(\rho^{+}_{(3)} \to u^{I} \bar{e}^{J}) &= \frac{1}{6} \Gamma_0 \left| X_{ul}^{IJ} \right|^2, \\
\Gamma(\rho^{-}_{(3)} \to d^{I} \bar{\nu}^{J}) &= \frac{1}{6} \Gamma_0 \left| X_{dl}^{IJ} \right|^2, \\
\Gamma(\rho^{3}_{(3)} \to d^{I} \bar{e}^{J}) &= \frac{1}{12} \Gamma_0 \left| X_{dl}^{IJ} \right|^2, \\
\Gamma(\rho^{3}_{(3)} \to u^{I} \bar{\nu}^{J}) &= \frac{1}{12} \Gamma_0 \left| X_{ul}^{IJ} \right|^2, \\
\Gamma(\rho^{0}_{(3)} \to d^{I} \bar{e}^{J}) &= \frac{1}{12} \Gamma_0 \left| X_{dl}^{IJ} \right|^2, \\
\Gamma(\rho^{0}_{(3)} \to u^{I} \bar{\nu}^{J}) &= \frac{1}{12} \Gamma_0 \left| X_{ul}^{IJ} \right|^2,
}
where we note that $(\rho^{+}_{(3)})^\ast \not= (\rho^{-}_{(3)})$.
The electromagnetic charges of $\rho^{+}_{(3)}$, $\rho^{-}_{(3)}$, $\rho^{3}_{(3)}$, $\rho^{0}_{(3)}$ are $+5/3$, $-1/3$ $2/3$, $2/3$, respectively.

%%%%%%%%%%%%%%%%%%%%%
\subsection{color-octet $\rho$}
%%%%%%%%%%%%%%%%%%%%%
\al{
\Gamma(\rho^{0}_{(8)} \to u^{I} \bar{u}^{J}) &= \frac{1}{12} \Gamma_0 \left| X_{uu}^{IJ} \right|^2, \\
\Gamma(\rho^{0}_{(8)} \to d^{I} \bar{d}^{J}) &= \frac{1}{12} \Gamma_0 \left| X_{dd}^{IJ} \right|^2, \\
\Gamma(\rho^{3}_{(8)} \to u^{I} \bar{u}^{J}) &= \frac{1}{12} \Gamma_0 \left| X_{uu}^{IJ} \right|^2, \\
\Gamma(\rho^{3}_{(8)} \to d^{I} \bar{d}^{J}) &= \frac{1}{12} \Gamma_0 \left| X_{dd}^{IJ} \right|^2, \\
\Gamma(\rho^{+}_{(8)} \to u^{I} \bar{d}^{J}) &= \frac{1}{6} \Gamma_0 \left| X_{dd}^{IJ} \right|^2.
}

%%%%%%%%%%%%%%%%%%%%%
\subsection{color-singlet $\rho$}
%%%%%%%%%%%%%%%%%%%%%
\al{
\Gamma(\rho^{3}_{(1)} \to u^{I} \bar{u}^{J}) &= \frac{1}{16} \Gamma_0 \left| X_{uu}^{IJ} \right|^2, \\
\Gamma(\rho^{3}_{(1)} \to d^{I} \bar{d}^{J}) &= \frac{1}{16} \Gamma_0 \left| X_{dd}^{IJ} \right|^2, \\
\Gamma(\rho^{3}_{(1)} \to e^{I} \bar{e}^{J}) &= \frac{1}{48} \Gamma_0 \left| X_{ll}^{IJ} \right|^2, \\
\Gamma(\rho^{3}_{(1)} \to {\nu^{I}} \bar{\nu}^{J}) &= \frac{1}{48} \Gamma_0 \left| X_{ll}^{IJ} \right|^2, \\
%%%%%%%
\Gamma(\rho^{3}_{(1)'} \to u^{I} \bar{u}^{J}) &= \frac{1}{48} \Gamma_0 \left| X_{uu}^{IJ} \right|^2, \\
\Gamma(\rho^{3}_{(1)'} \to d^{I} \bar{d}^{J}) &= \frac{1}{48} \Gamma_0 \left| X_{dd}^{IJ} \right|^2, \\
\Gamma(\rho^{3}_{(1)'} \to e^{I} \bar{e}^{J}) &= \frac{1}{16} \Gamma_0 \left| X_{ll}^{IJ} \right|^2, \\
\Gamma(\rho^{3}_{(1)'} \to {\nu^{I}} \bar{\nu}^{J}) &= \frac{1}{16} \Gamma_0 \left| X_{ll}^{IJ} \right|^2, \\
%%%%%%%
\Gamma(\rho^{0}_{(1)'} \to u^{I} \bar{u}^{J}) &= \frac{1}{48} \Gamma_0 \left| X_{uu}^{IJ} \right|^2, \\
\Gamma(\rho^{0}_{(1)'} \to d^{I} \bar{d}^{J}) &= \frac{1}{48} \Gamma_0 \left| X_{dd}^{IJ} \right|^2, \\
\Gamma(\rho^{0}_{(1)'} \to e^{I} \bar{e}^{J}) &= \frac{1}{16} \Gamma_0 \left| X_{ll}^{IJ} \right|^2, \\
\Gamma(\rho^{0}_{(1)'} \to \nu^{I} \bar{\nu}^{J}) &= \frac{1}{16} \Gamma_0 \left| X_{ll}^{IJ} \right|^2, \\
%%%%%%%
\Gamma(\rho^{+}_{(1)} \to u^{I} \bar{d}^{J}) &= \frac{1}{8} \Gamma_0 \left| X_{ud}^{IJ} \right|^2, \\
\Gamma(\rho^{+}_{(1)} \to \nu^{I} \bar{e}^{J}) &= \frac{1}{24} \Gamma_0 \left| X_{ll}^{IJ} \right|^2, \\
%%%%%%%
\Gamma(\rho^{+}_{(1)'} \to u^{I} \bar{d}^{J}) &= \frac{1}{24} \Gamma_0 \left| X_{ud}^{IJ} \right|^2, \\
\Gamma(\rho^{+}_{(1)'} \to \nu^{I} \bar{e}^{J}) &= \frac{1}{8} \Gamma_0 \left| X_{ll}^{IJ} \right|^2.
}

\bibliographystyle{JHEP}
\bibliography{VL-HC,dynamical_EWSB_strongly,add_ve2}

\end{document}